\documentclass[11pt,a4paper]{article}
\pdfoutput=1
\usepackage[utf8]{inputenc}
\usepackage{jheppub}
\usepackage{amsmath,amsfonts,bm}

\usepackage{verbatim}
\usepackage{graphicx}
\usepackage{color}
\usepackage{bigints}

\usepackage{calligra}
\usepackage[T1]{fontenc}

\newcommand{\beq}{\begin{equation}}
\newcommand{\eeq}{\end{equation}}
\newcommand{\bea}{\begin{eqnarray}}
\newcommand{\eea}{\end{eqnarray}}

\def\OMIT#1{{}}

\newcommand{\remove}[1]{}

\begin{document}

\author[a,b]{Joseph Karpie}
\author[a,b]{, Kostas Orginos}
\author[c]{, Alexander Rothkopf}
\author[d]{and Savvas Zafeiropoulos}
\affiliation[a]{Department of Physics, The College of William \& Mary, Williamsburg, VA 23187, USA}
\affiliation[b]{Thomas Jefferson National Accelerator Facility, Newport News, VA 23606, USA}
\affiliation[c]{Faculty of Science and Technology, University of Stavanger, 4021 Stavanger, Norway}
\affiliation[d]{Institute for Theoretical Physics, Heidelberg University, Philosophenweg 12, 69120 Heidelberg, Germany}
\emailAdd{jmkarpie@email.wm.edu }
\emailAdd{kostas@wm.edu}
\emailAdd{alexander.rothkopf@uis.no }
\emailAdd{ s.zafeiropoulos@thphys.uni-heidelberg.de}

\title{Reconstructing parton distribution functions from Ioffe time data: from Bayesian methods to Neural Networks}
\abstract{
The computation of the parton distribution
functions (PDF) or distribution amplitudes (DA) of hadrons from first principles lattice QCD constitutes a central open problem. In this study, we present and evaluate the efficiency of a selection of methods for inverse problems to reconstruct the full $x$-dependence of PDFs. Our starting point are the so called Ioffe time PDFs, which are accessible from Euclidean time calculations in conjunction with a matching procedure. Using realistic mock data tests, we find that the ill-posed incomplete Fourier transform underlying the reconstruction requires careful regularization, for which both the Bayesian approach as well as neural networks are efficient and flexible choices.}


\maketitle
\flushbottom

\section{ Introduction}
Parton distributions form the core of our theoretical understanding of the inner workings of hadrons~\cite{feynman1998photon}. They encode how the momentum and angular momentum is distributed among  quarks and gluons inside a hadron.  As such there exists an intense high precision experimental program devoted to their determination and it is the concurrent goal of theoretical nuclear physics to compute these quantities from first principles QCD. As they constitute genuinely non-perturbative objects, lattice QCD calculations appear as a very promising route in achieving this goal. 

All current methods for extracting parton distribution functions (PDF) or distribution amplitudes (DA) from lattice QCD require a Fourier transform or a quasi-Fourier transform of a certain class of hadronic position space matrix elements. These novel methods allow lattice QCD calculations to go beyond the traditional computation of the lowest moments~\cite{Best:1997qp, Guagnelli:2004ga, Alexandrou:2017oeh, Oehm:2018jvm}.
 Let us focus on two related methods, called the pseudo-PDF~\cite{Radyushkin:2017cyf} and the quasi-PDF~\cite{Ji:2013dva}.  Both involve the hadron matrix element of space-like separated quark fields connected with a Wilson line and a $\gamma$ matrix. For simplicity, we will use the example of the flavor iso-vector unpolarized quark PDFs. In this case the relevant matrix element 
has the following Lorentz decomposition
\begin{equation}
\langle h(p) | \bar\psi(z) \frac{\tau^3}{2}\gamma^\alpha W(z;0) \psi(0) |h(p)\rangle = p^\alpha \mathcal{M}(\nu,z^2) + z^\alpha \mathcal{N}(\nu,z^2) \, ,
\end{equation}
where $p$ is an arbitrary hadron momentum, $z$ is a space-like separation, $\tau^3$ denotes a flavor Pauli matrix, $\gamma^\alpha$ refers to a gamma matrix acting in spin space, $W(z;0)$ to the $0 \to z$ Wilson line, and $\nu=p\cdot z$ represents a Lorentz invariant quantity known as the Ioffe time. Through a choice of $p$, $z$, and $\alpha$ one is able to isolate the term containing leading twist contributions $\mathcal{M}(\nu,z^2)$~\cite{Orginos:2017kos}. 

Using pseudo-PDF formalism discussed in~\cite{Radyushkin:2018cvn,Zhang:2018ggy,Izubuchi:2018srq}, the real component of this term, $\mathcal{M}_R(\nu,z^2)$, is matched to the $\overline{MS}$ Ioffe time PDF, $\mathcal{Q}_R(\nu,\mu^2)$, via a perturbative kernel. The $\overline{MS}$ Ioffe time PDF on the other hand is related to the valence quark PDF, $q_v(x,\mu^2)$, through the quasi-Fourier transformation
 \begin{equation}
{\mathcal  Q}_R (\nu,\mu^2) \equiv    \int_0^1 dx \,  \cos (\nu x)  \, q_v  (x,\mu^2)  \,,
\label{MC}
 \end{equation}
 as was shown in~\cite{Orginos:2017kos,Karpie:2017bzm}. 
 
Using the quasi-PDF formalism, originally proposed in~\cite{Ji:2013dva}, the lattice calculated $\mathcal{M}_R(\nu,z^2)$ is related to the quasi-PDF, $\tilde q_v(y,p_3)$, via the Fourier transformation
 \begin{equation}
 \mathcal{M}_R ( \nu,\frac{\nu^2}{p_3^2}) \equiv \int_0^\infty d y \, \cos(\nu y) \tilde q_v(y,p_3) \,.
 \label{eq:M_R_FT}
 \end{equation}
The quasi-PDF can then be connected to the PDF using a perturbative kernel as discussed in~\cite{Xiong:2013bka,Stewart:2017tvs,Izubuchi:2018srq}.
Immediately after the presentation of the basic idea a plethora of works \cite{Lin:2014zya, Alexandrou:2015rja, Chen:2016utp, Alexandrou:2018pbm, Zhang:2017bzy, Chen:2017mzz,Broniowski:2017gfp} explored the properties of the new methodologies as well as of other approaches~\cite{Ma:2017pxb,Chambers:2017dov,Liu:1993cv,Bali:2018spj,Sufian:2019bol}. Possible doubts against this approach that emanate from the need to invert the Fourier transform in Eq.~\eqref{eq:M_R_FT} were raised in~\cite{Rossi:2017muf, Rossi:2018zkn}. These claims were refuted in~\cite{Ji:2017rah, Radyushkin:2018nbf, Karpie:2018zaz}, however as we discuss in this paper,
apart from the theoretical issues raised, the numerical inversion of the Fourier transform is not straight forward. 
 We refer the reader to~\cite{Lin:2017snn, Cichy:2018mum} for two detailed reviews of the topic.

This study will be concerned with the reconstruction of the PDF from the $\overline{MS}$ Ioffe time PDF, but many of the conclusions are also applicable to the reconstruction of quasi-PDFs from the lattice calculated matrix element. In inverting  Eq.\eqref{MC} to obtain $q_v  (x,\mu^2) $, there exist two challenges. First being that the required integral over $\nu$  does  not extend over {\it the full Brillouin zone}. The second  is that in practice only a {\it small number of points along $\nu$} can be computed. Only the second challenge exists for the calculation of the quasi-PDF. As we will discuss in more detail below, taken together these issues render the extraction highly ill-posed and we explore several regularization strategies on how to nevertheless reliably estimate the PDF from the data at hand. 


To assess the viability of different reconstruction approaches in practice and to elucidate their systematic uncertainties, we will carry out tests based on two different mock data sets. The first test scenario is based on experimentally determined PDF's for which it has been found that a simple ansatz is able to approximate their functional form quite well
\begin{equation}
p(x) =  \frac{\Gamma(a+b+2)}{\Gamma(a+1) \Gamma(b+1)} x^{a}  (1-x)^{b}\,  . 
\label{eq:p-func}
\end{equation}  
Based on phenomenological fits the expectation is that, for scales $\mu>2$GeV, the physical PDF shows a divergence close to $x=0$, while vanishing at $x=1$. This requires that $a<0$ while $b >0$. In mock scenario A, we insert into Eq.~\eqref{MC} such experimentally determined PDF's, which in turn tests the case $a<0$.

On the other hand at very low scales $\mu^2$, lattice results in the quenched approximation and with heavy pions~\cite{Orginos:2017kos} suggest that $a$ may become positive instead. Thus for mock scenario B we deform by hand the experimental PDF data so that it goes to zero at the origin. This scenario B with its different functional form serves as a gauge of the robustness of the methods we are testing. Also, scenario B is reminiscent of the quasi-PDF case which is known to converge in the limit of $y$ going to zero.

This article starts out with the study of the direct inversion in section 2, followed by the recapitulation of the main ideas of advanced reconstructions in section 3. In section 4 we present extensive numerical experiments of all methods employing mock data and section 5 summarizes our conclusions.

\section{Direct Inversion}

\begin{figure}[t]
\centering
\includegraphics[width=0.48\textwidth]{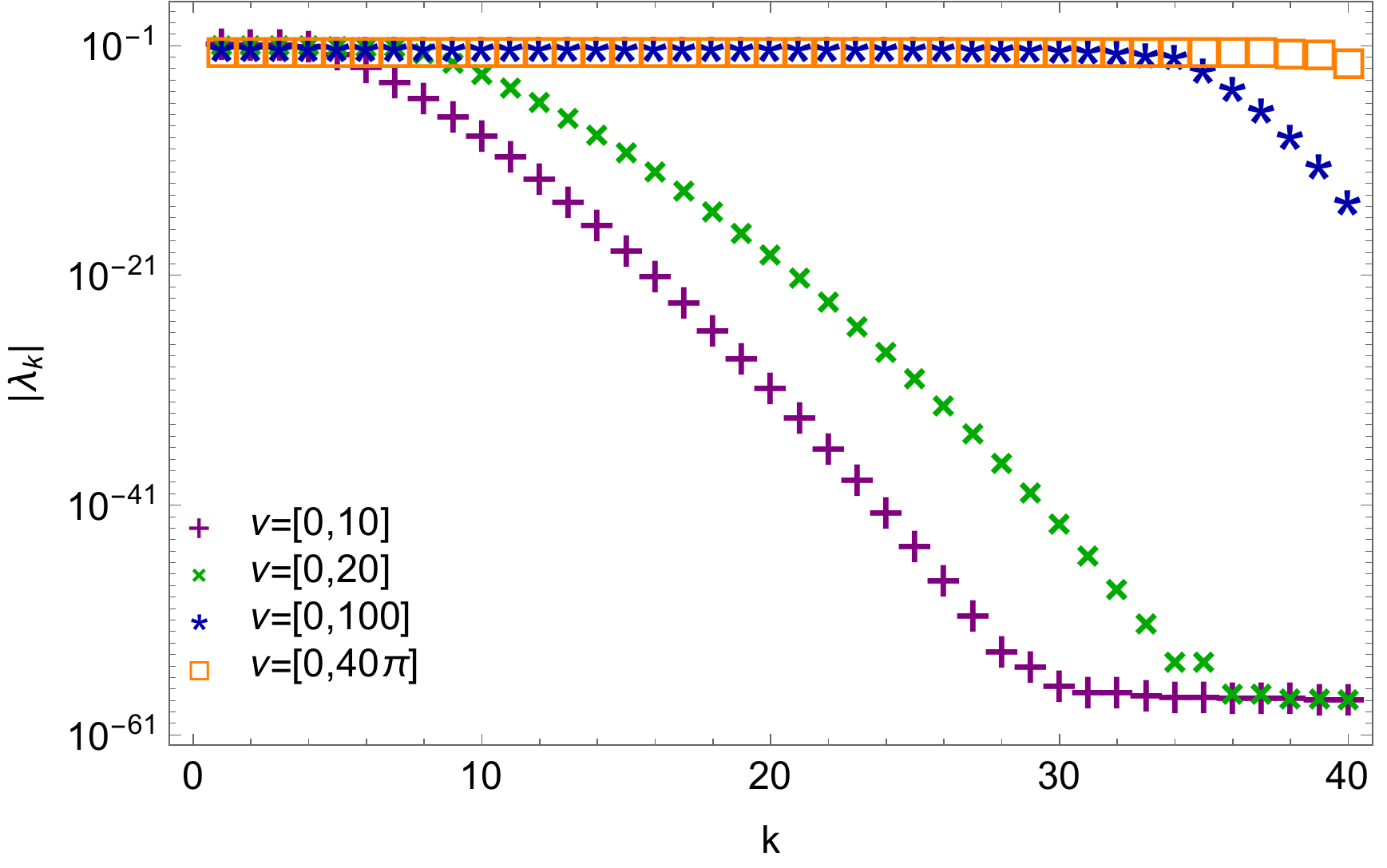}
\includegraphics[width=0.48\textwidth]{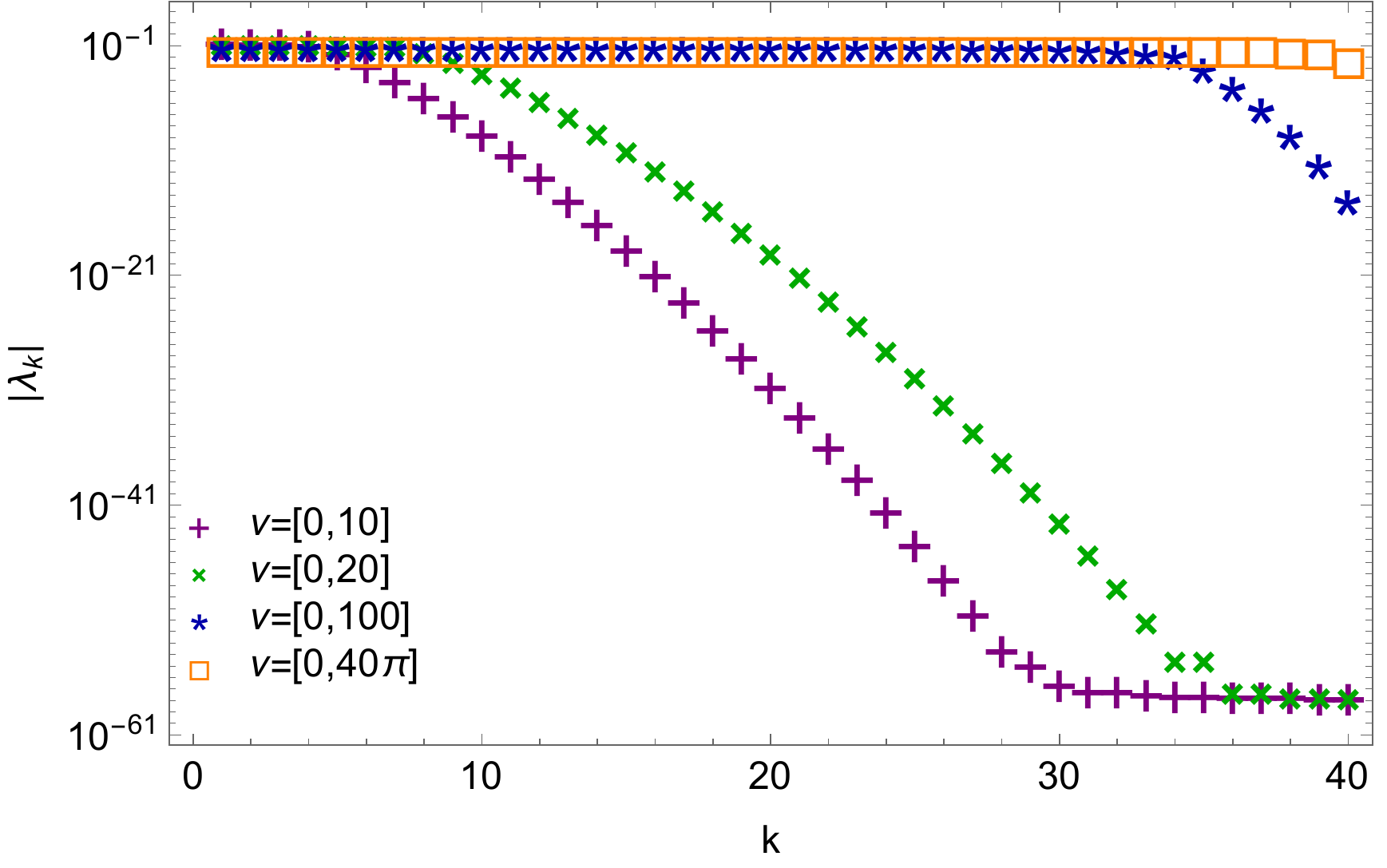}
\caption{Eigenvalues $\lambda_k$ of the Kernel matrix for the direct inversion method on the left and the derivative inversion method on the right calculated with different discretization intervals of $\nu$. Note that only for the case corresponding to a genuine discrete Fourier transform, $\nu=[0,40\pi]$, do all the eigenvalues remain of the same order. All other ranges show exponentially suppressed eigenvalues. The realistic case of $\nu=[0,20]$ already shows a significant degradation of the eigenvalue spectrum.}\label{Fig:Eigenvals}
\end{figure}

While Eq.~\eqref{MC} is a Fourier transform and inverting may be considered a simple task, the facts that the range of $\nu$ extends only over a finite interval different from the full Brillouin zone and that our data for ${\mathcal  Q}_R$ are discrete, makes its inversion highly ill-posed. Let us explore this issue in more detail by naively discretizing the integral in the interval $x\in[0,1]$, considering $N_x+1$ points in a trapezoid integration rule. In that case
 \begin{equation}
 \Delta x = \frac{1}{N_x}\,,\;\; x_k = k \Delta x = \frac{k}{N_x}
 \end{equation}
and
  \begin{equation}
{\mathcal Q}_R (\nu) =  \frac{\Delta x}{2} \cos (\nu x_0)  \, q_v  (x_0)  +    \sum_{k=1}^{N_x-1} \Delta x \,  
\cos (\nu x_k)  \, q_v  (x_k)   +  \frac{\Delta x}{2} \cos (\nu x_{N_x})  \, q_v  (x_{N_x}) \,.
\label{MC-trapezoid}
 \end{equation}
 Now let us further assume that we have an equal amount of $N_\nu+1$ data points for  ${\mathcal Q}_R (\nu)$. Then we can determine exactly the unknown values of the function $ q_v  (x_k)$ by solving a simple linear system of equations. We define the vector ${\mathfrak Q}$ with components
 \begin{equation}
 {\mathfrak  Q}_k = {\mathcal Q}_R (\nu_k) 
 \end{equation}
 where $\nu_k$ are the values of the Ioffe time for which data is available. Also let ${\mathfrak  q}$ be the vector with components the unknown values of $q_v(x_k)$ \textit{i}.\textit{e}.
 \begin{equation}
 {\mathfrak  q}_k = q_v (x_k).
 \end{equation}
 Then Eq.~\eqref{MC-trapezoid} can be written in matrix form as
 \begin{equation}
 {\mathfrak  Q} = {\mathfrak C} \cdot {\mathfrak  q},
 \label{eq:naive_inv}
 \end{equation}
 with ${\mathfrak C} $ being the coefficient matrix with matrix elements,
  \begin{eqnarray}
   {\mathfrak C} _{kl} = \Delta x \, \cos (\nu_k x_l)  = \frac{1}{N_x}  \,   \cos (\nu_k x_l)\;\;  &{\rm for } & \;\; l \in [1,N_x-1], \nonumber \\
    {\mathfrak C} _{kl} = \frac{1}{2} \Delta x \, \cos (\nu_k x_l)  =  \frac{1}{2} \frac{1}{N_x}  \,   \cos (\nu_k x_l)\;\;  &{\rm for } & \;\;  l=0,N_x\; .
  \end{eqnarray}
However, Eq.~\eqref{eq:naive_inv} may be badly conditioned rendering the computation of ${\mathfrak  q}$ a difficult task. 
As a concrete example we take an idealized situation of having in our possession $N_\nu=40$ data points for ${\mathcal Q}_R$ over the four different intervals $I_0=[0,10]$, $I_1=[0,20]$, $I_2=[0,100]$, $I_3=[0, 40\pi]$. Interval $I_3$ is special in the sense that with a redefinition of $x$ one may rewrite Eq.\eqref{MC} in the form of a genuine discrete Fourier transform, which we know is well conditioned. And indeed as shown in Fig.~\ref{Fig:Eigenvals}, for $\nu\in I_3$ the eigenvalues $\lambda_k$ of the matrix ${\mathfrak C}$ are all of the same  order (${\cal O}(0.1)$). Once the resolved $\nu$ region is shrunk below the full Brillouin zone, the ${\rm cosine}$ functions no longer constitute linearly independent basis functions and the columns of ${\mathfrak C}$ become linearly dependent, which manifest itself in eigenvalues that are exactly zero and eigenvalues that are exponentially suppressed.

The exponential decay of the eigenvalues tells us that the inversion problem is ill-conditioned and that a direct inversion will become impractical once the $\nu$ range is significantly smaller than the full Brillouin zone. To make this explicit we carry out a direct inversion of mock data. 


For this illustration we take Eq.~\eqref{eq:p-func} with parameters $a=-\frac{1}{4}$ and $b=3$. Ideal data for ${\mathcal Q}_R$ are obtained based on the three different discretization intervals $I_1,I_2$ and $I_3$. The ideal data are then distorted by Gaussian noise corresponding to constant and uncorrelated  relative errors on the averaged data of $\Delta {\mathcal Q}_R/{\mathcal Q}_R={\rm const.}$ The matrix inverse is computed via a singular value decomposition, where only singular values which are larger than $10^{-4}$ are inverted. 

\begin{figure}[h!]
\includegraphics[scale=0.5]{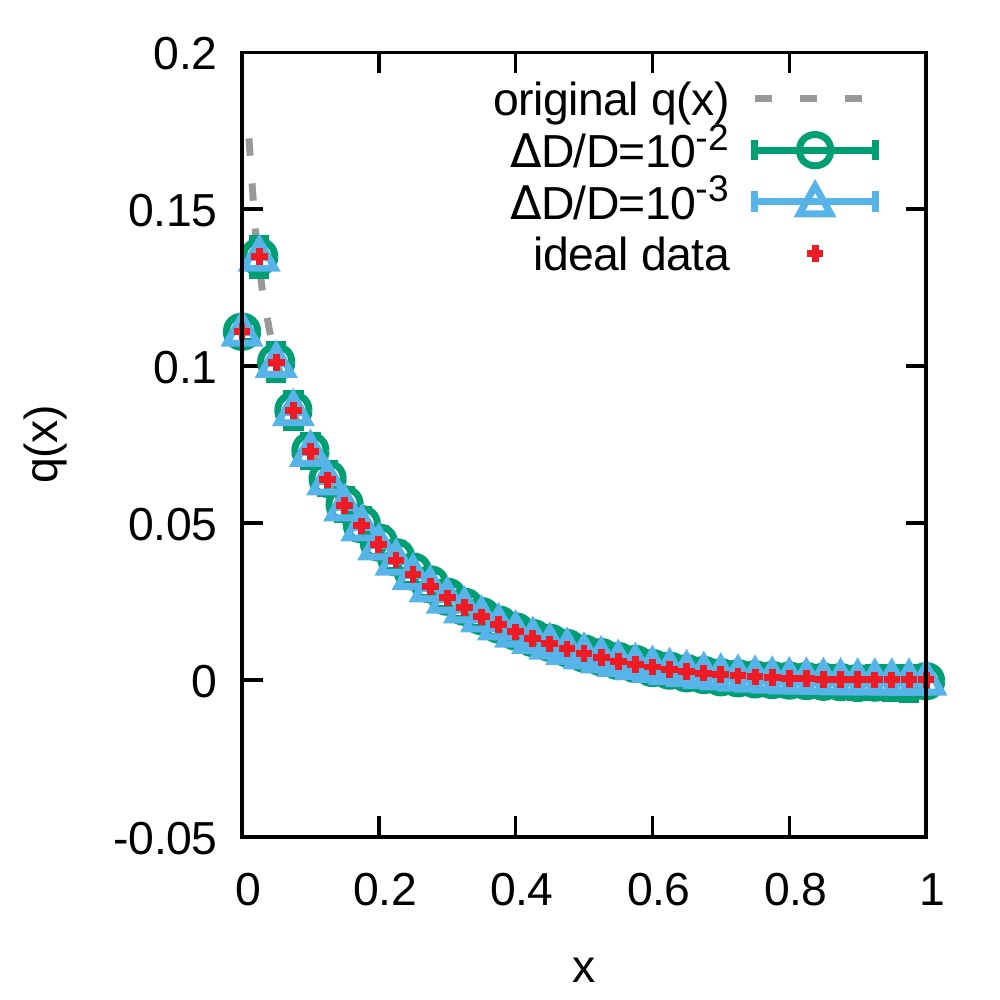}\hspace{-0.5cm}
\includegraphics[scale=0.5]{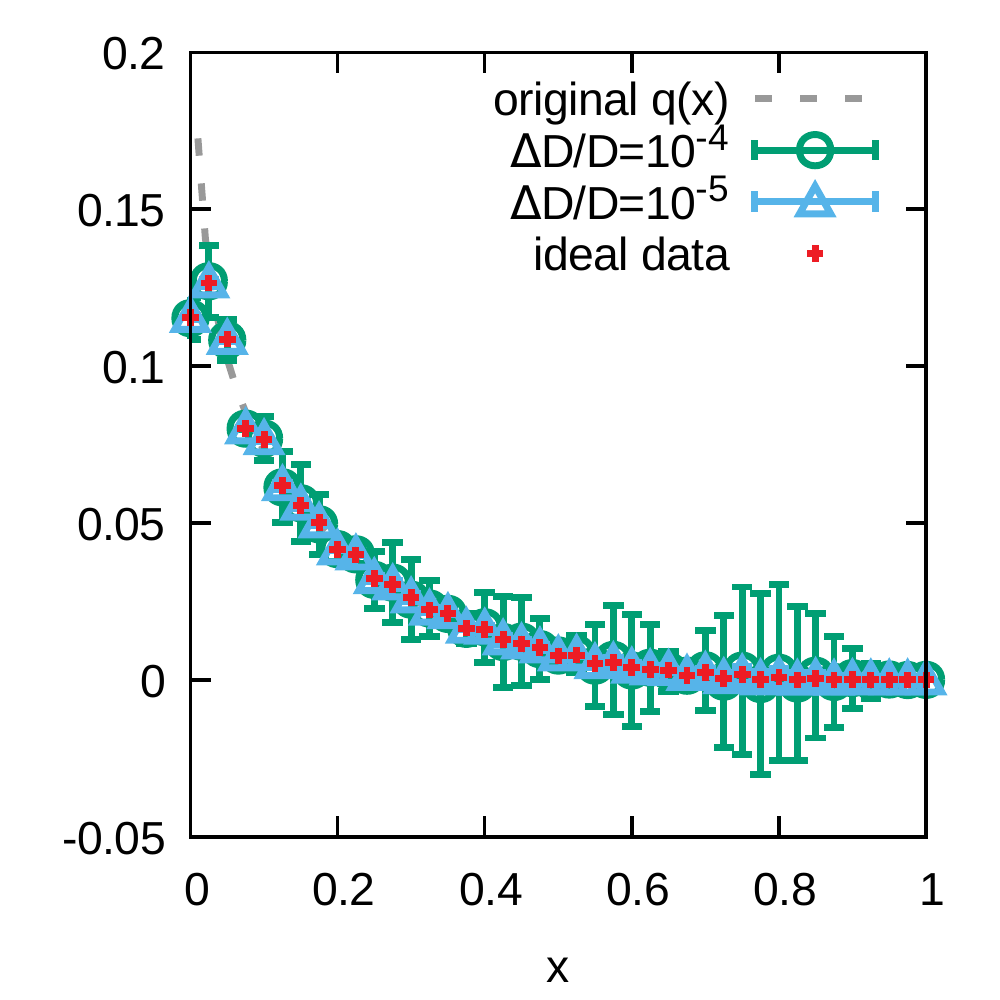}\hspace{-0.5cm}
\includegraphics[scale=0.5]{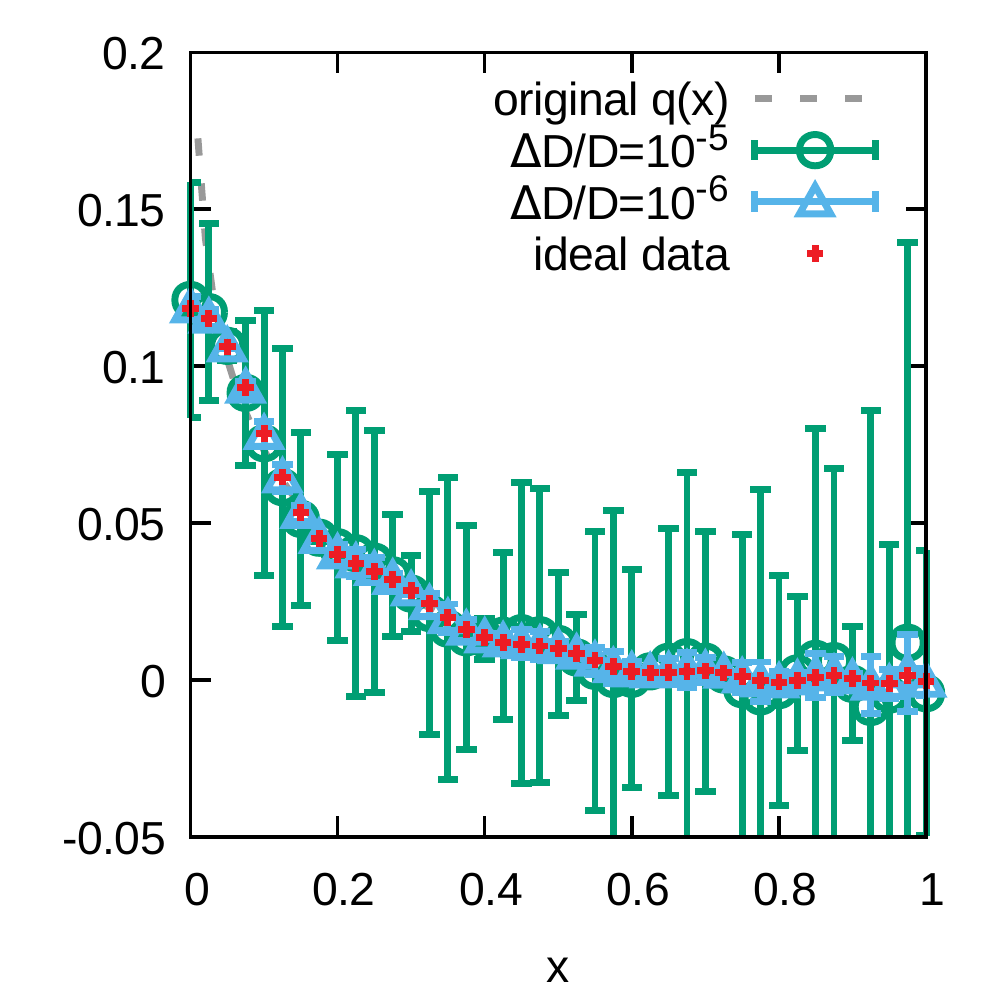}
\caption{Direct inversion results for different discretization intervals of $\nu$ (left $\nu=[0,40\pi]$, center $\nu=[0,100]$, right $\nu=[0,20]$). The matrix inverse is regularized by retaining only singular values of ${\mathfrak C}$ that are larger than $10^{-4}$. Note the different size of the relative errors needed, to obtain a well behaved result (left $\Delta {\mathcal Q}_R/{\mathcal Q}_R =10^{-2}$, center $\Delta {\mathcal Q}_R/{\mathcal Q}_R =10^{-5}$, right $\Delta {\mathcal Q}_R/{\mathcal Q}_R =10^{-6}$).}\label{Fig:DirectInv}
\end{figure}

In Fig.~\ref{Fig:DirectInv} the results of the direct inversion are shown, with the ideal data results depicted by red circles and the original $q(x)$ as gray dashed line. The leftmost panel corresponds to the well conditioned case of $I_3$, where the reconstruction based on ideal data works flawlessly and even produces accurate results already for a relative large error on the input data of $\Delta {\mathcal Q}_R/{\mathcal Q}_R =10^{-2}$. One exception is the point at $x=0$, which formally would have to diverge. 

The shorter the $\nu$ interval becomes the worse the reconstruction results, where even for ideal data having an error on the level of standard machine precision we obtain artificial fluctuations of the reconstructed PDF. At the same time, significantly smaller relative errors on the input data is required for a reasonably stable reconstruction to ensue. For $\nu=[0,100]$ we already need $\Delta {\mathcal Q}_R/{\mathcal Q}_R =10^{-5}$, while for $\nu=[0,20]$ even the $\Delta {\mathcal Q}_R/{\mathcal Q}_R =10^{-6}$ result only gives an approximate PDF with significant unphysical fluctuations.

One may attempt to improve the results of the direct inversion by considering a higher order integration scheme. At large values of $\nu$, the integrand is a highly oscillatory function due to the presence of $\cos(\nu x)$. As a result, an integration algorithm that approximates the integrand by a low degree polynomial is bound to fail at large values of $\nu$. We can improve on this by designing a better integration rule that performs similarly for all values of $\nu$. We know that the oscillatory nature of the integrand is due to the $\cos(\nu x)$ term and that the unknown function $q_v(x)$ is slowly varying in the interval $[0,1]$.
Therefore we will approximate $q_v(x)$ with a linear interpolation and perform the integral exactly leaving the result to linearly depend on the unknown values of the function  $q_v(x)$ on the grid points $x_k$.
A linear interpolation $f(x)$  for the function $q_v(x)$ is given by
\begin{equation}
f(x) =  \frac{x - x_k}{x_{k+1} - x_k}  q_v(x_{k+1})  +   \frac{x_{k+1} - x}{x_{k+1} - x_k}  q_v(x_k) \;\; {\rm for} \; \; x \in [x_k,x_{k+1}] \,.
\end{equation}
In order to compute the exact integral of the interpolating function we need to define
\begin{eqnarray}
I_0(a,b,\nu) &=& \int_a^b dx \, \cos(\nu x)  = \frac{1}{\nu} \left[\sin(\nu b) - \sin(\nu a)\right ], \nonumber \\
I_1(a,b,\nu) &=& \int_a^b dx\, x\, \cos(\nu x)  = \frac{1}{\nu} \left[b\sin(\nu b) - a\sin(\nu a)\right ] +  \frac{1}{\nu^2} \left[\cos(\nu b) - \cos(\nu a)\right ].
\end{eqnarray}
Note that both integrals are finite for $\nu=0$,
\begin{eqnarray}
I_0(a,b,0) &=& b - a, \nonumber \\
I_1(a,b,0) &=& \frac{1}{2}\left[ b^2 - a^2\right].
\label{eq:I_lims}
\end{eqnarray}
With these results at hand we can now write down the improved integration rule,
\begin{equation}
{\mathcal Q}_R (\nu) = \int_0^1 dx \, \cos(\nu x) q_v(x) \approx \int_0^1 dx \, \cos(\nu x) f(x).
\end{equation}
The approximate integral is now a sum of integrals on the intervals $ [x_k,x_{k+1}]$.
To simplify expressions we introduce the notation
\begin{equation}
 I_i(k,\nu)  \equiv I_i(x_k,x_{k+1},\nu)  \;\; {\rm for}\;\; i=0,1\,.
\end{equation}
With this notation we have
\begin{eqnarray}
{\mathcal Q}_R (\nu)  \approx   \sum_{k=0}^{N_x-1}  \left[  \frac{I_1(k,\nu) - x_k\,I_0(k,\nu) }{x_{k+1} - x_k}  q_v(x_{k+1})  
+   \frac{x_{k+1}\,I_0(k,\nu) - I_1(k,\nu) }{x_{k+1} - x_k}  q_v(x_{k}) 
\right],\,
\end{eqnarray}
or
\begin{eqnarray}
{\mathcal Q}_R (\nu)  &\approx &  \frac{x_{1}\,I_0(0,\nu) - I_1(0,\nu) }{x_{1} - x_0}  q_v(x_{0})  +\nonumber \\
&+&\sum_{k=1}^{N_x-1}  \left[  \frac{I_1(k-1,\nu) - x_{k-1}\,I_0(k-1,\nu) }{x_{k} - x_{k-1}}  +  \frac{x_{k+1}\,I_0(k,\nu) - I_1(k,\nu) }{x_{k+1} - x_k}  \right] q_v(x_{k})  \nonumber \\
&+ &  \frac{I_1(N_x-1,\nu) - x_{N-1}\,I_0(N-1,\nu) }{x_{N} - x_{N-1}}  q_v(x_{N}) \,.
\end{eqnarray}
Note that for $\nu=0$ the above expression simplifies to the trapezoid rule for the parton density $q_v(x)$.

Using the same notation as before, we can now compute the matrix elements of the  coefficient matrix ${\mathfrak C} $ as
  \begin{eqnarray}
   {\mathfrak C} _{lk} = \frac{I_1(k-1,\nu_l) - x_{k-1}\,I_0(k-1,\nu_l) }{x_{k} - x_{k-1}}  +  \frac{x_{k+1}\,I_0(k,\nu_l) - I_1(k,\nu_l) }{x_{k+1} - x_k}  \;\;  &{\rm for } & \;\; k \in [1,N-1], \nonumber \\
    {\mathfrak C} _{lk} =    \frac{x_{1}\,I_0(0,\nu_l) - I_1(0,\nu_l) }{x_{1} - x_0}   &{\rm for } & \;\;  k=0, \nonumber \\
      {\mathfrak C} _{lk} =    \frac{I_1(N-1,\nu_l) - x_{N-1}\,I_0(N-1,\nu_l) }{x_{N} - x_{N-1}}  &{\rm for } & \;\;  k =N,
  \end{eqnarray}

We have tested this improved integration method and compared it to the trapezoid rule. It turns out that the relative integration error for typical functions $q_v(x)$ with 10 interpolating points 
continuously grows with $\nu$ for the trapezoid rule and reaches 100\%  at $\nu=15$. On the other hand the improved integration, which performs similarly with the trapezoid rule at small $\nu$,
at $\nu=15$ has 65\% relative error which remains almost unchanged up to $\nu=100$. The improved integration achieves a constant relative error  as a function of $\nu$ at large $\nu$
for any number of interpolating points. On the other hand the trapezoid rule has a relative error that always increases with $\nu$ for any number of interpolating points.
Therefore, this improved integration scheme achieves its design goal of having a constant integration error for all values of $\nu$.
We could further improve the integration by using a second order interpolation formula therefore producing a generalized Simpson's rule. Perhaps this is needed because the number of points we have is very limited. However, it may be that the largest systematic in our problem is the truncation at relatively small values of $\nu$ and further improvement of the integration scheme will not affect this systematic. The proposed integration scheme can prove valuable because it can significantly reduce the number of points required to discretize the integral resulting a smaller maximum value of $\nu$ for which the problem is no longer ill-defined. 

It has been proposed in \cite{Lin:2017ani}, that the unphysical oscillations in the related quasi-PDF inverse problem can be controlled by considering the derivative of the integral equation with respect to $\nu$ or $z_3$. Even if the derivative method results in a smoother curve, it does not alleviate the ill-posed nature of the problem. Assuming that the derivative can be calculated explicitly or accurately, Eq.~\eqref{MC} becomes

\begin{equation}
\partial \mathcal{Q}(\nu) = - \int_0^1 dx x \sin(\nu x) q(x),
\end{equation}
which can be discretized as before to
\begin{equation}
\partial {\mathfrak  Q} = (\partial {\mathfrak C}) \cdot {\mathfrak  q},
 \end{equation}
 with $\partial {\mathfrak C} $ being the coefficient matrix with matrix elements,
  \begin{eqnarray}
   \left[\partial{\mathfrak C}\right]_{kl} = -\Delta x \, x_l \, \sin (\nu_k x_l)  = - \frac{1}{N_x}  \, x_l  \, \sin (\nu_k x_l)\;\;  &{\rm for } & \;\; l \in [1,N_x-1], \nonumber \\
   \left[\partial {\mathfrak C}\right]_{kl} = -\frac{1}{2} \Delta x \, x_l \, \sin (\nu_k x_l)  =  -\frac{1}{2} \frac{1}{N_x}  \, x_l  \, \sin(\nu_k x_l)\;\;  &{\rm for } & \;\;  l=0,N_x\; .
  \end{eqnarray}
Using the same intervals of $\nu$ and x mentioned above, one can find that the eigenvalues of this matrix, shown in Fig.~\ref{Fig:Eigenvals}, have the same pattern of exponential decay which characterizes an ill-conditioned inverse problem.

Our numerical experiments indicate  that for realistic scenarios encountered in lattice studies, for which we foresee $\nu\in[0,20]$ and $\Delta  {\mathcal Q}_R/ {\mathcal Q}_R=10^{-2}-10^{-3}$, a direct inversion is not feasible and more sophisticated regularization schemes of the ill-defined inversion problem are needed. In particular we have observed that the truncation in $\nu$, results in a reconstructed PDF with unphysical fluctuations. This is an effect similar to that observed in~\cite{Lin:2017ani}. In addition we also see that any non-analytic behavior of $q(x)$, e.g. a divergence at $x=0$ will pose a difficulty for (direct) inversion methods and needs to be considered with care. Therefore in the main part of this paper we will explore
several modern methods for treating inverse problems and compare their efficiency in dealing with the uncertainties introduced by the truncation of the integration regime.

\section{Advanced PDF reconstructions}

The fundamental difficulty of solving an ill-posed inverse problem lies in the fact that the input data by themselves do not single out a unique answer. To give meaning to the PDF extraction the inversion needs to be regularized, i.e. we need to provide criteria on how to choose a single PDF from an infinite number of possible solutions reproducing the discrete and noisy input.

Na\"ive methods, such as the  direct inversion and the derivative method, introduce a regularization by removing small singular values of the kernel or by setting data to zero at large distances. However, these approaches  introduce uncontrolled systematic errors. A straightforward example of a stable solution to the inverse problem, which was used in~\cite{Orginos:2017kos,Karpie:2017bzm}, is to parametrize the solution with a functional form, such as~\eqref{eq:p-func}, containing a small number of parameters. One can then use this functional form to fit the data using a $\chi^2$ minimization. 
This approach typically results in smaller statistical errors, however, it introduces a model systematic that one may control by varying the functional form used in the model. As the flexibility of the parametrization increases the model systematic is reduced  and the ill-posed nature of the inverse problem becomes more apparent. 
More sophisticated methods often use some additional prior information to constrain the results in order to improve the robustness of the result and provide  controlled systematic error estimates.

In the following, we discuss three extraction methods that do not presuppose a functional form of the encoded PDFs or have a very flexible functional form parametrizing the solution. Each of these methods regulates the inverse problem  in different ways. The first, the Backus-Gilbert method, regulates the problem by minimizing the statistical variance of the solution. The second, a neural network parametrization, that provides a flexible parametrization given by specific  choices of network geometry and activation functions. The third class of methods, is  based on Bayesian inference, that rely on Bayes theorem to systematically incorporate prior information, such as e.g. positivity or  smoothness of the solution. In addition, the Bayesian inference  methods introduce a default model, which represent the prior information in that  this model is the correct solution in the absence of any data. Non Bayesian approaches, while not explicitly mentioning prior information often incorporate additional knowledge about the problem at hand in an indirect fashion. One such way is to apply a preconditioning to the inverse problem, in case that a relatively good guess of the form of the solution is available. In a Bayesian method such information would instead be supplied in the form of a default model.

Let us discuss the preconditioning, which we will employ in order to improve the results of the Backus-Gilbert and neural network methods. The main ingredient is to define a rescaled kernel and rescaled PDF $h(x)$
\begin{equation}
K_j(x) \equiv cos(\nu_j x)p(x) \;\;{\rm and \,,}\; h(x)\equiv \frac{q_v(x)}{p(x)}\,,\label{eq:precond}
\end{equation}
where $p(x)$ corresponds to an appropriately chosen function that makes the problem easier to solve. In particular, we wish to incorporate into $p(x)$ most of the non-trivial structure of $q(x)$, such that $h(x)$ is a slowly varying function of $x$ and contains only the deviation of $q(x)$ from $p(x)$. 

As discussed in the introduction, it has been found that the Ansatz Eq.~\eqref{eq:p-func} describes phenomenological PDFs rather well. Integrated over the cosine, it yields the function
\begin{align}
\mathcal{Q}_p(\nu,a,b)&\!=\!\!\pi  2^{-a-b-1} \Gamma (a+1) \Gamma (b+1) {}_2\tilde{F}_3\!\left(\frac{a+1}{2},\frac{a+2}{2};\frac{1}{2},\frac{1}{2}
   (a+b+2),\frac{1}{2} (a+b+3);-\frac{\nu^2}{4}\right)\label{Eq:FitFunc}
\end{align}
where ${}_2\tilde{F}_3$ is the generalized hypergeometric function. One can find a good choice for the parameters $a$ and $b$ by first fitting the Ioffe-time PDF data with Eq.~\eqref{Eq:FitFunc}.  
The choice of preconditioning function or default model will influence the result of the reconstruction and it is therefore necessary to explore the stability of the end result on that choice.


\subsection{Backus-Gilbert Method}

One approach to inverse problems, which has been used in a number of engineering and physics applications, is the Backus-Gilbert method~\cite{Backus:1968hl,Press:1992zz,Brandt:2015sxa,Tripolt:2018xeo}. Like many approaches to the inverse problem, the Backus-Gilbert method provides a unique solution to the ill posed problem given some condition. This method differs from other regularizations, which impose a smoothness criterion on the resulting function $q(x)$, by imposing a minimization condition of the variance of the resulting function.  The Backus-Gilbert method has been studied previously in~\cite{Liang:2017mye} as a solution to an inverse problem for extracting PDFs from other types of lattice calculated hadronic matrix elements.

Let us start from the preconditioned expression~\eqref{eq:precond} with a rescaled PDF $h(x)$ that is only a slowly varying function of $x$. Hence our inverse problem becomes
\begin{equation}
d_j \equiv  {\mathcal Q}_R (\nu_j)  = \int_0^1 dx K_j(x) h(x)\,.
\end{equation}
The Backus-Gilbert method introduces a function $\Delta(x-\bar x)$ that is written as 
\begin{equation}
\Delta(x-\bar x) = \sum_j a_j(\bar x) K_j(x) \,,
\label{eq:BG_delta}
\end{equation}
where $a_j(\bar x)$ are unknown functions to be determined.
It then estimates the unknown function as a linear combination of the data,
\begin{equation}
\hat h(\bar x) = \sum_j a_j(\bar x) d_j \,,  \quad {\rm or}\quad\; \hat q_v(\bar x) = \sum_j a_j(\bar x) d_j p(\bar x) \,.
\label{eq:BGsol}
\end{equation}
Given the above definitions, if $\Delta(x-\bar x)$ were to be a Dirac $\delta$ function then $\hat h(\bar x)$ would be equal to $h(\bar x)$. If $\Delta(x-\bar x)$ approximates a $\delta$-function with a width $\sigma$, then the smaller $\sigma$ is, the better the approximation of $\hat h(x)$ to $h(x)$. In other words, if $\hat h_\sigma(x)$ is the approximation resulting from $\Delta(x)$ with a width $\sigma$ then
\begin{equation}
\lim_{\sigma\rightarrow 0} \hat h_\sigma(x) = h(x)\,.
\end{equation}
With this in mind the Backus-Gilbert method minimizes the width $\sigma$ given by
\begin{equation}
\sigma = \int_0^1 dx (x - \bar x)^2 \Delta(x - \bar x)^2\,.
\end{equation}

Note that other choices for the definition of the width can be used. This choice makes the resulting integrals easy to compute and the minimization problem becomes quadratic in the unknown values $a_j$.
Furthermore, if $\Delta(x)$ is to approximate a $\delta$-function then one has to impose the constraint
\begin{equation}
\int_0^1 dx\, \Delta(x-\bar x) = 1\,.
\label{eq:BG_delta_norm}
\end{equation}
Using a Lagrange multiplier $\lambda$ one can minimize the width and impose the constraint by minimizing
\begin{equation}
\chi[a] =  \int_0^1 dx (x - \bar x)^2 \sum_{j,k} a_j(\bar x) K_j(x) K_k(x) a_k(\bar x)  + \lambda \int_0^1 dx  \sum_j K_j(x) a_j(\bar x) \,.
\end{equation}
By setting the derivative to zero with respect to $ a_j(\bar x)$ we get
\begin{equation}
\frac{\partial \chi[a]}{\partial  a_j(\bar x)} = 2 \int_0^1 dx (x - \bar x)^2 \sum_{k}  K_j(x) K_k(x) a_k(\bar x) +  \lambda \int_0^1 dx  K_j(x) = 0 \,,
\end{equation}
which results to
\begin{equation}
 \int_0^1 dx (x - \bar x)^2 \sum_{k}  K_j(x) K_k(x) a_k(\bar x) = - \frac{1}{2}  \lambda \int_0^1 dx K_j(x)  \,.
\end{equation}
Let's now define the matrix $\bf M$  with matrix elements
\begin{equation}
M_{jk} =  \int_0^1 dx (x - \bar x)^2  K_j(x) K_k(x)\,,
\end{equation}
 the vector ${\bf  u}$  with components 
\begin{equation}
 {u }_j = \int_0^1 dx\, K_j(x) \,,
\end{equation}
and promote $a_j(\bar x)$ to a vector $\bf a$. With these definitions the minimization condition takes the matrix equation form
\begin{equation}
{\bf M} {\bf a} = -  \frac{1}{2} \lambda  {\bf  u} 
\end{equation}
or
\begin{equation}
 {\bf a} =  - \frac{1}{2} \lambda  {\bf M}^{-1}  {\bf  u}\,.
\end{equation}
Imposing the normalization condition for $\Delta(x)$ we get
\begin{equation}
\lambda =  - 2 \left[ {\bf u}^T {\bf M}^{-1} {\bf u} \right]^{-1}\,,
\end{equation}
therefore,
\begin{equation}
 {\bf a} =  \frac{ 1}{ {\bf u}^T {\bf M}^{-1} {\bf u} } {\bf M}^{-1}  {\bf  u}\,. 
\end{equation}
We can now obtain an estimate of  the unknown function using Eq.~\eqref{eq:BGsol}. We can also compute the width of the function $\Delta(x)$ which is otherwise known as the resolution function.
The width  of the distribution $\Delta(x-\bar x)$ is given by
\begin{equation}
\sigma(\bar x) = {\bf a}^T {\bf M} {\bf a}\,.
\end{equation}
Note that the width is dependent on $\bar x$ because $\bf a$ depends on $\bar x $ and represents the resolution of the method at $x = \bar x$. Features of the unknown function at
scales shorter than $\sigma(\bar x) $ will not be resolved and thus the method works better when the unknown function is smooth. 
This approach  is known to fail when the matrix $\bf M$ becomes singular, and thus various techniques for regularizing the inversion of $\bf M$ have been proposed. As mentioned before, minimization of the variance of the resulting solution leads to a regularization of the matrix $\bf M$ with the covariance matrix of the data. 
This is implemented by the substitution 
\begin{equation}
\bf M \rightarrow \bf M + \rho \bf C\,,
\end{equation}
where $\bf C$ is the covariance matrix of the data and $\rho$ is a small free parameter. The larger $\rho$ the better conditioned the matrix $\bf M$ and the lower the resolution of the method is. A similar method is the Tikhonov regularization~\cite{Ulybyshev:2017szp, Ulybyshev:2017ped}, where one makes the substitution
\begin{equation}
\bf M \rightarrow \bf M + \rho {\bf 1}\,,
\end{equation}
where $\bf 1$ is the identity matrix and $\rho$ a small adjustable parameter. In this case all the singular values of $\bf M$ are lifted resulting in a well-defined matrix $\bf M$.
However, the Tikhonov regularization does not minimize the variance of the resulting solution of the inverse problem as the covariance matrix regularization does. Again larger values of $\rho$ result in better regulated matrices at the cost of reduced resolution.

Another regularization method similar to Tikhonov is the method we adopted for our experiments. We just project out the singular vectors with singular values smaller than a given cut-off  $\rho$ which we choose. For our problem we have noticed that the Tikhonov regularization and the approach we chose result in nearly identical solutions of the inverse problem.
Furthermore, the covariance matrix regularization for appropriate values of the parameter $\rho$ produces very similar results as the other two methods for our problem. 
Therefore, we decided to use our simple approach for regulating the matrix $\bf M$ and present results of our experiments using this SVD cutoff method.

The Backus-Gilbert method provides a unique solution to the inverse problem demanding that the solution maximizes a stability criterion. The method has a tunable free parameter, $\rho$, which provides a trade off between stability and resolution. This freedom can lead to a bias from the user, but the Backus-Gilbert method provides the variance and the resolution function as quantitative measures to assure a proper analysis is performed. It should be noted that the forms of preconditioning functions, $p(x)$, are restricted in the Backus-Gilbert method. The preconditioning function must be chosen such that integrals of the preconditioned kernels which define $\bf{M}$ and $\bf{u}$ remain finite. Furthermore, it should be noted that a good preconditioning function makes the remainder smooth, as a result the demand for a small resolution is smaller the better the preconditioning is.

\subsection{Neural Network Reconstruction}

 Neural networks can be used as a rather flexible parametrization of functions and thus have been used in the literature to address various inverse problems. Neural networks do not provide any explicit constraints such as a smoothness condition or minimization of a variance, as  the Backus-Gilbert method does but instead it is the depth and structure of their layers that limits what functional forms may be encoded. Hence, neural networks provide a very general parametrization of the unknown function, without forcing a model dependent form such as Eq.~\eqref{Eq:FitFunc}, for a statistical regression. Neural networks have e.g.\ been used in the literature to extract PDFs from experimental data. The pioneering work of~\cite{Forte:2002fg} is now one of the established approaches in obtaining PDFs from cross section data by the NNPDF collaboration~\cite{Ball:2017nwa,Ball:2014uwa,Ball:2010de,Ball:2010gb}.
In this work, we explore the use of neural networks in solving the inverse problem at hand. The specific neural network implementation used in this work is known as a multilayer feedforward neural network.

A neural network is composed of a system of an interconnected nodes, called the neurons. The connection between neuron pairs is called the synapsis. The output of each neuron, called the activation, is  typically a non-linear  transformation of its input,  which is called the activation function of that neuron. The input for each neuron  is a weighted  sum  of the activations of the connected neurons shifted by some real number, which is called the threshold or bias of the neuron. The weights and the thresholds are the free parameters which will be determined by some process which is called the training procedure.

\begin{figure}[h!]
\centering\includegraphics[scale=0.6]{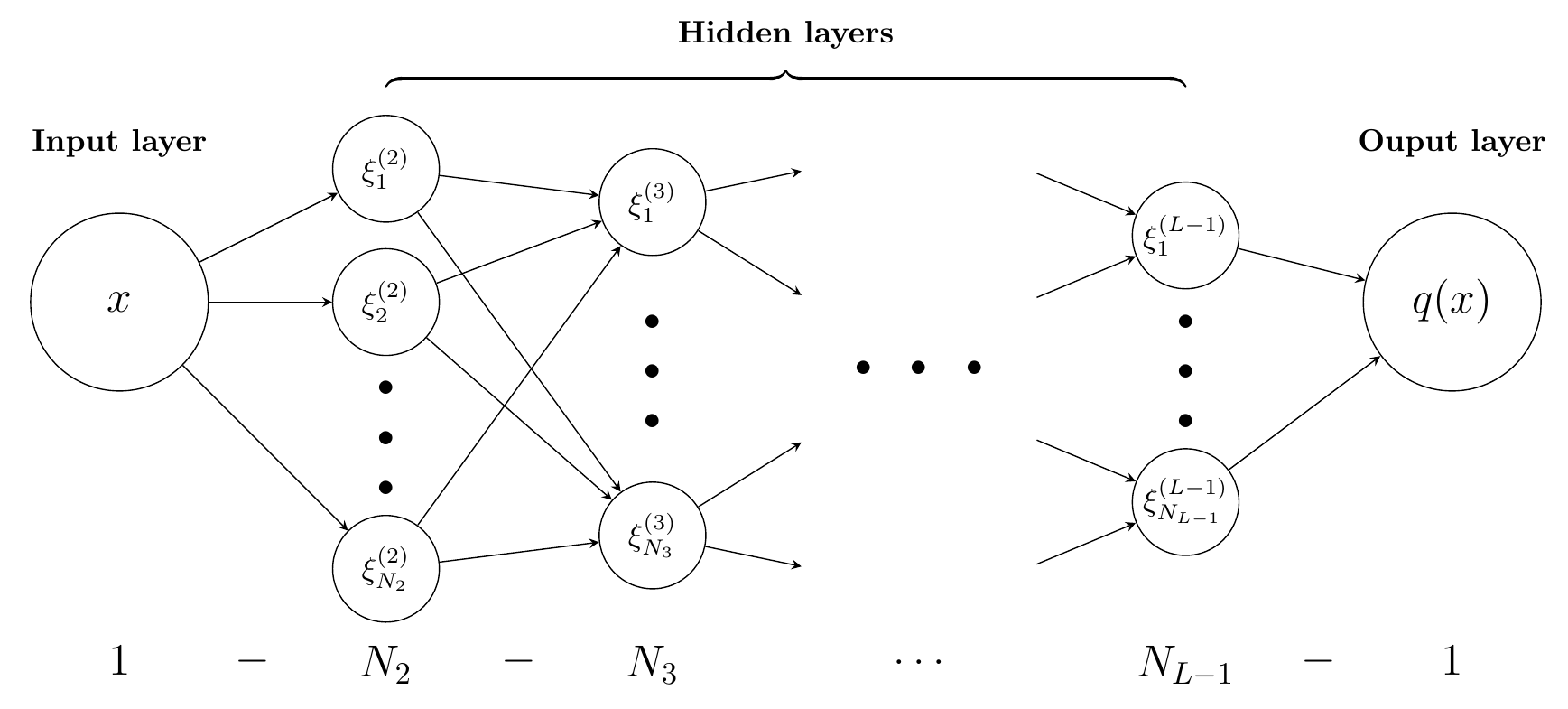}
\caption{A neural network can be used as a general parametrization of an unknown function from $\rm I\!R^{N_1} \rightarrow \rm I\!R^{N_L}$. For the case of a PDF, a single valued function of a single argument, the input and output layers have only one neuron. }\label{Fig:NNGeom}
\end{figure}

In a multilayer neural network, the neurons are organized into distinct layers whose neurons are not directly connected to each other. In a  multilayer feedforward network, the layers are ordered and each neuron on a given layer is only connected to the neurons on the previous and subsequent layers. The geometry of these networks, shown in Fig.~\ref{Fig:NNGeom}, is described by a list of numbers, $N_1$, ..., $N_L$, giving the number of neurons on each of the $L$ layers. The first layer, $\xi^{(1)}$, is a given input vector of length $N_1$ and the final layer, $\xi^{(L)}$ is the output vector of length $N_L$, also called the response of the neural network. All other layers are referred to as hidden layers. The activation of the $i$-th neuron on the $l$-th layer of the network is given by the recursive relationship

\begin{equation}
 \xi_i^{(l)} = g_i^{(l)}\left(\sum_j^{N_{l-1}} w^{(l)}_{ij} \xi_j^{(l-1)} -\theta^{(l)}_i\right) \,.
\end{equation}

Specified by the geometry, activation functions, thresholds, and weights, a multilayer feedforward neural network can be considered a parametrization of a function from $\rm I\!R^{N_1} \rightarrow \rm I\!R^{N_L}$. For a given geometry and a set of activation functions, the thresholds and weights can be chosen for the network to approximate a given continuous function. For the case of reconstructing the rescaled PDF $h(x)$, which is a single valued function with a single argument, the geometry is restricted to have $N_1 = 1$ and $N_L = 1$. It should be noted that NNPDF has also used $\log \frac 1x$ as a second argument in the input layer~\cite{Rojo:2018qdd}, which is important to reproduce the small $x$ behavior of the PDF. However, in our case our data are not sensitive to small $x$ and thus using  $\log \frac 1x$ at the input layer is not essential.

A neural network can be used to perform a regression by choosing the thresholds and weights with a  training procedure. During the training procedure, the weights and thresholds are modified to minimize some error function, which describes the difference between the response of the neural network and some desired output. When using a neural network to perform a statistical regression, a common choice of error function is the $\chi^2$ function, e.g.

\begin{equation}
\chi^2(\{ w\}, \{\theta\}) = \sum_{k=1}^N \big (\mathcal{Q}_k - \int_0^1dx\, K_k(x) h(x;\{ w\}, \{\theta\})  \big)^2 / \sigma_k^2\,,
\end{equation}
where $\mathcal{Q}_k$ are $N$ data points with standard deviations $\sigma_k$ and $h$ is the output layer of the neural network given an input layer $x$, weights $\{ w\}$, and thresholds $\{\theta\}$. Using a neural network to parametrize the unknown function may result in a $\chi^2(\{ w\}, \{\theta\})$ with a large number of local minima. Some these local minima are trivial multiplicities due to  symmetries a neural network has under permutations of the weights and thresholds. However, the possibility of multiple non-trivial local minima exists resulting in many realizations of the network that reproduce the data equally well. In these cases special care has to be taken to avoid ``over-fitting'' and several methods to do so have been developed in the literature~\cite{Rojo:2018qdd}.

These roughly equivalent minima can be found  by a training procedure such as a genetic algorithm. A genetic algorithm is an iterative process based upon the idea of natural selection. Each iteration, also called a generation, begins with a sample of possible networks, called a population. A fitness function is evaluated for each of the networks, which in this case  is the error function $\chi^2(\{ w\}, \{\theta\}) $. Those networks which are the ``least fit'', i.e. largest $\chi^2$, are removed from the population. The surviving population is then ``mutated'' by randomly changing their parameters, i.e. weights and thresholds, to create the starting population for a new generation. This procedure is iterated for enough generations that a final population covers a sufficient number of minima with sufficiently small values of the error function.

The genetic algorithm used in this study is based upon simulated annealing. The initial population is created from $N^{0}_{\textrm{rep}}$ sets of initial weights and thresholds which are generated from a random normal distribution with a wide initial search width, $\sigma_0$.
\begin{equation}
\{w\}_i^{(0)}, \{\theta\}_i^{(0)} \sim \mathcal{N}(0,\sigma_0)
\end{equation}
where $\{w\}_i^{(0)}$ and $ \{\theta\}_i^{(0)}$ are the parameters of the $i$-th neural network in the initial population. The weights and thresholds of this initial population of neural networks is the starting point for minimizing with respect to $\chi^2$ to some new values $\{w'\}_i^{(0)}$ and $ \{\theta'\}_i^{(0)}$. The resulting neural networks which are the least fit, i.e. that have the largest $\chi^2$, are removed from the population and not used in the next generation. The surviving neural networks are then mutated by adjusting their weights and thresholds by some small, Gaussian distributed amount with a mutation width $\sigma$,
\begin{equation}
\{w\}_i^{(g)}\sim \{w'\}_i^{(g-1)}( 1  + \eta  \mathcal{N}(0,\sigma)) \qquad \{ \theta\}_i^{(g)}  \sim \{ \theta'\}_i^{(g-1)} ( 1  + \eta  \mathcal{N}(0,\sigma)), 
\end{equation}
where $\eta$ is the relative size of the mutation, $\{w\}_i^{(g)}$ and $ \{\theta\}_i^{(g)}$ are the initial parameters of the $i$-th neural network in the $g$-th generation and $\{w'\}_i^{(g)}$ and $ \{\theta'\}_i^{(g)}$ are the parameters of the $i$-th surviving neural network in the $g$-generation after minimization. This procedure of minimization and removal is repeated for $N_{\textrm{gen}}$ generations. After enough generations have passed, the population of neural networks will all exist in various minima of $\chi^2$ which can be used to estimate the PDF. 

Many other training procedures exist which all have different methods of identifying minima, both stochastic and deterministic. An example of such a procedure is the approach the NNPDF collaboration is taking in obtaining PDFs from experimental data~\cite{Rojo:2018qdd}. Though different training procedures vary in their efficiency for finding minima, they should eventually all converge on the same set of minima of the error function.

Neural networks provide a rather general solution to the inverse problem. The large number of highly interconnected parameters allows for model independent result, but the complexity of the hidden layers do not give any insight to what effects underlie that result.

\subsection{Bayesian PDF reconstruction}

The third approach to inverse problems we discuss, which has proven to be versatile in practice, is Bayesian inference. It acknowledges the fact that the inverse problem is ill-defined and a unique answer may only be provided once further information about the system has been made available. This method does not require any explicit constraints, though it may benefit from them, nor does it require any functional form, even one as flexible as a neural network. This method finds the most probable value of $q(x)$ given the data and whatever prior information is provided. Other regularization methods may often be rewritten in terms of Bayesian reconstruction with the constraints treated as prior information.

Formulated in terms of probabilities, one finds in the form of Bayes theorem that
\begin{align}
P[q | {\mathcal Q}, I] =\frac{P[{\mathcal Q}|q,I] P[q|I]}{P[{\mathcal Q}|I]}.
\end{align}
It states that the so called posterior probability, $P[q |{\mathcal Q}, I] $, for a test function $q$ to be the correct $x$-space PDF, given lattice calculated Ioffe time PDF $\mathcal Q$ and additional prior information $I$ may be expressed in terms of three quantities. The likelihood probability $P[{\mathcal Q}|q,I]$ denotes how probable it is to find the data $\mathcal Q$ if $q$ were the correct PDF. Obtaining the most probable $q$ by maximizing the likelihood is nothing but a $\chi^2$ fit to the ${\mathcal Q}$ data, which as we saw from the direct inversion is by itself ill-defined. The second term of importance is the prior probability $P[q|I]$, which quantifies, how compatible our test function $q$ is with respect to any prior information we have. In particular such information can be related to the appearance of non-analytic behavior of $q(x)$ at the boundaries of the $x$ interval or the positivity of the PDF. $P[{\mathcal Q}|I]$, the so called evidence represents a $q$ independent normalization.

For sampled (lattice) data, due to the central limit theorem, the likelihood probability may be written as the quadratic distance functional $P[\mathcal{Q}|q,I]={\rm exp}[-L]$,
\begin{align}
L=\frac{1}{2}\sum_{k,l} ( {\mathcal Q}_k - {\mathcal Q}^q_k ) C_{kl}^{-1} ( {\mathcal Q}_l - {\mathcal Q}^q_l ),
\end{align}
where ${\mathcal Q}^q_k$ are the Ioffe-time data arising from inserting the test function $q$ into Eq.\eqref{MC} and 
\begin{align}
C_{kl}=\frac{1}{N_m(N_m-1)}\sum_h \big( {\mathcal Q}^h_k - {\mathcal Q}_k \big)\big( {\mathcal Q}^h_l - {\mathcal Q}_l \big),
\end{align}
denotes the covariance matrix of the $N_m$ measurements of simulation data ${\mathcal Q}^h_k$.

For the regularization of the inversion task, we now have to specify an appropriate prior probability $P[q|I]={\rm exp}[\alpha S[q,m]]$. Prior information enters in two ways here: on the one hand the (unique) extremum of the $S$ functional is given by the so called default model $m$, which, by means of Bayes theorem, represents the most probable answer in the absence of data. On the other hand the functional form of $S$ encodes which solutions are admissible to the inverse problem, e.g. enforcing positivity.

In a Bayesian approach the preconditioning discussed in the beginning of this section is naturally incorporated not via a modification of the Kernel but via the specification of an appropriate default model. Actually dividing out a fitted Ansatz $p(x)$ from the Kernel and using a constant default model is equivalent to working with the original Kernel and simply assigning $m(x)=p(x)$. In other words, in the Bayesian approach we select the most appropriate default model from a best fit of Eq.~\eqref{Eq:FitFunc} to the Ioffe time data and estimate the dependence of the end result $q_{\rm Bayes}(x)$ on the choice of $m$ by repeating the reconstruction with default models arising from a slight variation of the best fit parameters in Eq.~\eqref{Eq:FitFunc}.

Based on different sets of prior information, different regularization functionals have been derived in the literature. Note that due to Bayes theorem, in the "Bayesian continuum limit" i.e. the combined limit of the number of supplied datapoints becoming infinite and the uncertainty on the data going to zero, all reconstructions will lead to the same solution. A good choice of regulator functional and the availability of valid prior information can help us approach this solution more closely even for data that are coarse and noisy. Indeed if an accurate approximate solution is already available we can use Bayesian methods that imprint this prior information strongly onto the end result (steep $S[q]$), while in cases where only unreliable prior information is available one wishes to use a method that let's the data ``speak" as freely as possible without distorting them through the default model (weaker $S[q]$). 

The reconstruction of PDF's from Ioffe-time PDF data benefits from the availability of good prior information in the form of the empirically obtained Eq.\eqref{eq:p-func}. This situation is quite different from other inverse problems, e.g. the reconstruction of hadronic spectral functions, where little to no relevant information about the spectral structures of interest is known. We therefore foresee that methods with a steep prior probability will fare better than those that are explicitly designed to minimize the impact of the default model. 

The first example is the well known Maximum Entropy Method \cite{skilling, Asakawa:2000tr}, which features the Shannon-Jaynes entropy as regulator
\begin{align}
S_{SJ}[q,m]=\sum_n \Delta x_n \Big( q_n-m_n-q_n{\rm log}\big(\frac{q_n}{m_n}\big)\Big).\label{Eq:SJreg}
\end{align}
Its functional form has been derived using arguments from two-dimensional image reconstruction and it is designed to introduce as least as possible additional correlations into the end result, beyond what is encoded in the input data. The standard MEM implementation restricts the space of solutions to those close to the default model, meaning that for a small number of available input points its result will lie close to the function $m$. While it has been shown \cite{Rothkopf:2011ef} that this implementation of the MEM in general fails to recover the global extremum of the posterior, for a default model that is close to the correct result, the outcome is often satisfactorily accurate.

We contrast the MEM to another Bayesian approach simply named Bayesian reconstruction (BR) \cite{Burnier:2013nla}, which has been derived by requiring positive definiteness of the resulting $q$, smoothness of $q$, where the data ${\mathcal Q}$ do not provide constraints on its form, as well as independence of the resulting $q$ on the units used to express ${\mathcal Q}$. The resulting regulator reads 
\begin{align}
S_{BR}[q,m]=\sum_n \Delta x_n \Big( 1-\frac{q_n}{m_n}+{\rm log}\big(\frac{q_n}{m_n}\big)\Big).\label{Eq:BRreg}
\end{align}
It has been shown that this regulator also leads to a unique extremum of the posterior and since in contrast to the MEM no flat direction appears in $S_{BR}$, the search space does not need to be restricted. The BR method had been developed in particular for inverse problems, where only scarce or unreliable prior information is available, imprinting the form of $m$ as weakly as possible onto the end result. We expect thus that in the present case with good prior information available, this method will produce less accurate results than the MEM, as it does not leverage the information in $m$ as strongly.

Note that in the definition of $P[q|I]$ we introduced a further parameter $\alpha$, a so called hyperparameter, which weighs the influence of simulation data and prior information. It has to be taken care of self-consistently. In the MEM it is selected, such that the evidence has an extremum. In the BR method, we marginalize the parameter $\alpha$ a priori, i.e. we integrate the posterior with respect to the hyperparameter, assuming complete ignorance of its values $P[\alpha]=1$.

Up to this point we have only considered regulators for the reconstruction of positive definite functions. It is fathomable though that one has to deal with PDF reconstructions where positive definiteness does not hold, such as for spin dependent PDFs. In that case the regulator \eqref{Eq:BRreg} is not applicable. A choice of regulator often employed in the literature in such a situation is the quadratic one, which corresponds to a modified form of Tikhonov regularization
\begin{align}
S_{QDR}[q,m]=\sum_n \Delta x_n \Big(q_n - m_n\Big)^2.\label{Eq:QDRreg}
\end{align}
It is a comparatively strong regulator and imprints the form of the default model significantly onto the end result. Marginalizing the hyperparameter $\alpha$ with a flat prior probability $P[\alpha]=1$ is hence not possible in this case. Indeed integrating over the normalized prior probability with respect to $\alpha$ leads to a delta function that fixes the end result to $\rho=m$ everywhere. Instead one uses the "historic MEM" approach~\cite{skilling, Asakawa:2000tr}, where $\alpha$ is chosen such that for the corresponding reconstruction $q^\alpha$ the likelihood takes on the value $L[q^\alpha]=N_\nu$. As in the following we are able to utilize default models, which are already close to the correct result, the use of this regulator is justified.

Again we can compare this approach to one specifically designed to keep the influence of the default model to a minimum, i.e. to imprint its functional form as weakly as possible on the end result. To this end we resort to an extension of the BR method \eqref{Eq:BRreg} to non-positive functions. The corresponding
\begin{align}
S_{BRg}[q,m]=\sum_n \Delta x_n \Big(-\frac{|q_n-m_n|}{h_n}+{\rm log}\big(\frac{|q_n-m_n|}{h_n}+1\big)\Big),\label{Eq:BRgreg}
\end{align}
keeps the advantageous properties of the original BR prior, e.g. smoothness and scale invariance at the price of having to introduce another default model related function $h$. It may be thought of as the confidence we have in our default model. Note that for $h=m$ one obtains a regulator, which for $q>0$ and $q>m$ takes on the form of the standard BR prior and mirrors it for $q<m$. In order to obtain the full uncertainty budget of the end result obtained with the generalized BR prior, one has to vary not only the default model but also the function $h$. Eq.~\eqref{Eq:BRgreg} also admits for an a priori integrating out of $\alpha$.

Once $L$, $S$ and $m$ have been provided, the most probable PDF $q$, given simulation data and prior information is obtained by numerically finding the extremum of the posterior
\begin{align}
\left. \frac{\delta  P[q | {\mathcal Q}, I] }{\delta q}\right|_{q=q_{\rm Bayes}} = 0.
\end{align}
It has been proven that if the regulator is strictly concave, as is the case for all the regulators discussed above, there only exists a single unique extremum in the space of functions $q$ on a discrete $\nu$ interval.

Note that as long as positive definiteness is imposed on the end result, the space of admissible solutions is significantly reduced. On the other hand, regulators admitting also $q$ functions with negative contributions have to distinguish between a multitude of oscillatory functions, which if integrated over, resemble a monotonous function to high precision. We will observe the emergence of ringing artifacts for the quadratic and generalized BR prior below.

Other regularizations may be reformulated in the spirit of the Bayesian strategy. To give meaning to the inversion task they introduce in addition to the likelihood another functional which encodes prior information or prior expectations on the end result. In the standard Tikhonov approach e.g. the condition is that the values of $q$ shall have the smallest possible magnitude. While they do not rely on an explicitly formulated default model, they contain the prior information implicitly in the form of their regularization functional. Otherwise they could not provide a unique regularized answer to the task. The benefit of the Bayesian method is that all prior information is made explicit, so that the dependence of the end result on the regularization may be thoroughly and fully tested. 

\section{Mock Data Tests}

\begin{figure}
     \centering
     \includegraphics[width=0.48\textwidth]{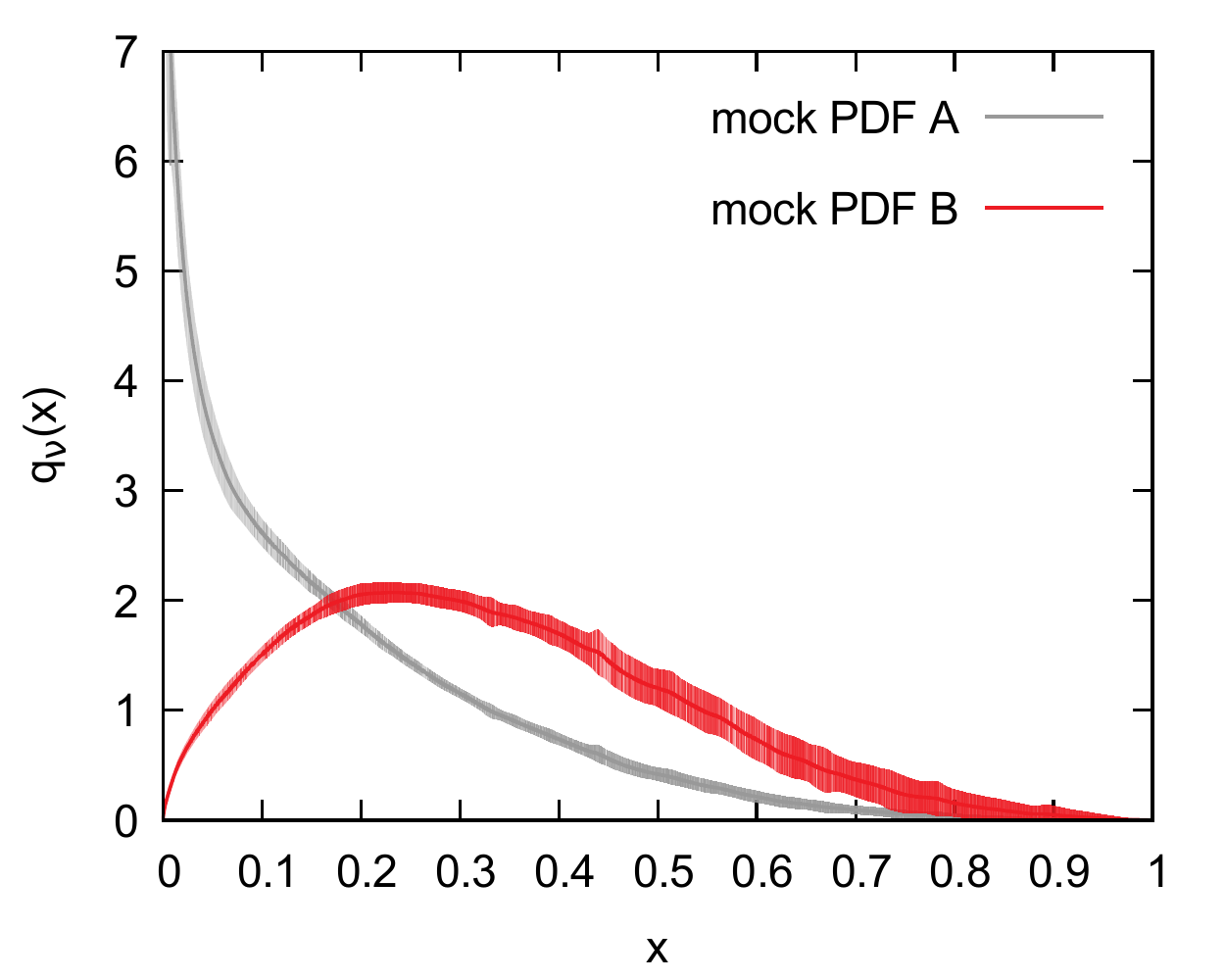}
     \includegraphics[width=0.48\textwidth]{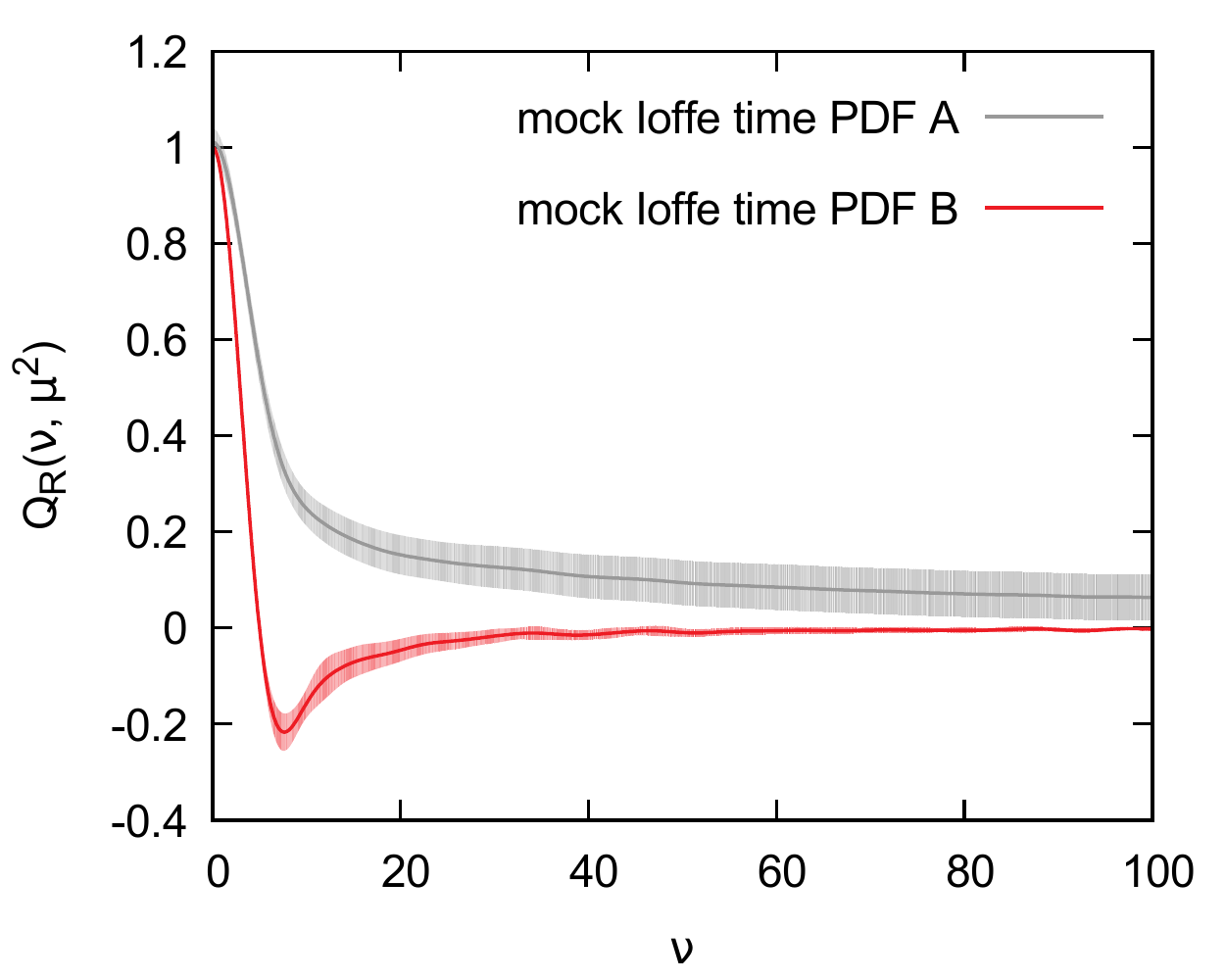}
     \caption{The two mock datasets used for testing the methods. The PDF on the left. The Ioffe time PDF on the right. Scenario A (red band) will test the case of a PDF which diverges at small $x$ and scenario B (blue band) will test the case of a PDF which converges at small $x$.}
\label{fig:orig_data}
\end{figure}

In this section, the methods constructed above are tested using data for the Ioffe time PDFs constructed from the phenomenological PDFs. 
Using the software package LHAPDF~\cite{Buckley:2014ana} and the dataset  {\tt NNPDF31\_nnlo\_as\_0118} from the NNPDF collaboration~\cite{Ball:2017nwa}, we Fourier transform numerically the iso-vector quark PDFs and obtain data for the ${\mathcal Q}_R(\nu)$ at $Q^2=2\, \textrm{GeV}^2$. This data set will be called scenario A. A second dataset, scenario B, is created by multiplying the NNPDF data by $N x$, where $N$ is a normalization to fix the valence quark sum rule. This modification forces the PDF to vanish as $x$ approaches zero. This scenario is reminiscent of the quasi-PDF case which is finite at low $y$ before the matching procedure is applied. These data sets contain a set of replicas that can be used to obtain errors on the PDFs. The standard deviation over the ensemble of replicas represent the 68\% confidence interval. The PDFs and Ioffe time PDFs for these two cases are plotted in Fig.~\ref{fig:orig_data}. For the various tests, the Ioffe time PDF will be sampled in the ranges mentioned above to study the dependence on the maximum available Ioffe time. 

\subsection{Backus-Gilbert}

The first advanced reconstruction technique to be tested is the Backus-Gilbert method. All integrals are computed numerically in double precision to tolerance $1.0\times10^{-16}$. In order to regularize the matrix $\bf M$ an SVD cut-off of  $\rho=1\times 10^{-8}$ is used as discussed in pervious sections~\footnote{We  have also tested using the Moore-Penrose pseudo inverse instead and got similar results.}. With this set up, a resolution function width, $\sigma(\bar x)$, of ${\cal O}(10^{-2})$ is obtained for most of $x<0.75$  gradually increasing after that reaching ${\cal O}(10^{-1})$ for $x=1.0$. With the ensemble of replicas, one can also compute the covariance matrix of the data and therefore use the covariance matrix regularization for the Backus-Gilbert method. We have done that as well as employ the Tikhonov regularization and obtained results that are similar with the results we discuss here where the SVD cutoff method was used to regularize the matrix inverse. In order to  obtain the statistical error-band and the mean for the reconstructed curve we take the standard deviation and the mean of the ensemble of reconstructed curves. The resulting statistical error-band represents a 68\% confidence level.  

\begin{figure}
     \centering
     \includegraphics[width=0.48\textwidth]{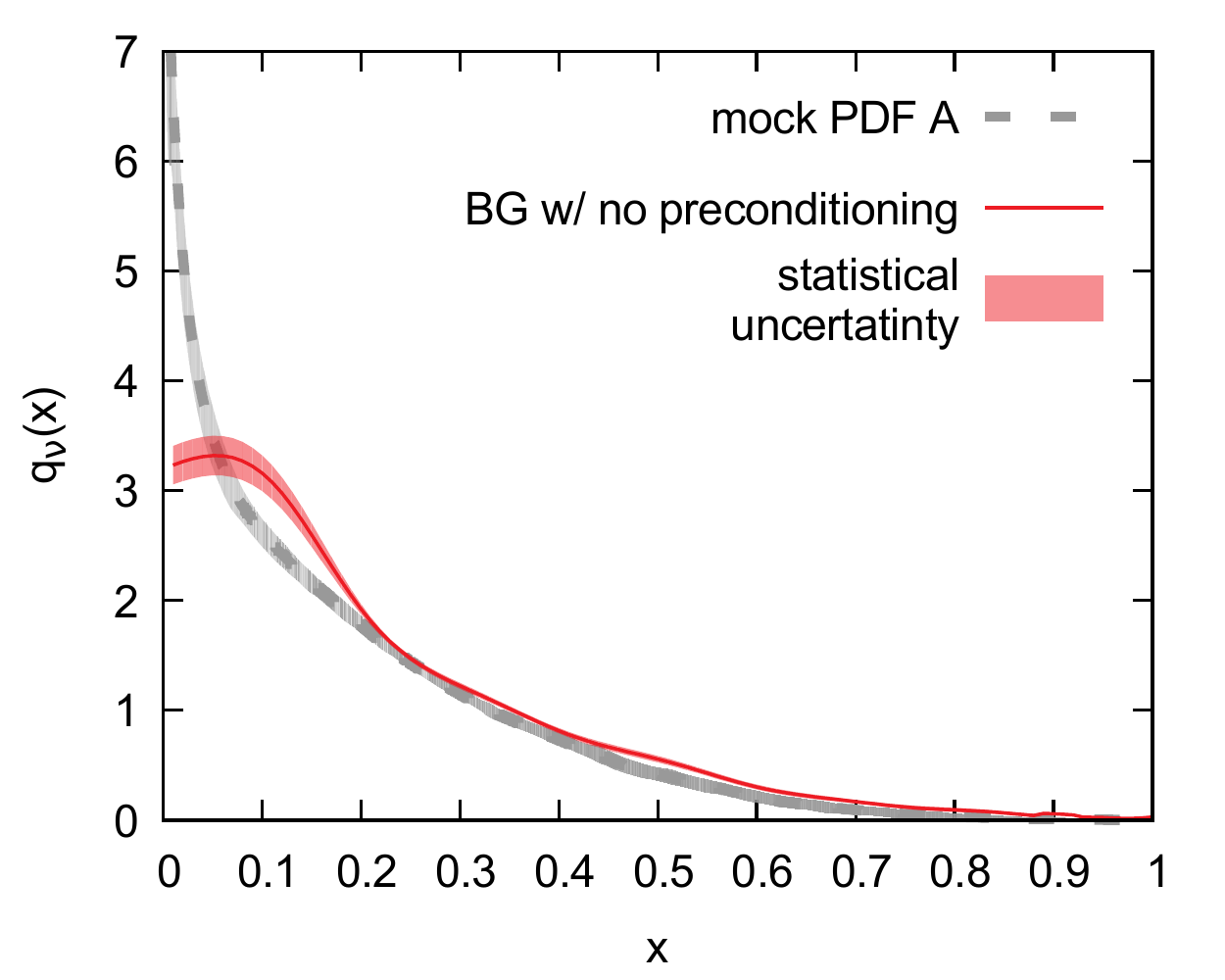}
     \includegraphics[width=0.48\textwidth]{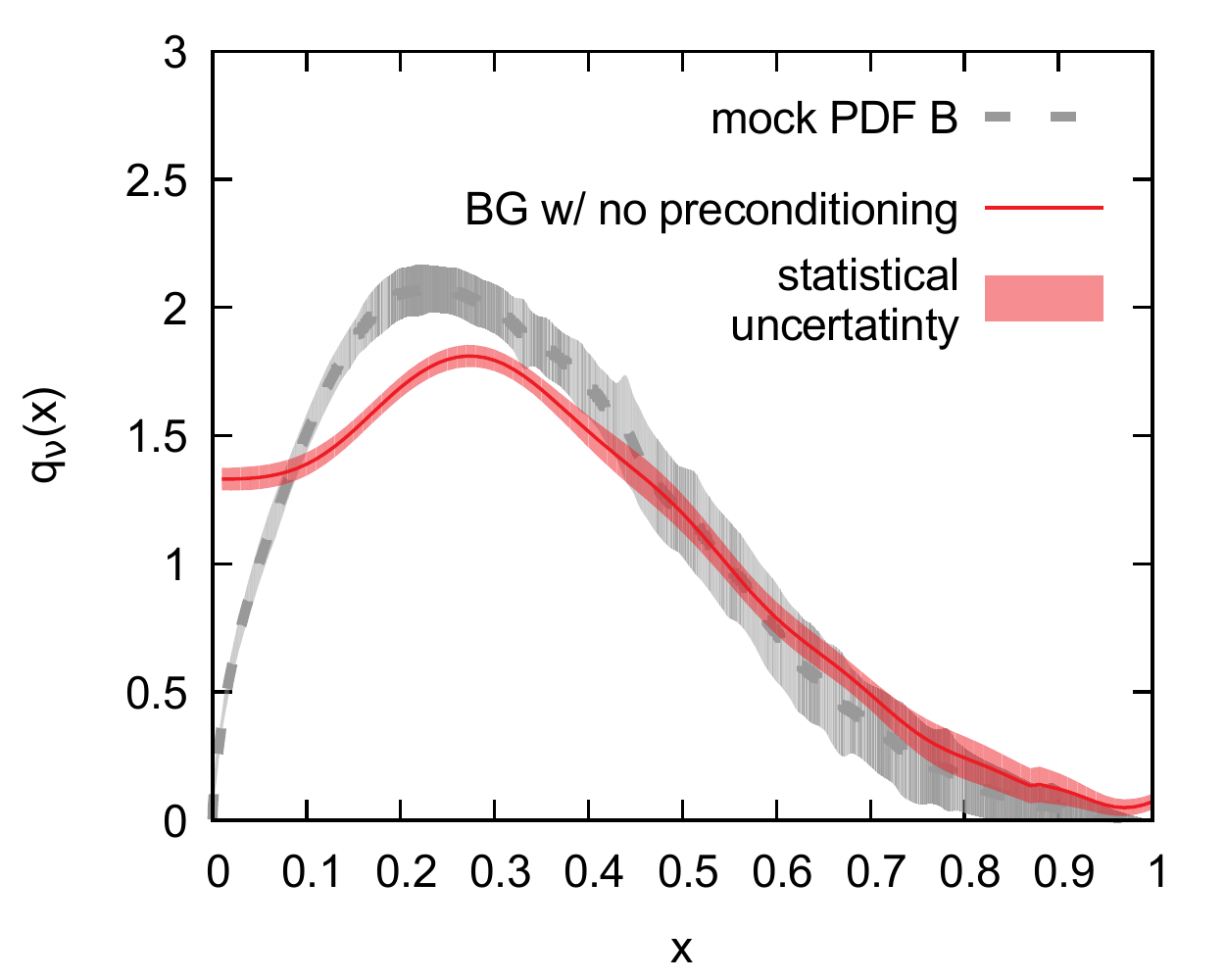}
          \includegraphics[width=0.48\textwidth]{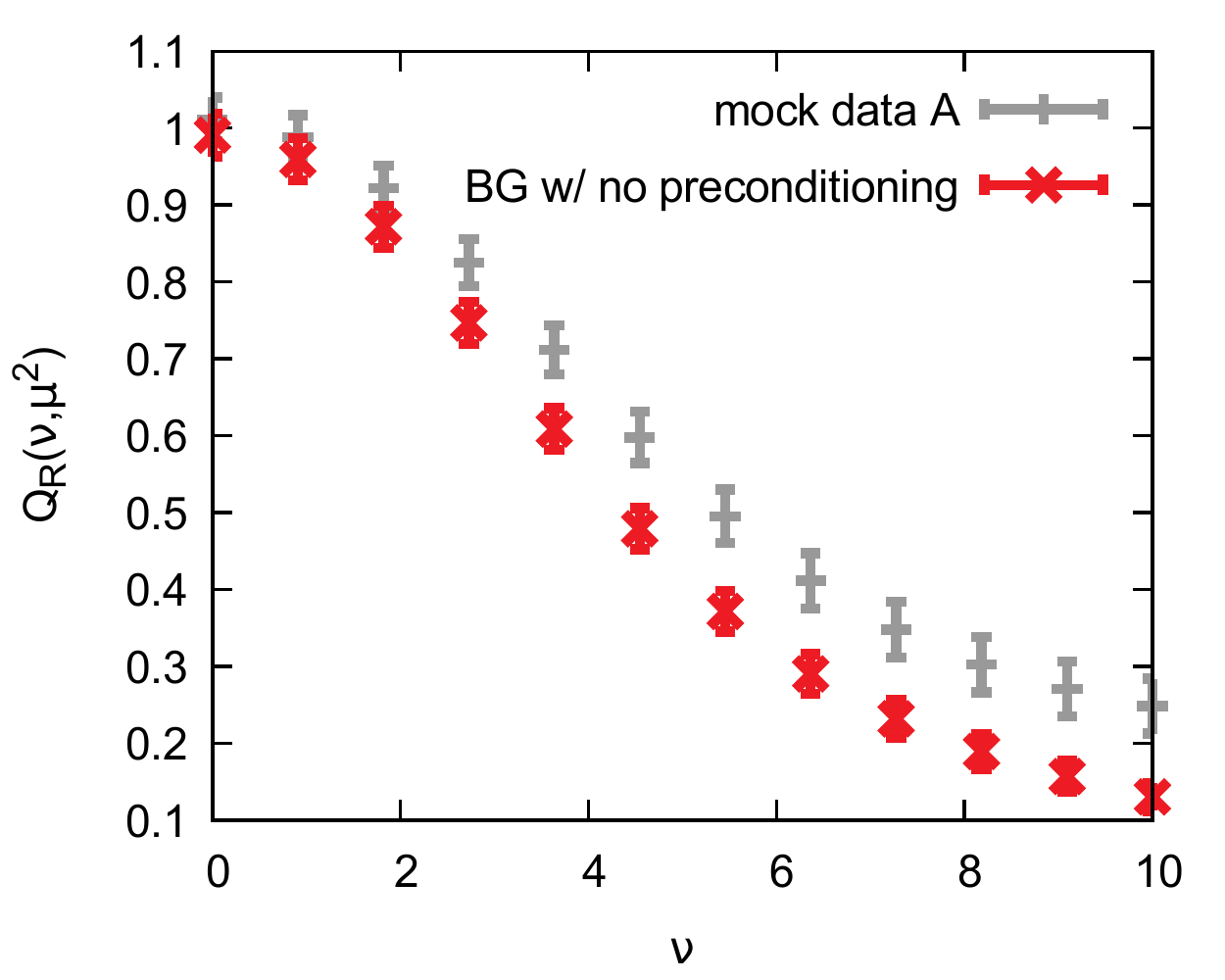}
     \includegraphics[width=0.48\textwidth]{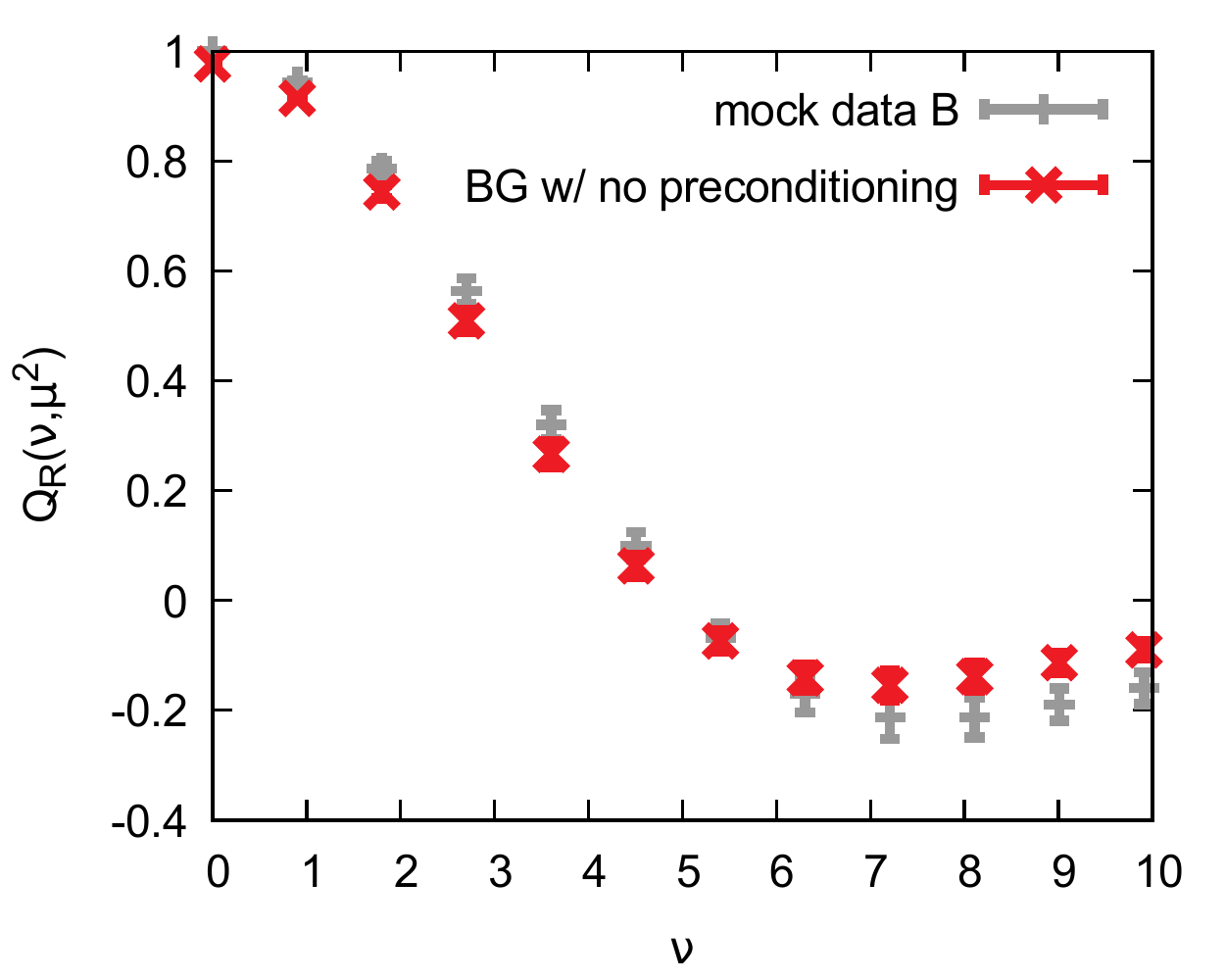}
    \caption{ $x$-space PDF's reconstructed using the Backus-Gilbert  (BG) method from $N_\nu=12$ Ioffe-time data points on the interval $\nu=[0,10]$ (top) as well as the input data (gray crosses) compared to the data arising from the reconstructed PDF (red crosses) in the bottom panels. The plots in the left column show the results for mock data based on a phenomenological PDF, while the right column from the modified scenario where the PDF vanishes at the origin. In both scenarios  no preconditioning has been employed.
}
    \label{fig:BG_RECON_NOPRECON}
 \end{figure}

After performing several numerical experiments on the Backus-Gilbert method, we have concluded that the use of a preconditioning function $p(x)$ is essential in obtaining an accurate determination of the PDF. The reconstructions of the PDF from data with $\nu_{max}=10$ and no preconditioning function are shown in Fig.~\ref{fig:BG_RECON_NOPRECON}. The reconstructed PDF is consistent with the mock PDF for $x >0.3$, but deviates from it in the low $x$ region. This result is expected, because the low $x$ region is dominated by the Ioffe time distribution at large Ioffe times which are not present in the input data. Also in Fig.~\ref{fig:BG_RECON_NOPRECON}, the fidelity of the reconstruction is tested by taking the Fourier transform of the reconstructed PDF and comparing to the data used to generate it which shows agreement across the range of Ioffe time. 

As argued earlier, preconditioning is essential in improving the Backus-Gilbert extraction of the PDF. In order to test its effectiveness, the preconditioning function defined in Eq.~\eqref{eq:p-func} is used with  a range of exponents $a$ and $b$  that allows for all the integrals that define  $\bf M$ and $\bf u$ to be convergent.
For scenario A we achieved the best reproduction with $a=-0.35$ and $b=2$ while for scenario B with $a=0.3$ and $b=2$. In order to get a handle of possible systematics due to preconditioning we varied the exponents $a$ and $b$ in the intervals $[-0.25, -0.4]$ and $[1, 4]$ respectively for scenario A.  Correspondingly for scenario B we varied the exponents in the intervals $[0.2, 0.35]$ and $[1.5, 3]$.

 %
%
%
 \begin{figure}
     \centering

           \includegraphics[width=0.48\textwidth]{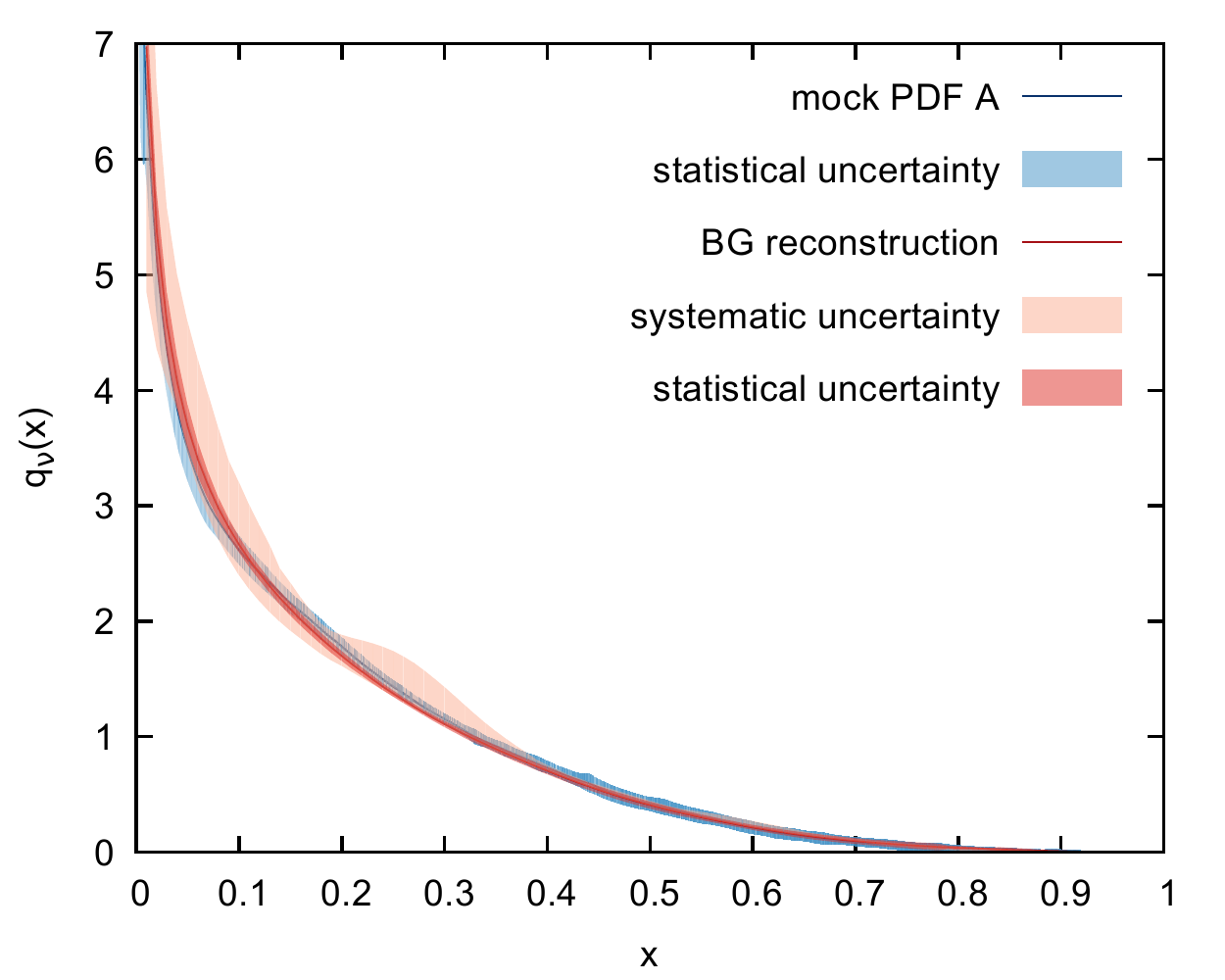}
                       \includegraphics[width=0.48\textwidth]{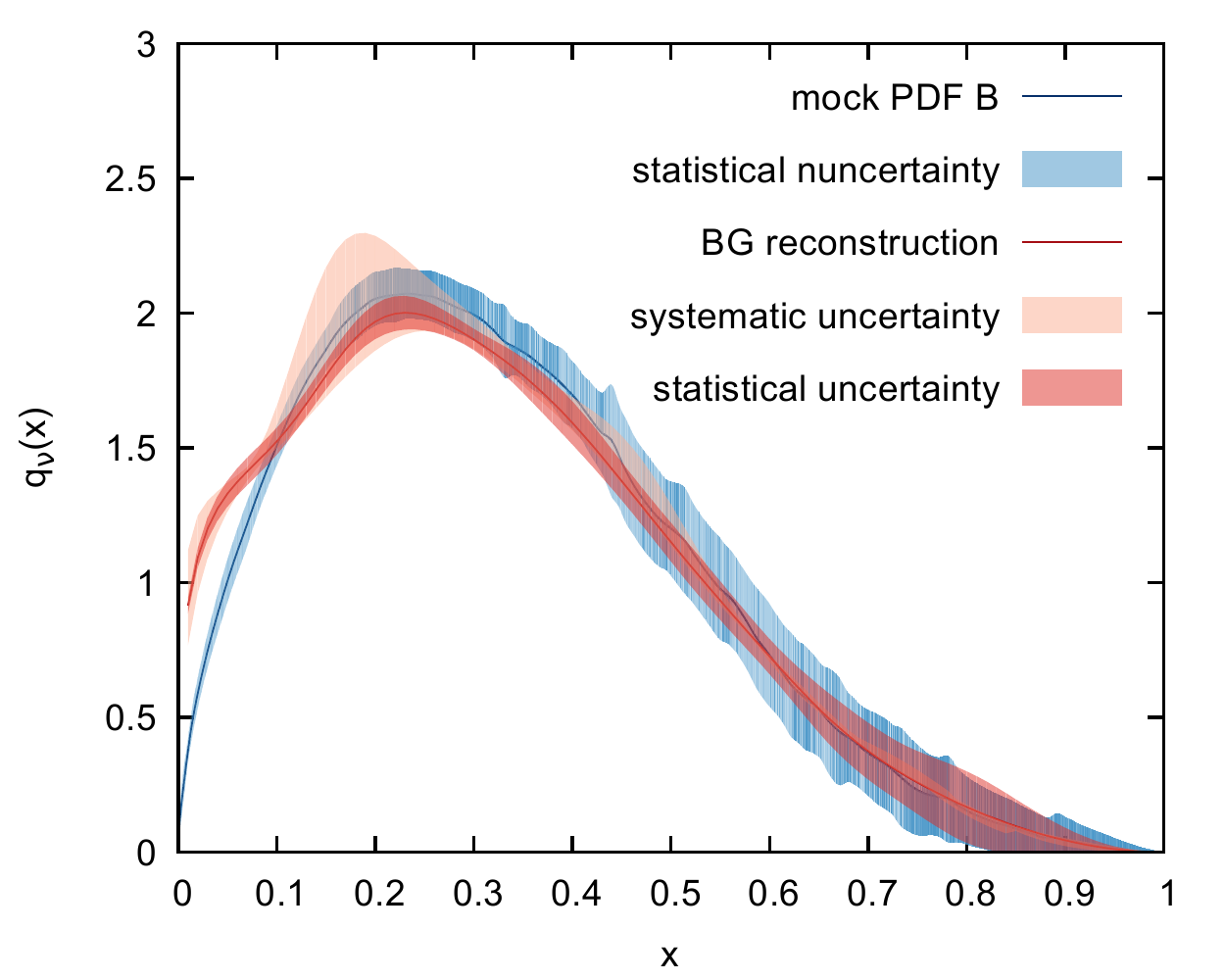}
     \includegraphics[width=0.48\textwidth]{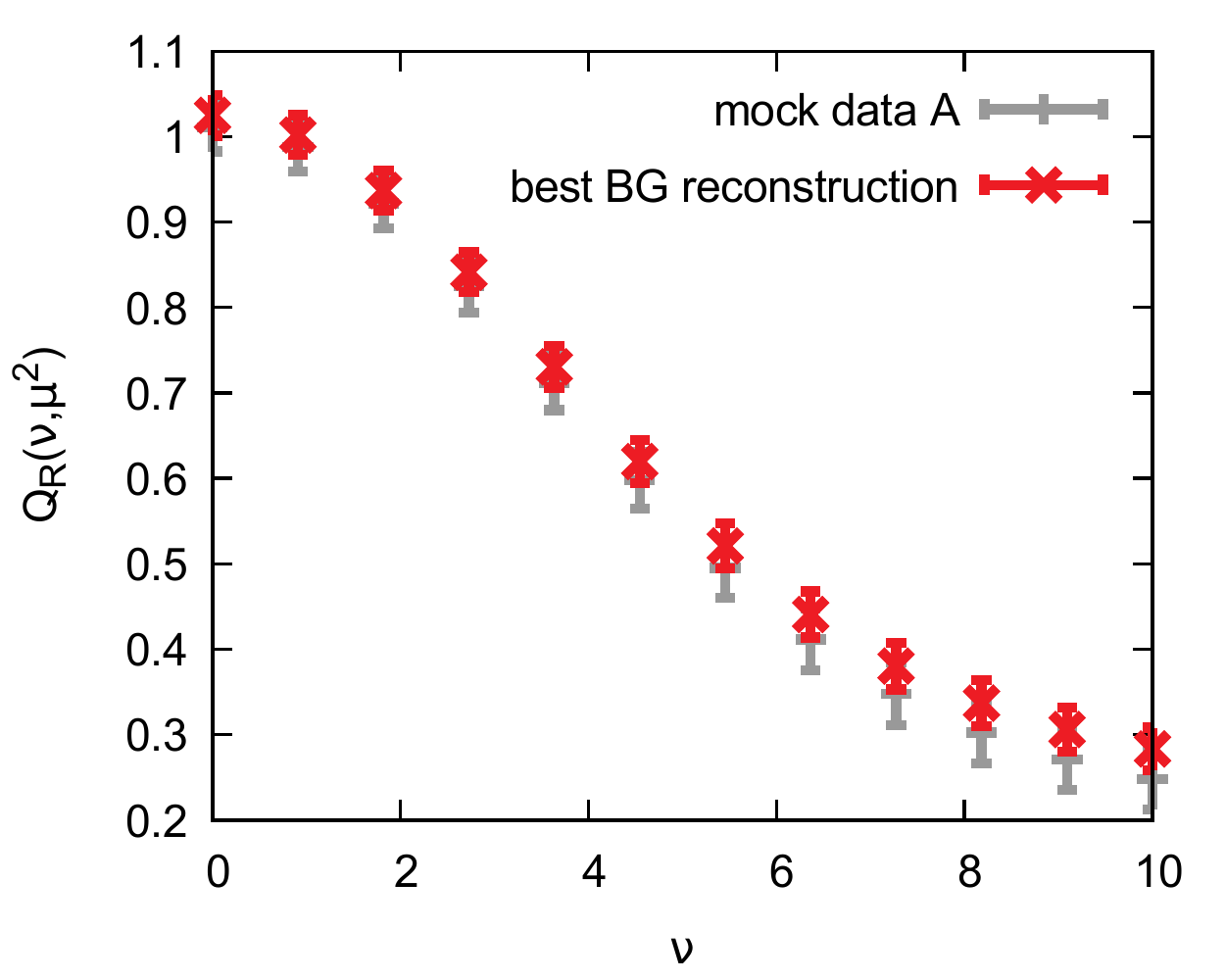}
     \includegraphics[width=0.48\textwidth]{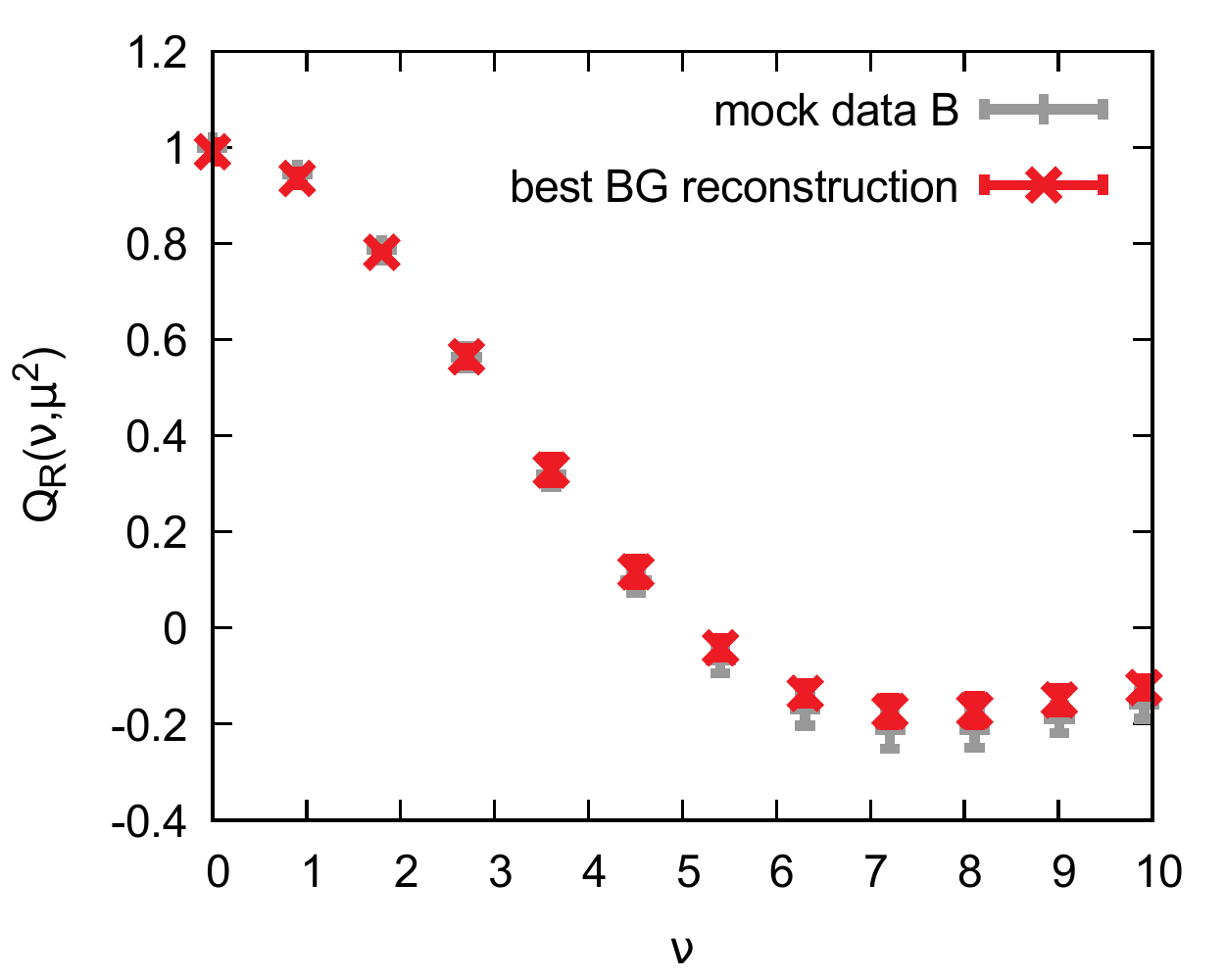}
    \caption{$x$-space PDF's reconstructed using the Backus-Gilbert  (BG) method from $N_\nu=12$ Ioffe-time data points on the interval $\nu=[0,10]$ (top) as well as the input data (gray crosses) compared to the data arising from the reconstructed PDF (red crosses) in the bottom panels. The plots in the left column show the results for mock data based on a phenomenological PDF, while the right column from the modified scenario where the PDF vanishes at the origin. The reconstruction was performed with preconditioning exponents $a=-0.35$ and $b=2$ for scenario A   and $a=0.3$ and $b=2$ for scenario B.}
    \label{fig:NNPDF_BG_RECON_prec}
 \end{figure}
 
 \begin{figure}
     \centering
                \includegraphics[width=0.48\textwidth]{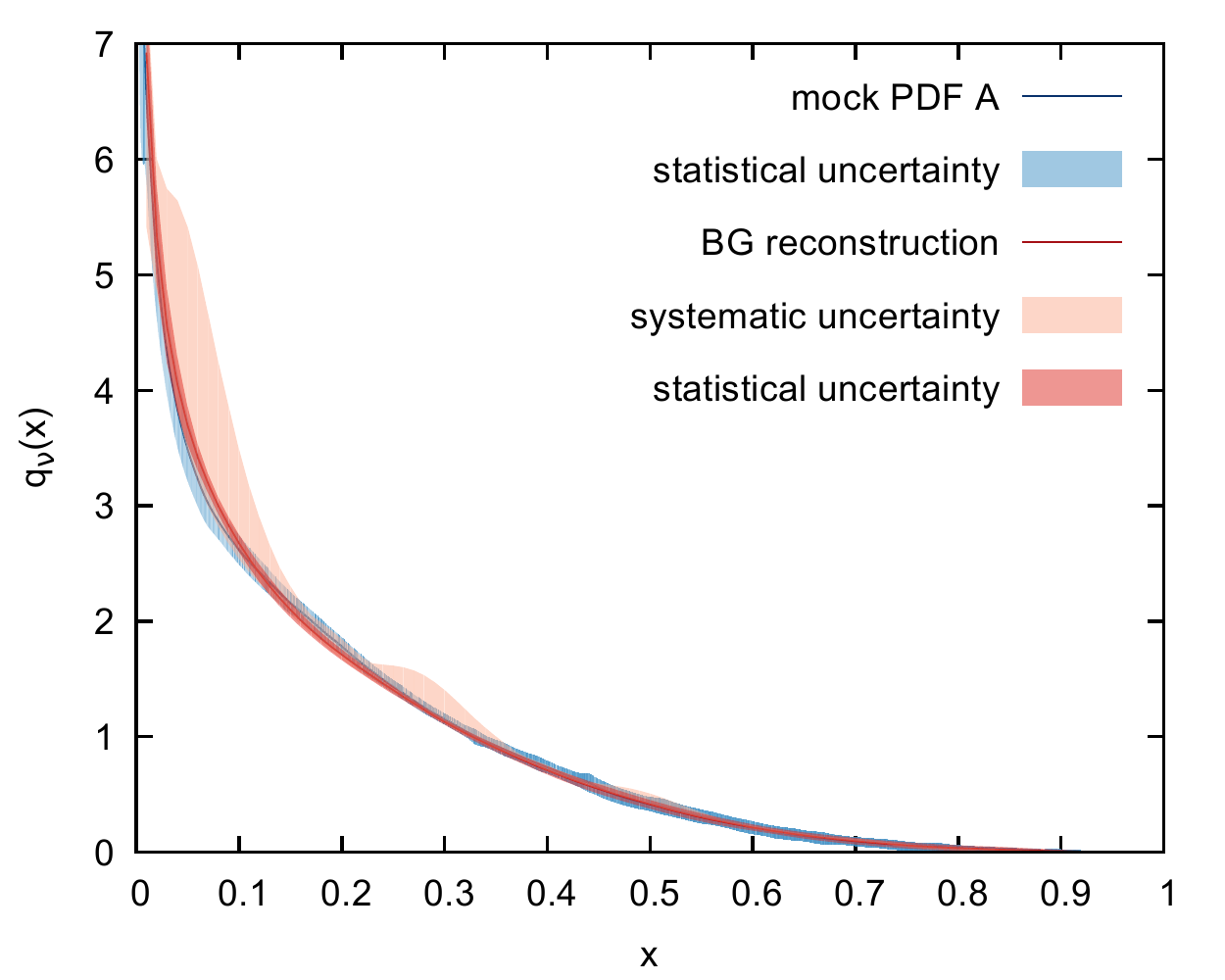}
                                       \includegraphics[width=0.48\textwidth]{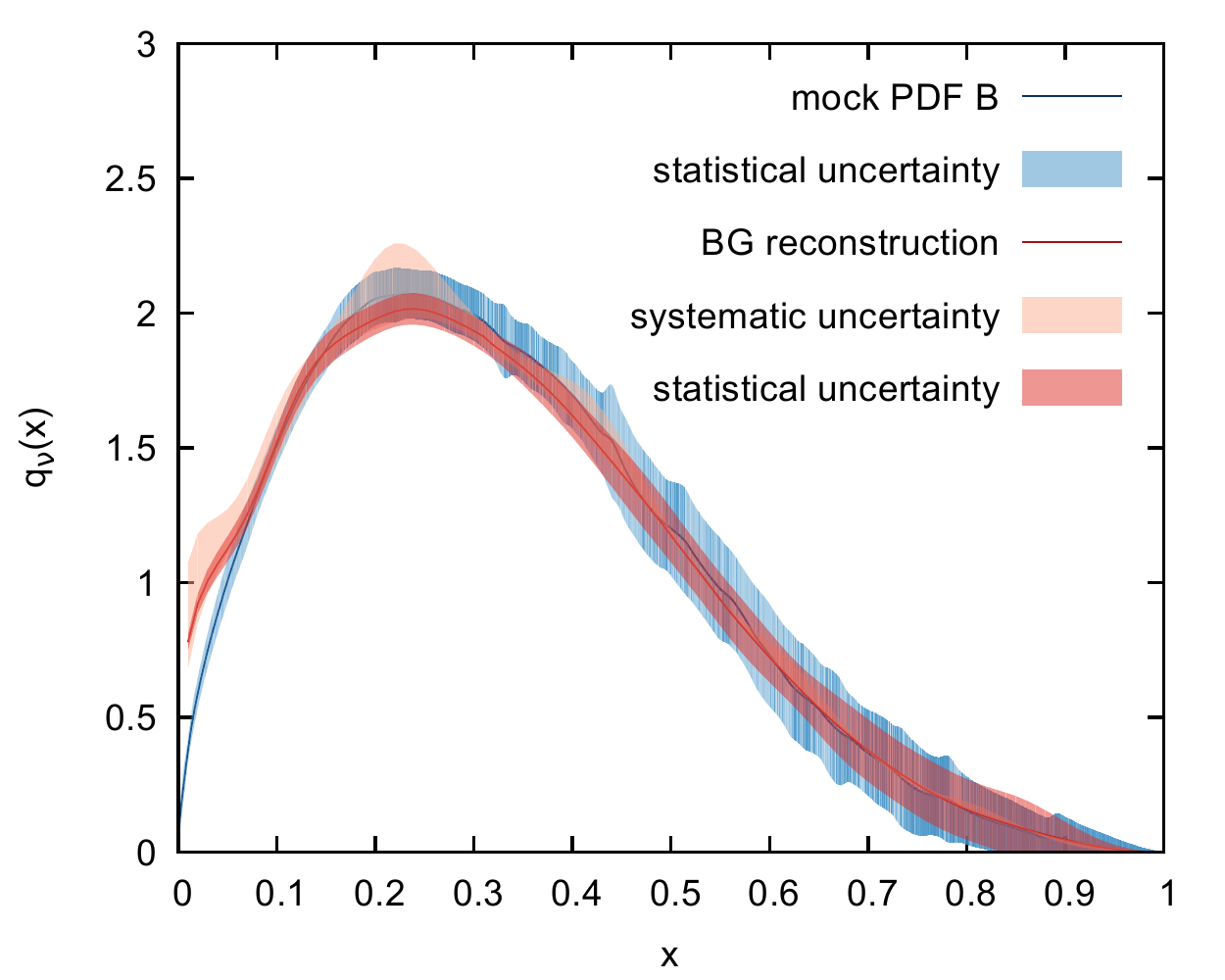}

           \includegraphics[width=0.48\textwidth]{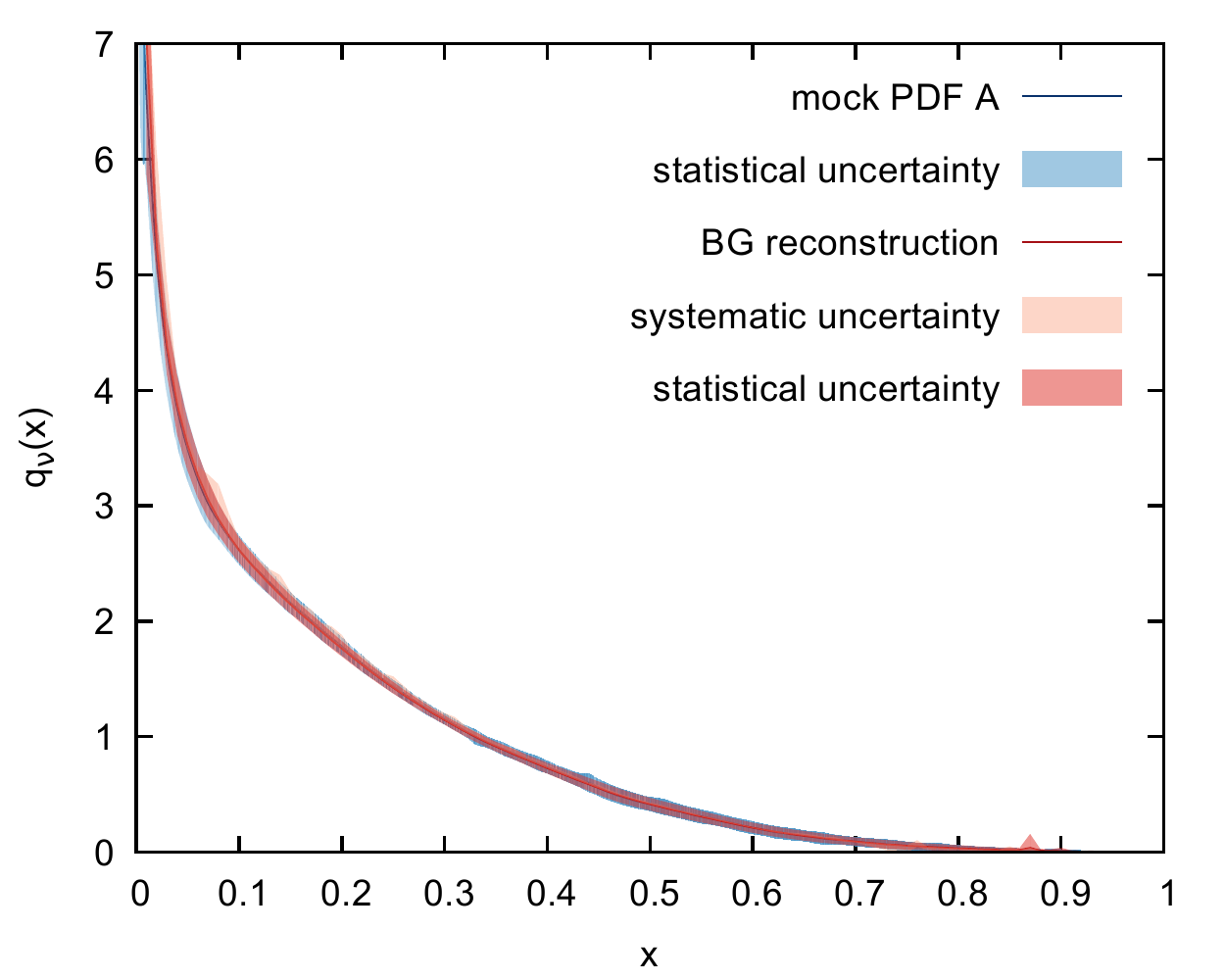}
                       \includegraphics[width=0.48\textwidth]{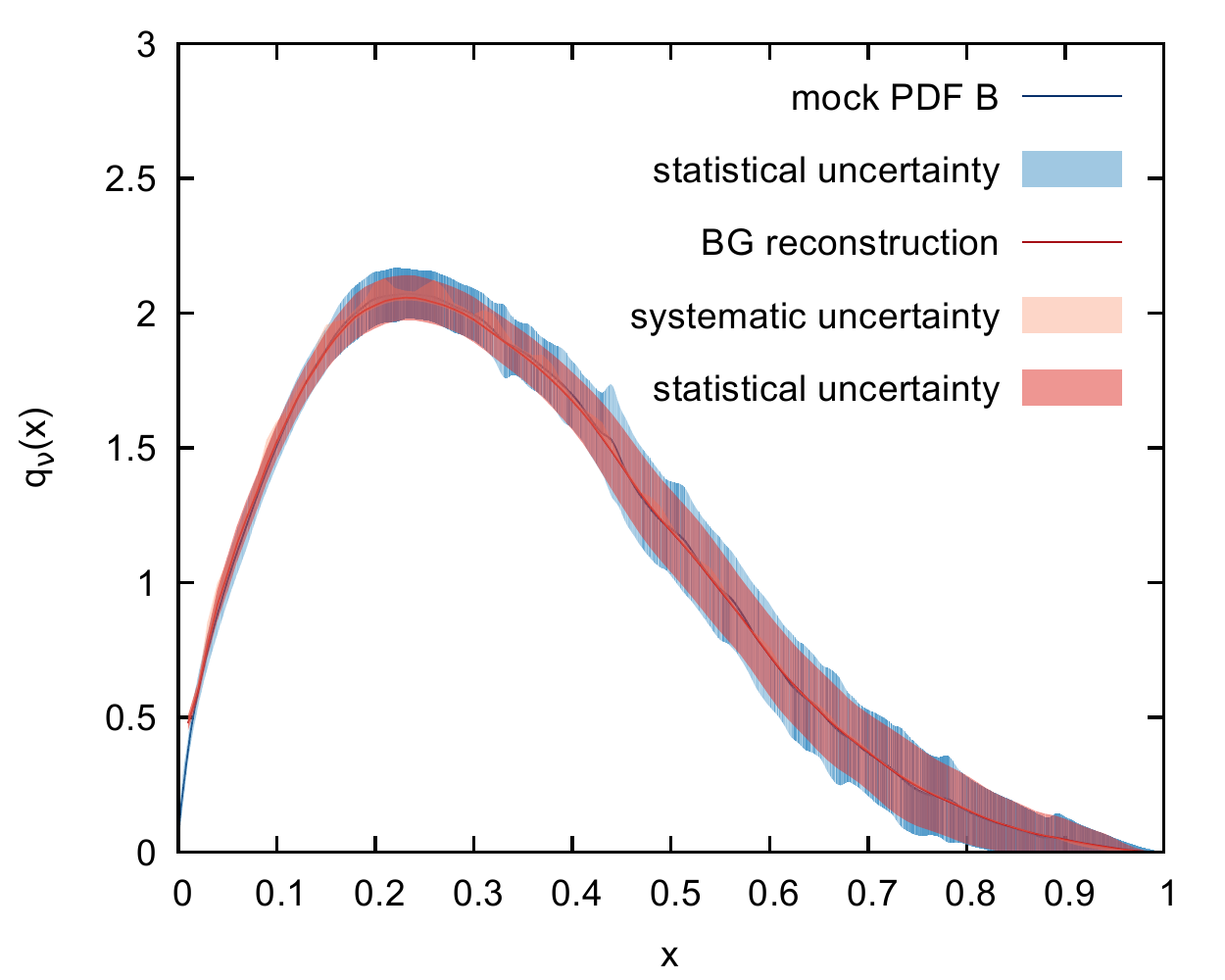}

    \caption{$x$-space PDF's reconstructed using the Backus-Gilbert  (BG) method from $N_\nu=23$ Ioffe-time data points on the interval $\nu=[0,20]$ (top) and from $N_\nu=112$ Ioffe-time data points on the interval $\nu=[0,100]$ (bottom). The plots in the left column show the results for mock data based on a phenomenological PDF, while the right column from the modified scenario where the PDF vanishes at the origin. The reconstruction was performed with preconditioning exponents $a=-0.35$ and $b=2$ for scenario A   and $a=0.3$ and $b=2$ for scenario B. It is noteworthy that by having lattice data up to $\nu_{\rm max}=20$ one can achieve a very good reconstruction with the BG method.}
    \label{fig:NNPDF_BG_RECON_prec_nus}
 \end{figure}

 As one can see, for both scenarios, a wide range in $x$ is well reproduced and deviations from the original data begins to appear for $x<0.2$. Perhaps, these deviations are expected due to the rather small maximum value of $\nu$ used in this example, $\nu_{\textrm{max}} =10$. However, this aggressive cutoff for $\nu$ is realistic for lattice calculations.
In addition, the number of selected points, which is 12, is also a number that is plausible for present lattice QCD calculations.
 From these results we can conclude that the fidelity of the reconstruction can be improved significantly by choosing appropriate preconditioning functions.
 The best reconstruction is obtained using a preconditioning function  that is chosen to roughly fit the data. In this case the resulting reconstruction is indistinguishable from the original data for nearly all the range of $x$ with the largest deviations occurring for $x< 0.1$. 
 Therefore, we conclude that the Backus-Gilbert reconstruction with an appropriate preconditioning function is well suited for the reconstruction of the PDFs from the limited data for the Ioffe time PDFs that are provided by present day lattice QCD calculations.

Finally, we explore the dependence of the reconstruction on the maximum value of $\nu$. Experiments are performed with $\nu_{max} = 20$ and $\nu_{max} = 100$. In all cases the data points were selected to be uniformly spread in the interval  with a separation $\delta \nu=10/11$. 
  The results are shown in Fig.~\ref{fig:NNPDF_BG_RECON_prec_nus}. The best result that is in agreement with the input PDF within errors across the whole range of $x$ is obtained from the experiment with $\nu_{max}=100$. However, this is an unrealistically large range in $\nu$ unlikely to be accessible in contemporary lattice QCD calculations. As $\nu_{max}$ is lowered then the reconstruction deteriorates. However,
even for $\nu_{max}=20$ the  results seem to agree with the input with the exception of one small region  $x=[0.05,0.1]$ where we observe a one $\sigma$ deviation. Overall, the $\nu_{max}=20$ result is only very slightly better than the case of $\nu_{max}=10$ which was presented in Fig.~\ref{fig:NNPDF_BG_RECON_prec}. Therefore, we conclude that extending the range of $\nu$ slightly beyond $\nu=10$ will have only small improvement in the determination of the PDF. 
Our conclusion from this analysis is that 12 points  for $\nu \in [0,10]$ seems to be adequate to obtain a good reconstruction of the underlying PDF. This finding is rather encouraging for present day calculations as it is possible to cover the range of $\nu \in [0,10]$  with about 12 points in lattice QCD calculations.

 \subsection{Neural network reconstruction}
  
In this section, the neural network method described above is tested to reconstruct the mock PDFs. The data used for this reconstruction is the mock scenario A and B in the smallest range of $\nu$, $\nu \in [0,10]$, discretized into $N_\nu=12$ points.  For this study, we chose the hyperbolic tangent as activation function  for all the nodes in the hidden layers. The activation function for the final layer is linear with the threshold value fixed to zero. For a single hidden layer of size $N_2$, the neural network parametrizes the PDF as
\begin{equation}
q(x) = \sum_{i=1}^{N_2} w^{(3)}_{1,i} \tanh(w^{(2)}_{i,1} x + \theta^{(2)}_i)) \,,
\label{eq:NNparam}
\end{equation}
while for two hidden layers of sizes $N_2$ and $N_3$
\begin{equation}
q(x) = \sum_{j=1}^{N_3} w^{(4)}_{1,j} \tanh\left(\sum_{i=1}^{N_2} w^{(3)}_{j,i} \tanh(w^{(2)}_{i,1} x + \theta^{(2)}_i) +  \theta^{(3)}_j\right) \,.
\label{eq:NNparam2}
\end{equation}

The neural network behavior is governed by the collective interactions of all the weights, thresholds, and connections, not on any individual neurons. Slight modifications of the geometry, such as adding or removing a neuron, will have little change in the final result. To test this point, 1-3-1, 1-4-1, and 1-2-2-1 geometries are tried, all resulting in similar PDFs. No preconditioning function will be used in this test. A preconditioning function of the form in Eq.~\eqref{eq:p-func} can be set to prefer certain features, such as the PDF vanishing in the limit $x \to 1$ by using $b>0$ or a divergence as $x \to 0$ by using $a<0$. Remarkably the neural network was able to reproduce both these high and low $x$ features without the need of any constraint or prior information.

\begin{figure}[h!]
\centering
\includegraphics[scale=0.58]{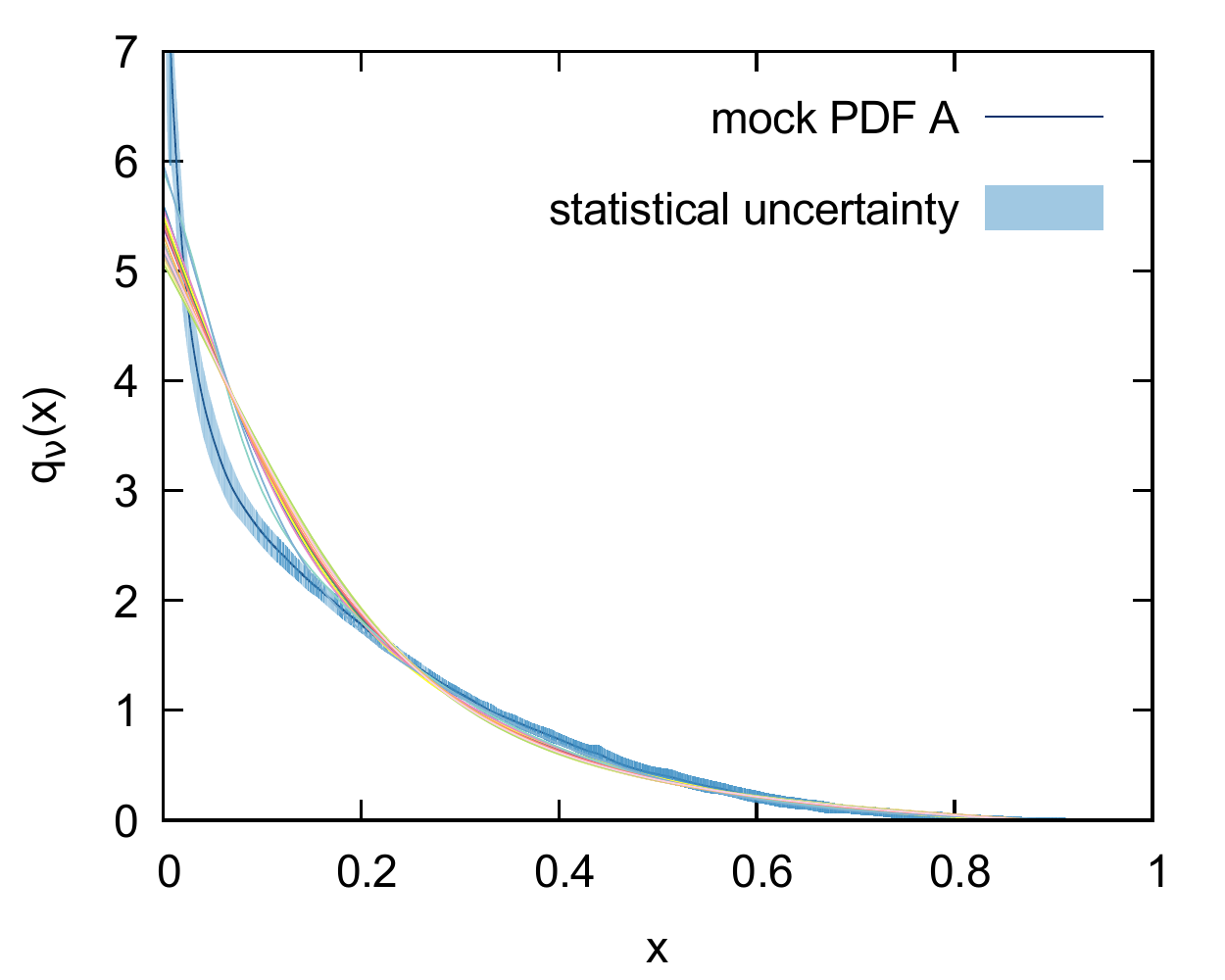}
\includegraphics[scale=0.58]{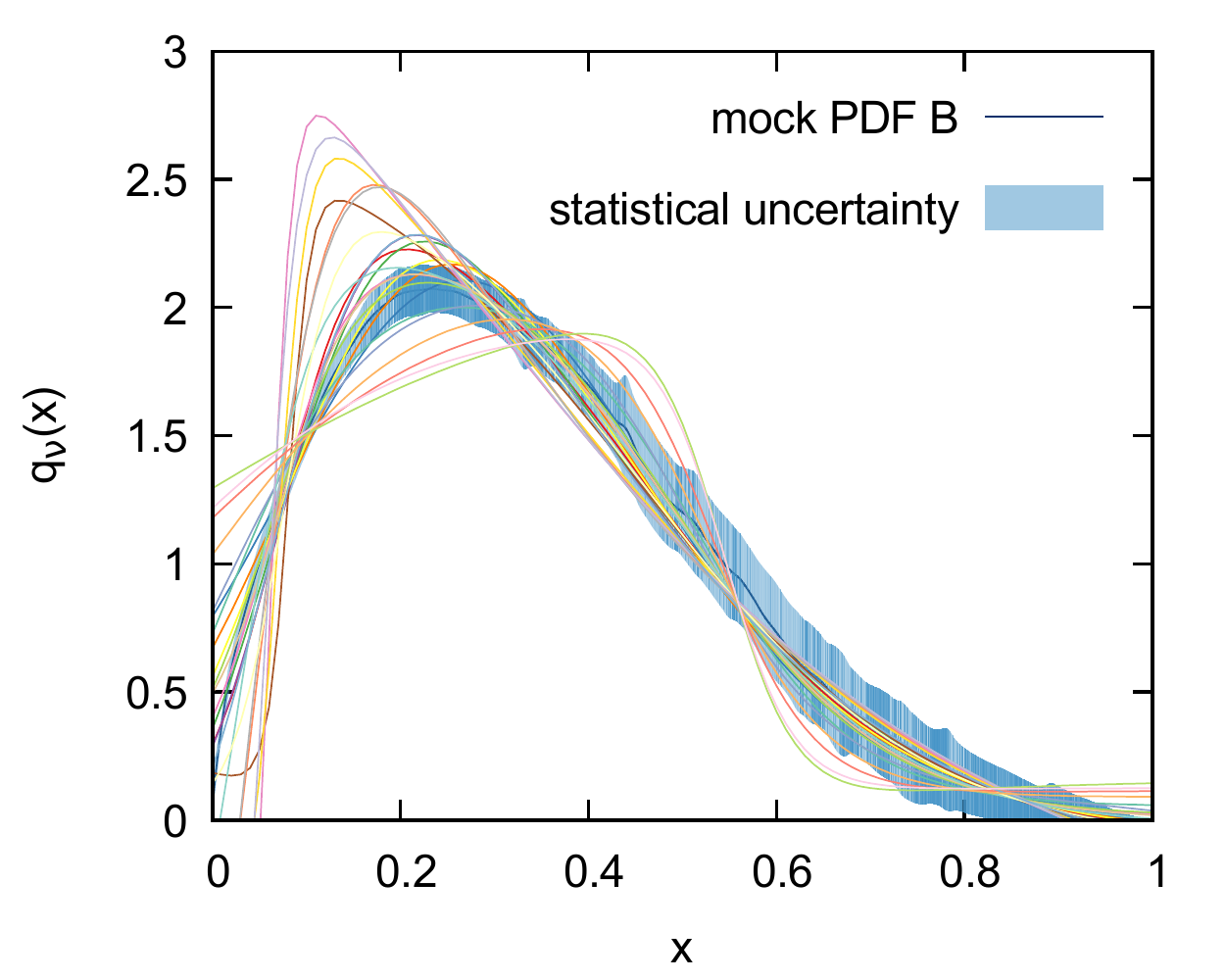}
\includegraphics[scale=0.58]{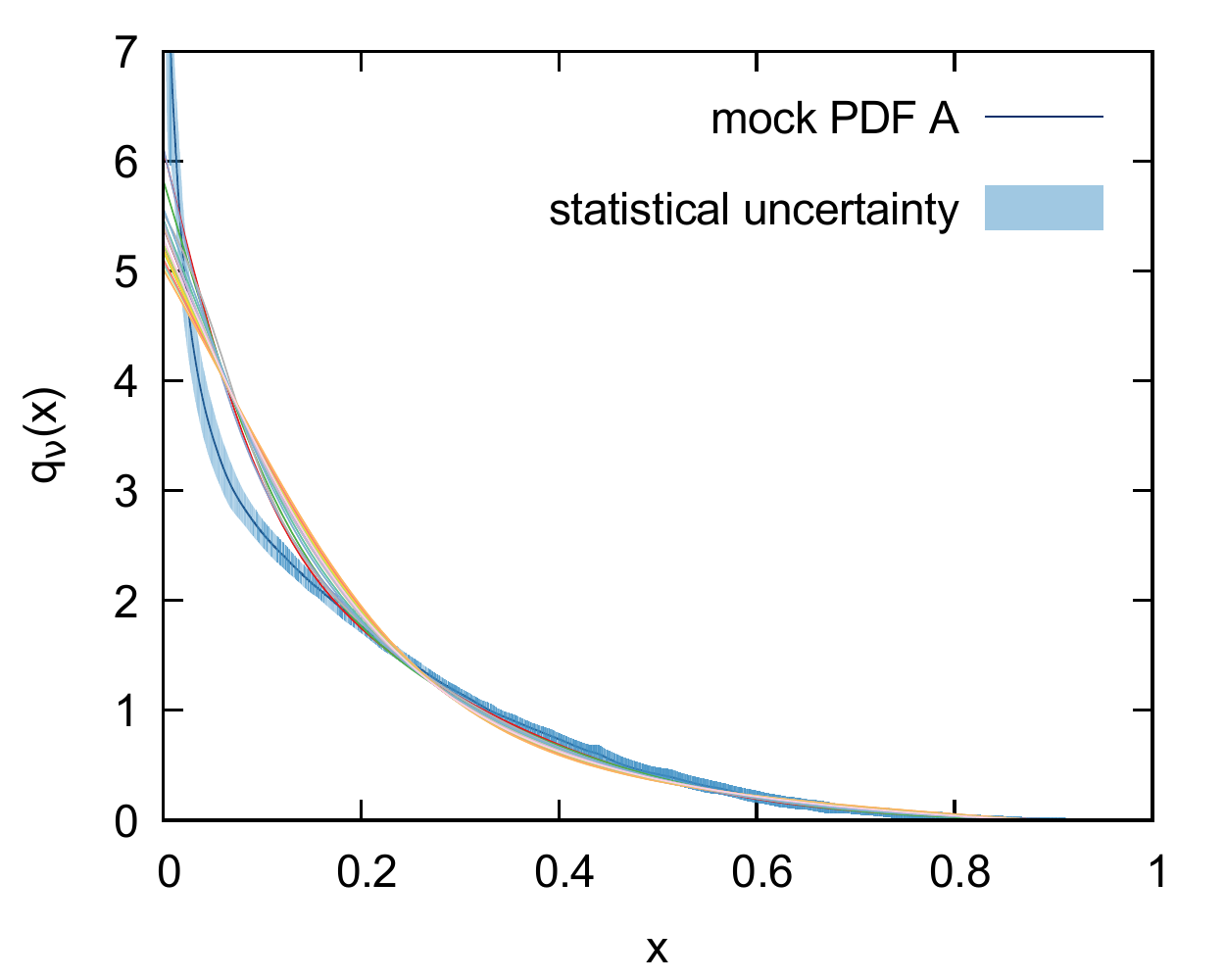}
\includegraphics[scale=0.58]{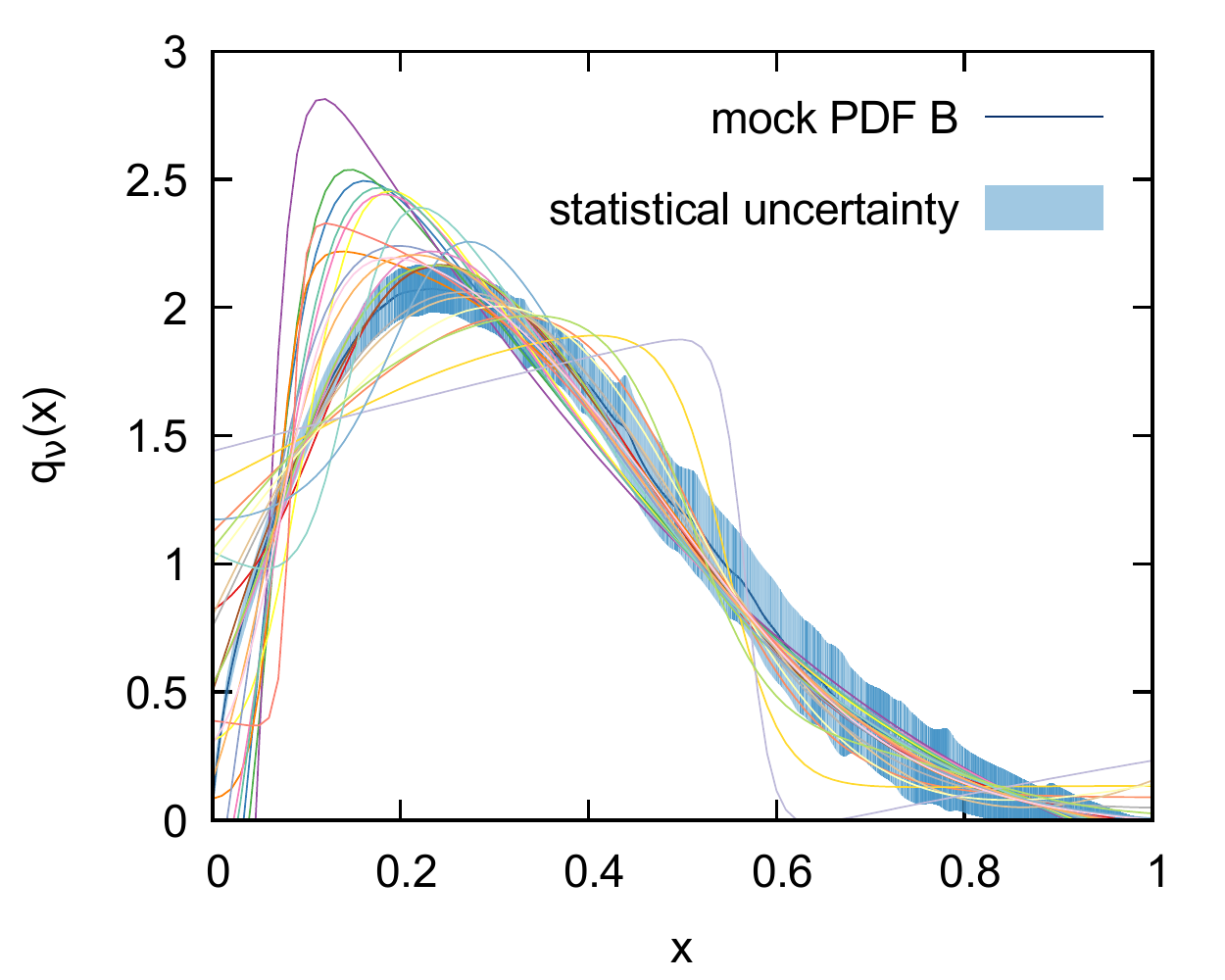}
\includegraphics[scale=0.58]{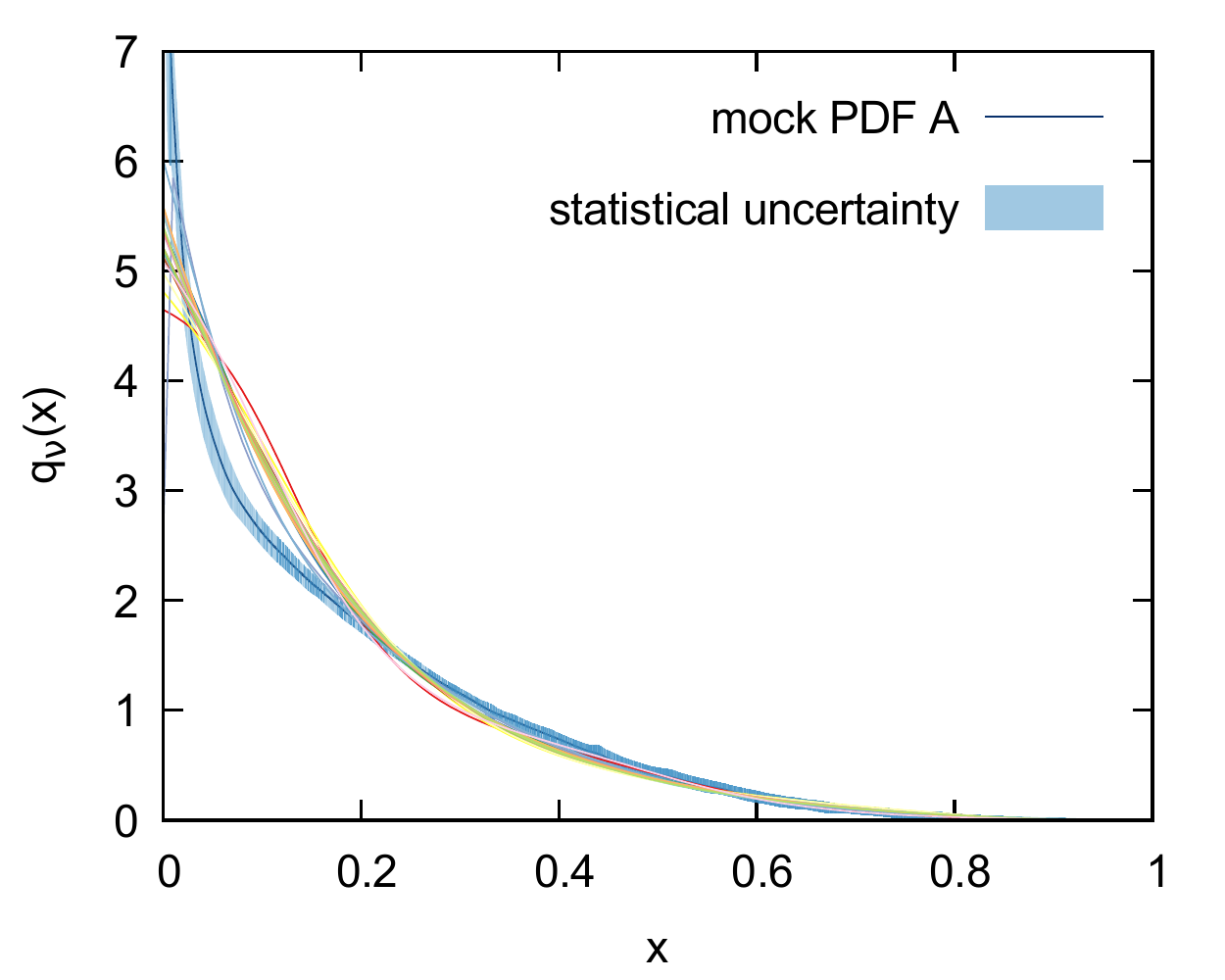}
\includegraphics[scale=0.58]{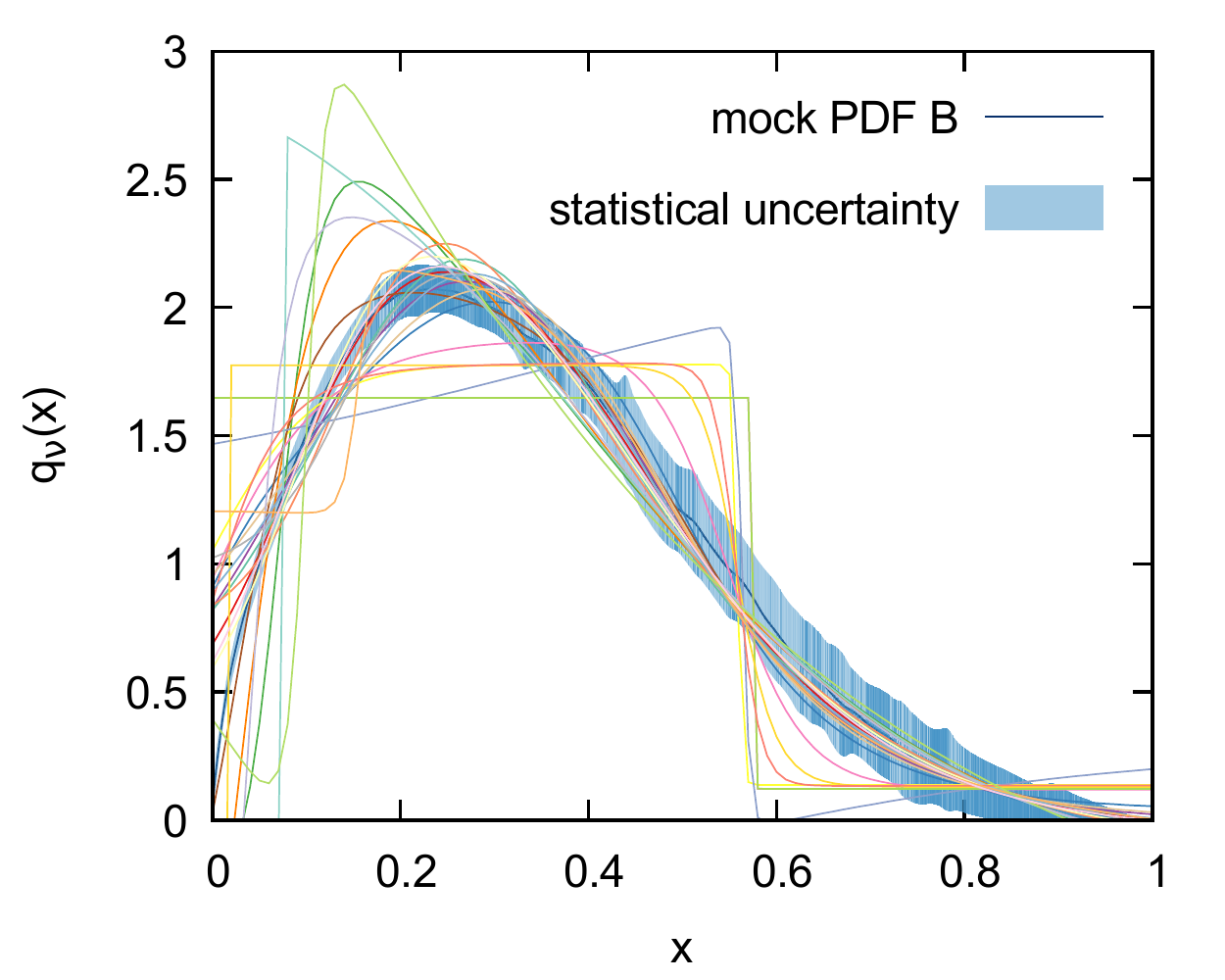}
\caption{The genetically trained neural nets. The blue band is the original data. The curves are the responses for the final population of genetically trained neural networks. The left column is with NNPDF data. The right column is with modified data. The first row has a network geometry of 1-3-1. The second row has a geometry of 1-4-1. The third row has a geometry of 1-2-2-1. }\label{Fig:TrainedNets}
\end{figure}

The training procedure described above has a number of tunable parameters, but the end result does not depend very strongly on any particular parameter. For this study, the initial search width, $\sigma_0$, is set to 5, the mutation width, $\sigma$, is set to 1, and the mutation size, $\eta$, is set to 0.05. From most starting points in any given generation, the minimization procedure sends the $\chi^2$ to very small values. In case this does not occur, the minimization is restarted. Three checks on $\chi^2$ are performed and if the value is too large then the mutation and minimization steps are performed again. First if the $\chi^2$ is greater than $10^{-2}$, then the mutation and minimization steps are repeated with the same mutation width and size. Second if the $\chi^2$ is greater than $10^{-1}$, then the mutation and minimization steps are repeated with the same mutation width and a size of 0.025. Third if the $\chi^2$ is still greater than $10^{-1}$, then the mutation and minimization steps are repeated with the same mutation width and a size of 0.0125. The above minimization strategy was used in two stages. First we used the global average of all available data and began with 45 initial neural networks and removed 2 every generation for 10 generations until 25 neural network replicas remained, $\{q_i(x)\}$. The response of the surviving population is shown in Fig.~\ref{Fig:TrainedNets} for each of the geometries studied. The standard deviation of the response of these neural networks is used to estimate the systematic error in this method of regression. 
\begin{equation}
\sigma_{sys}  = \textrm{StdDev} \big[ q_i(x)\big] \quad ; \qquad i = 1\dots N_{rep}
\end{equation}

\begin{figure}[h!]
\centering
\includegraphics[scale=0.58]{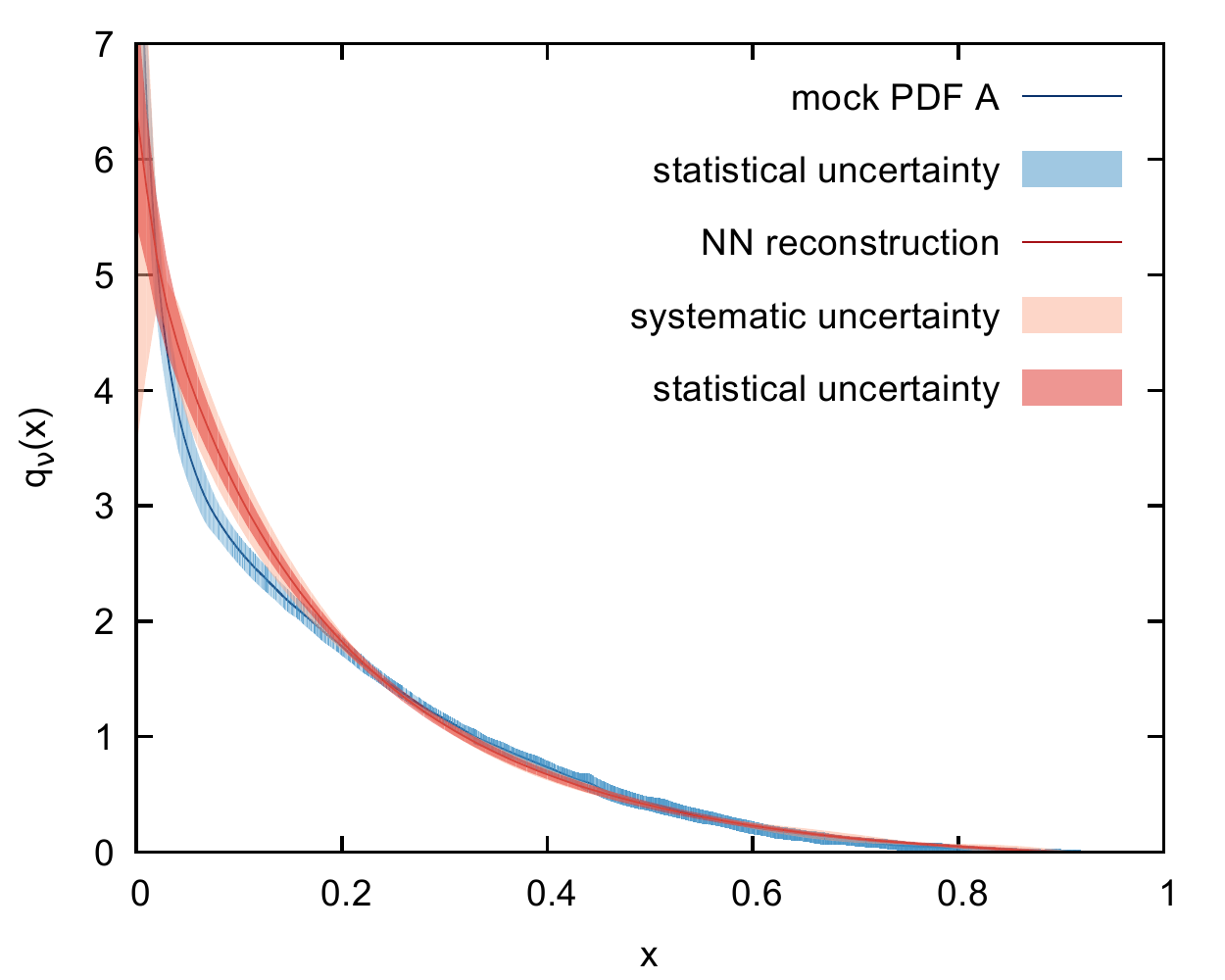}
\includegraphics[scale=0.58]{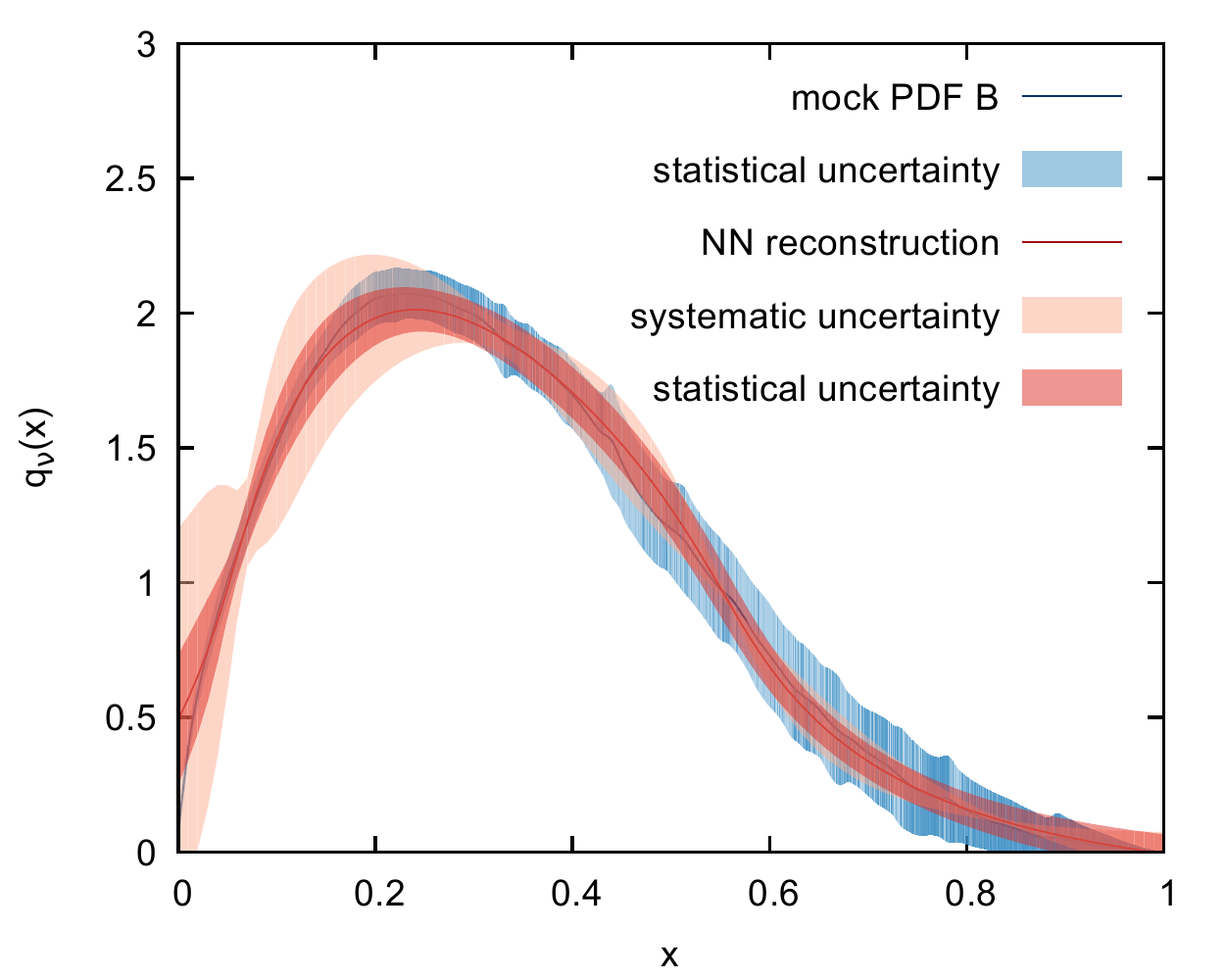}
\includegraphics[scale=0.58]{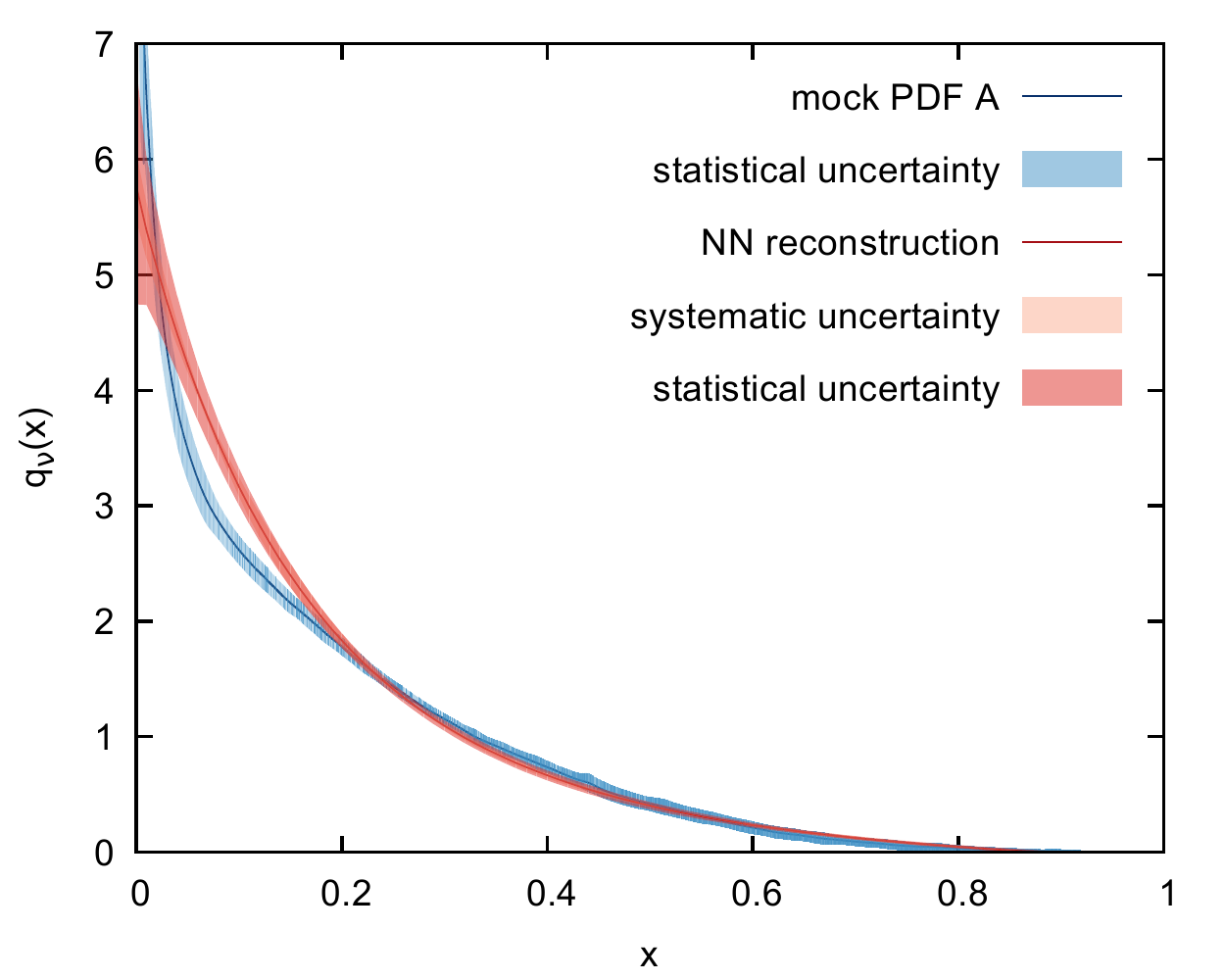}
\includegraphics[scale=0.58]{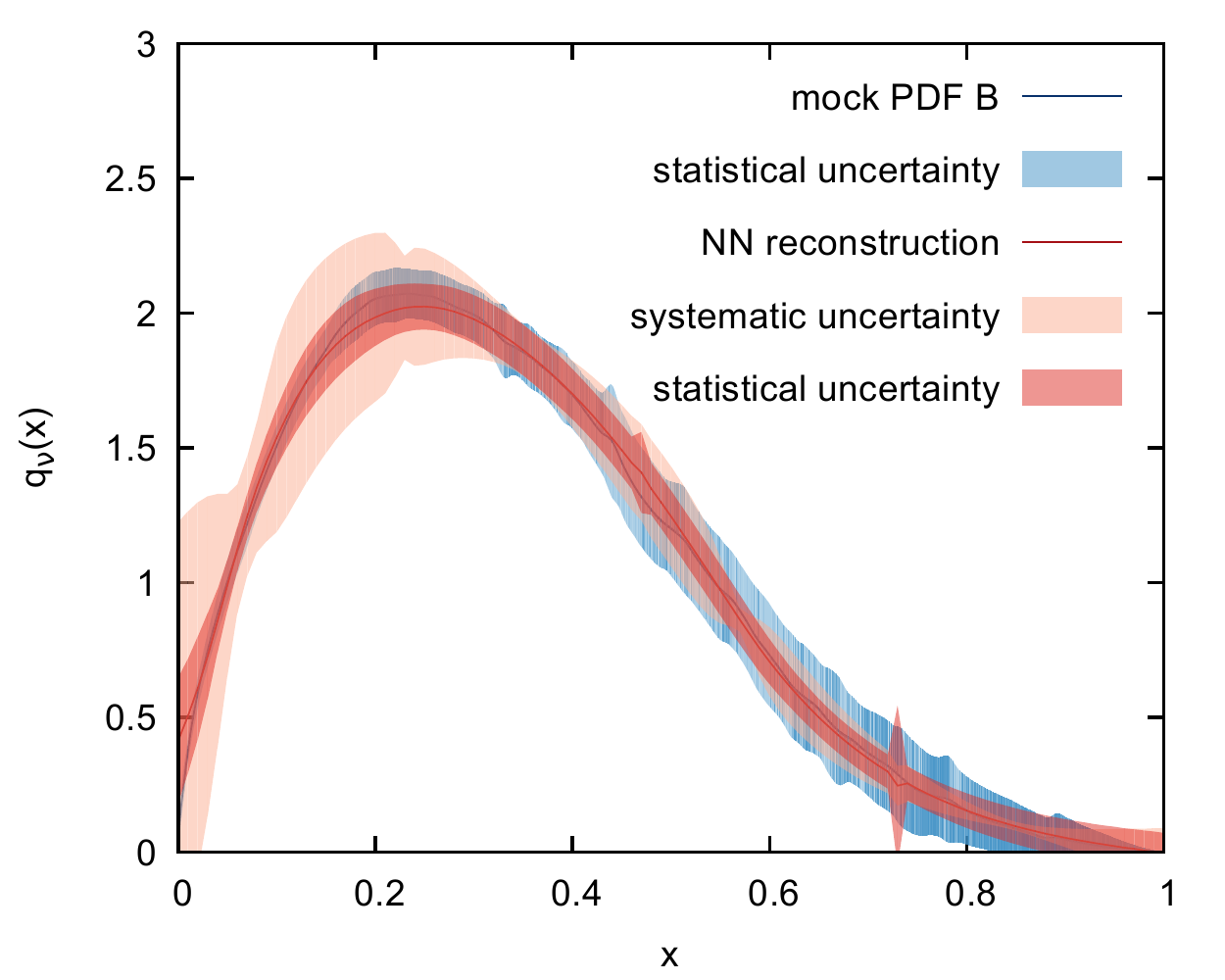}
\includegraphics[scale=0.58]{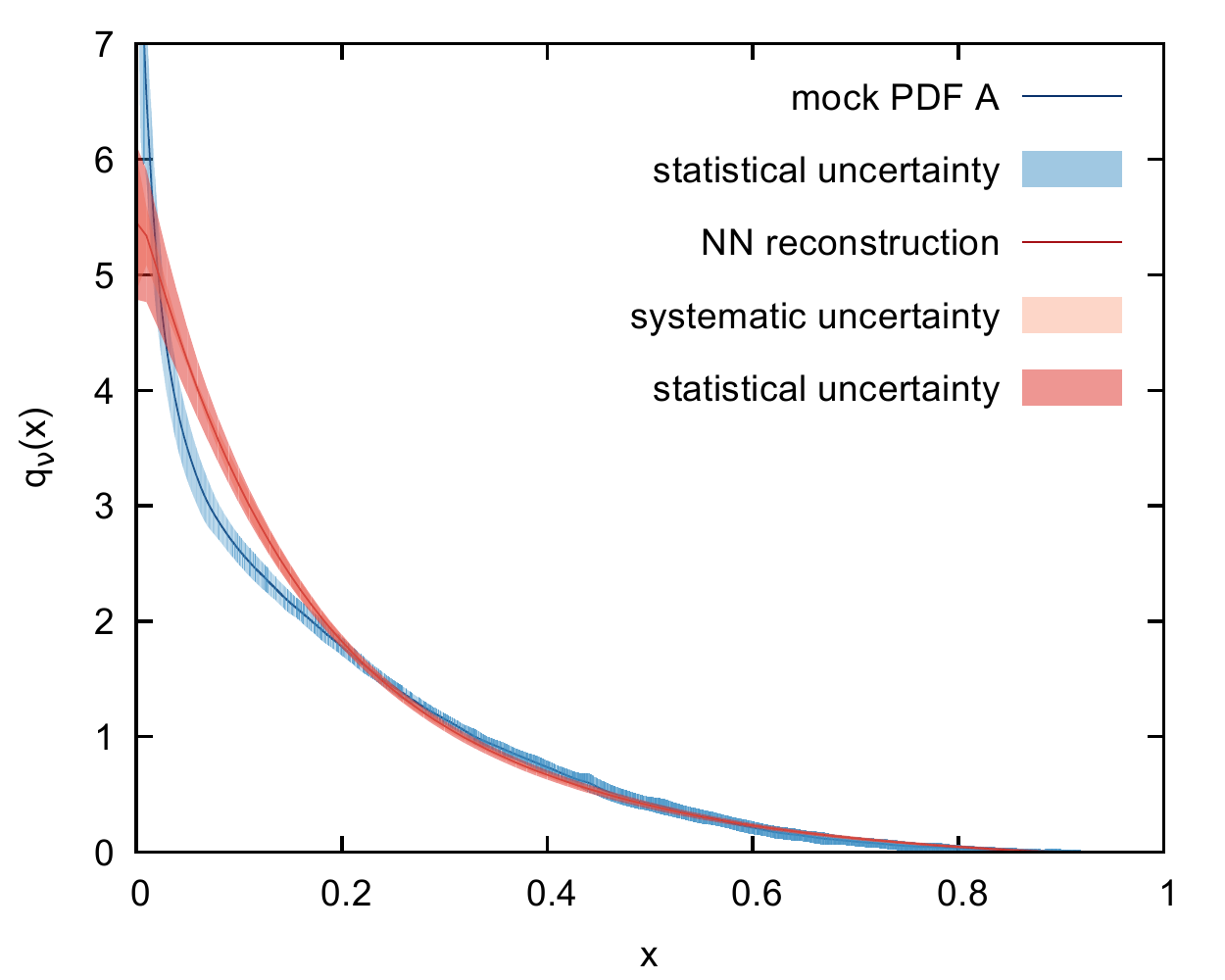}
\includegraphics[scale=0.58]{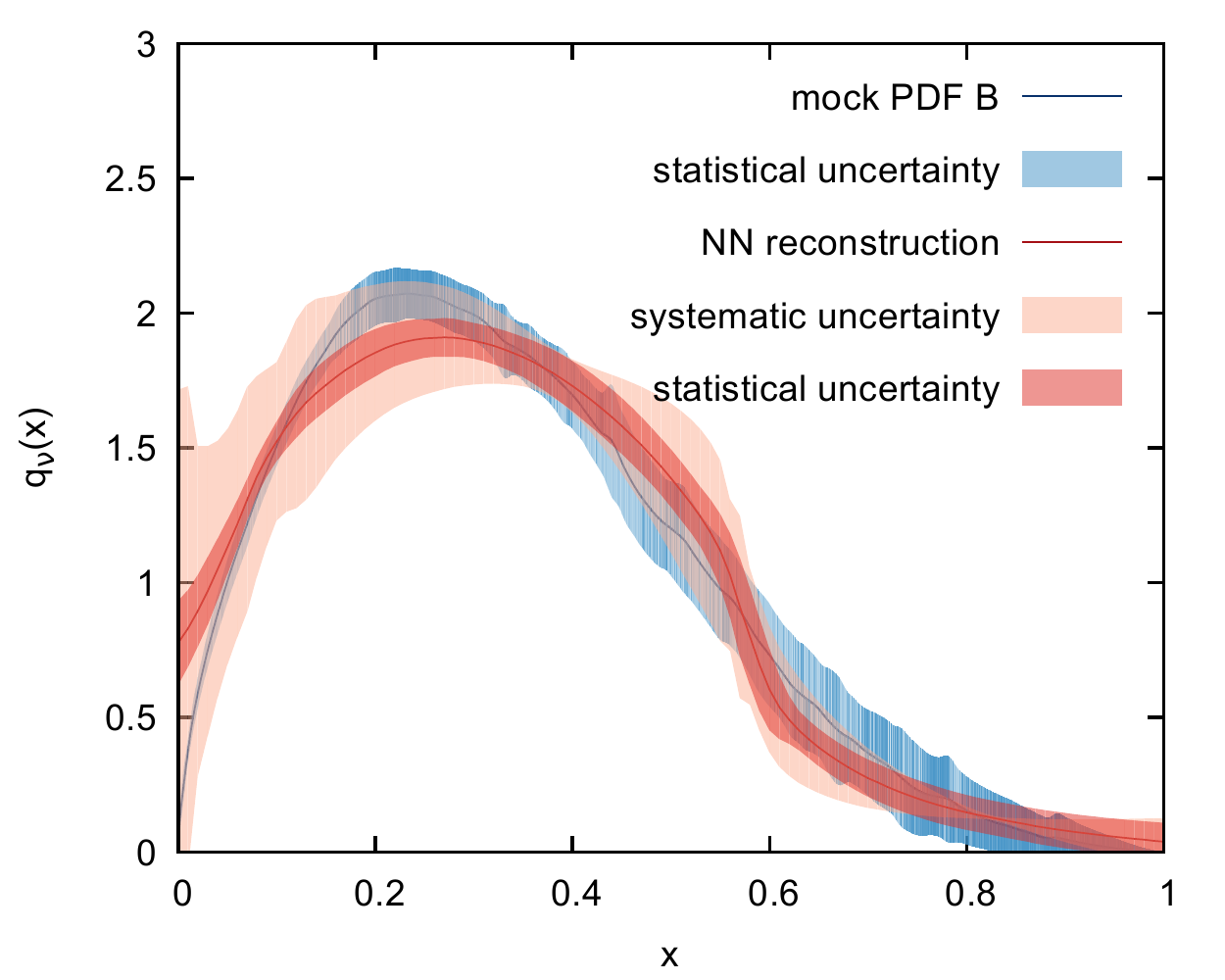}
\caption{The genetically trained neural nets. The blue band is the original data. The red band is the reconstructed PDF with statistical and systematic errors. The left column is with NNPDF data. The right column is with modified data. The first row has a network geometry of 1-3-1. The second row has a geometry of 1-4-1. The third row has a geometry of 1-2-2-1.}\label{Fig:NNPDFResults}
\end{figure}

In the second stage, we perform the minimization on each of the $N=100$ mock Ioffe time PDF replicas~\footnote{These the replicas the NNPDF collaboration provides in their data set. In a realistic lattice calculations these replicas would be a set of jackknife or bootstrap samples of the matrix elements.}  in order to estimate the statistical fluctuations of the resulting function. Starting with each of the networks in the final population, each PDF replica is used as the training data for another minimization to create the PDF replica dependent networks of the final population, $\{q^{(b)}_i(x)\}$. The variance of the response averaged over the members of the final population across the different PDF replica dependent networks is used to estimate the statistical error,
\begin{equation}
\sigma_{stat}  = \textrm{StdDev} \big[ \frac{1}{N_{\textrm{rep}}} \sum_{i=1}^{N_{\textrm{rep}}} q^{(b)}_i(x)\big] \quad ; \qquad b = 1\dots N.
\end{equation}
The average response of the PDF replica dependent networks and the combined systematic and statistical errors are shown in Fig.~\ref{Fig:NNPDFResults}. As anticipated, the errors tend to grow at small-$x$. Since the input data is over a truncated region of $\nu$, the information of the low-$x$ behavior is lost. The neural networks are able to precisely reconstruct the large-$x$ region while in the low-$x$ region the flexible parametrization allows for a wide range of functions that can reproduce the data. One can also see that the various network geometries all give comparable results for the reconstructed PDF.

\begin{figure}[h!]
\centering
\includegraphics[scale=0.58]{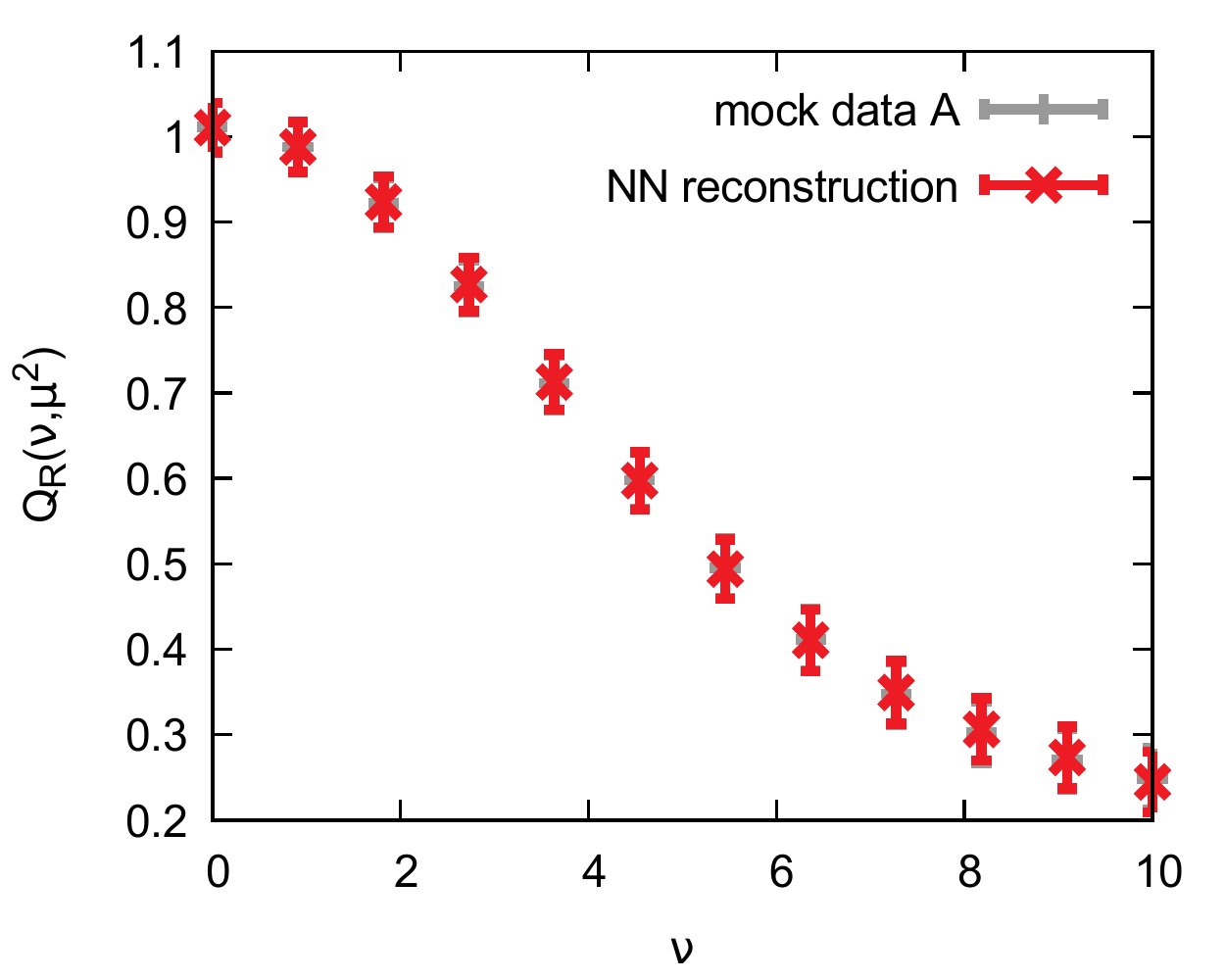}
\includegraphics[scale=0.58]{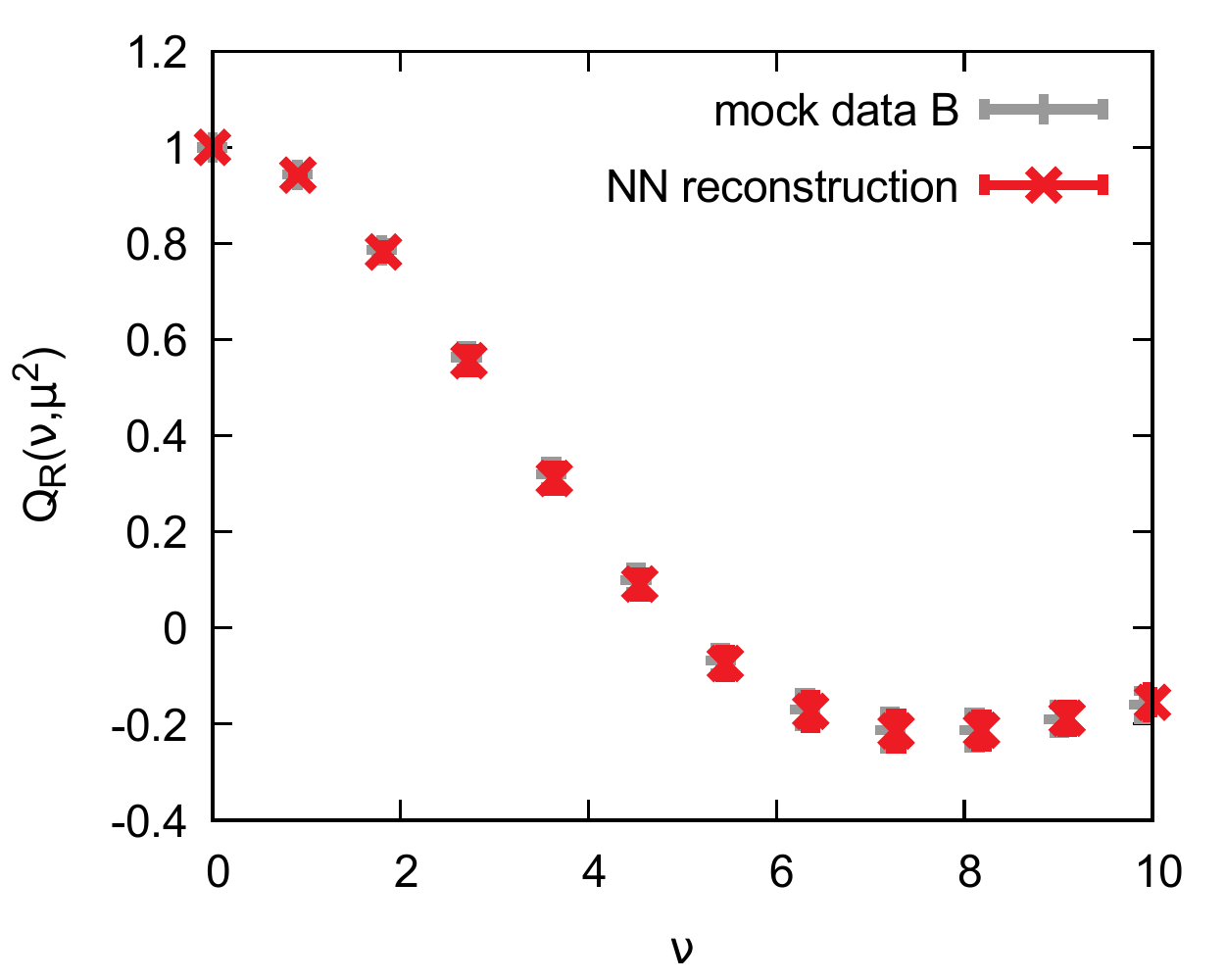}
\includegraphics[scale=0.58]{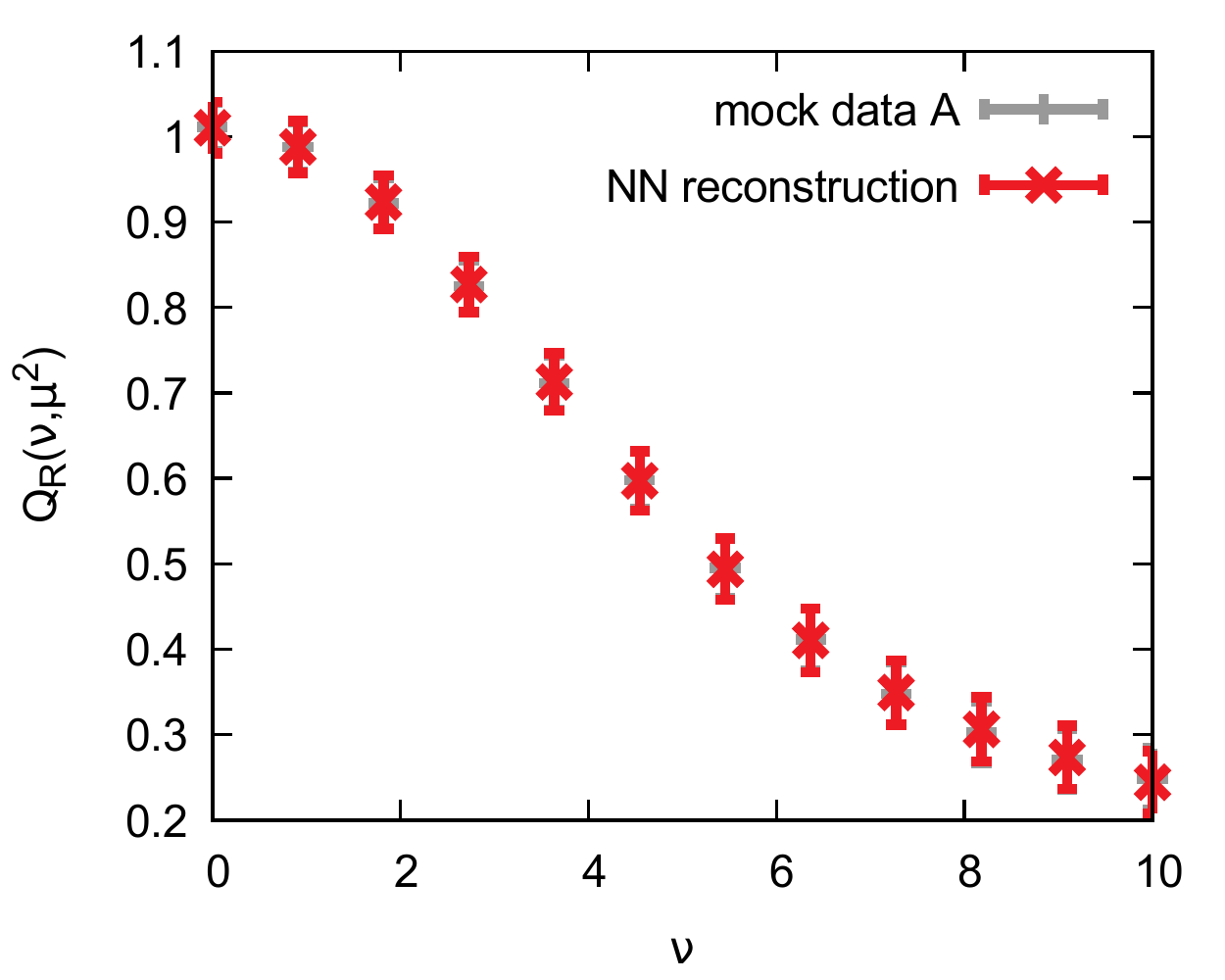}
\includegraphics[scale=0.58]{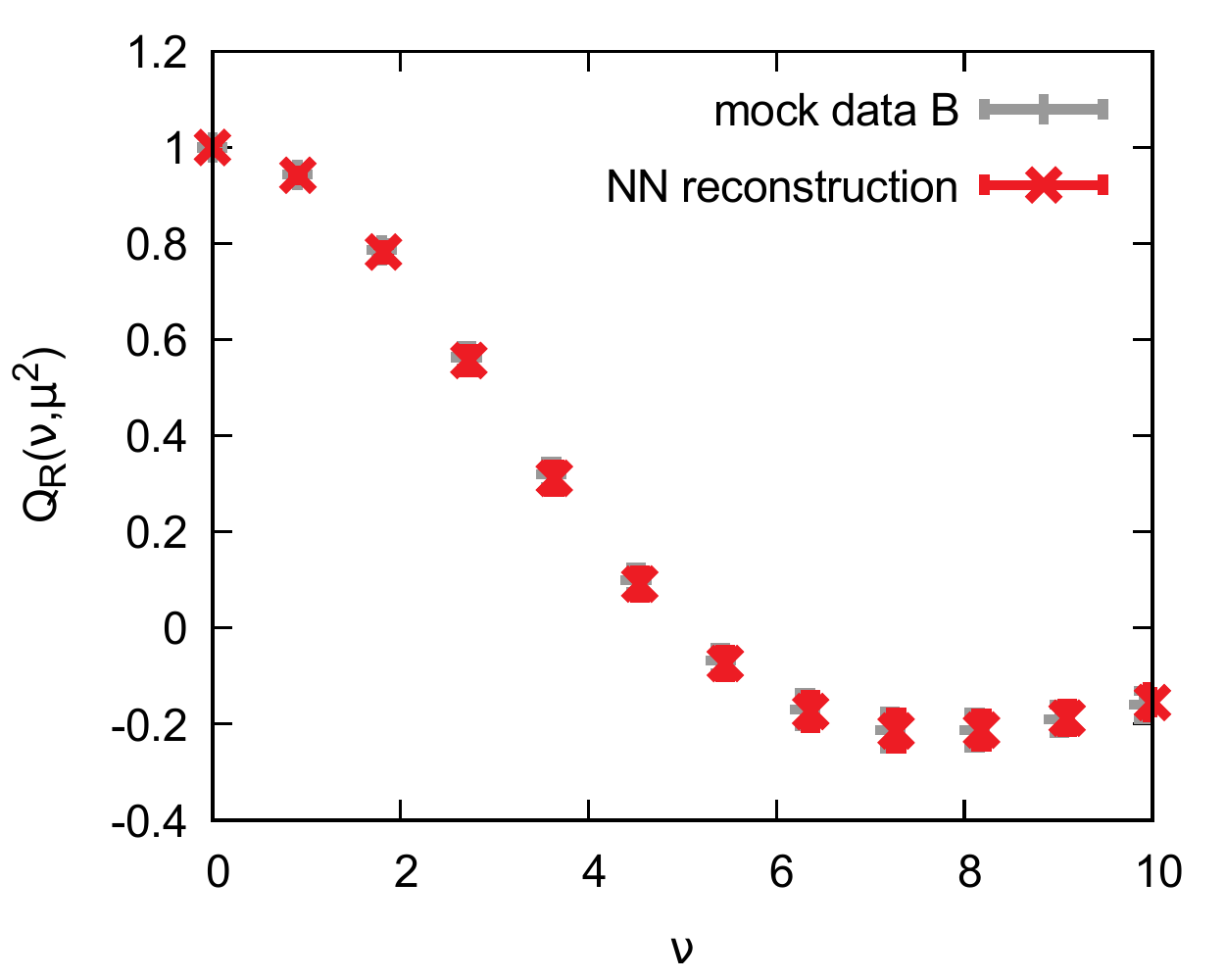}
\includegraphics[scale=0.58]{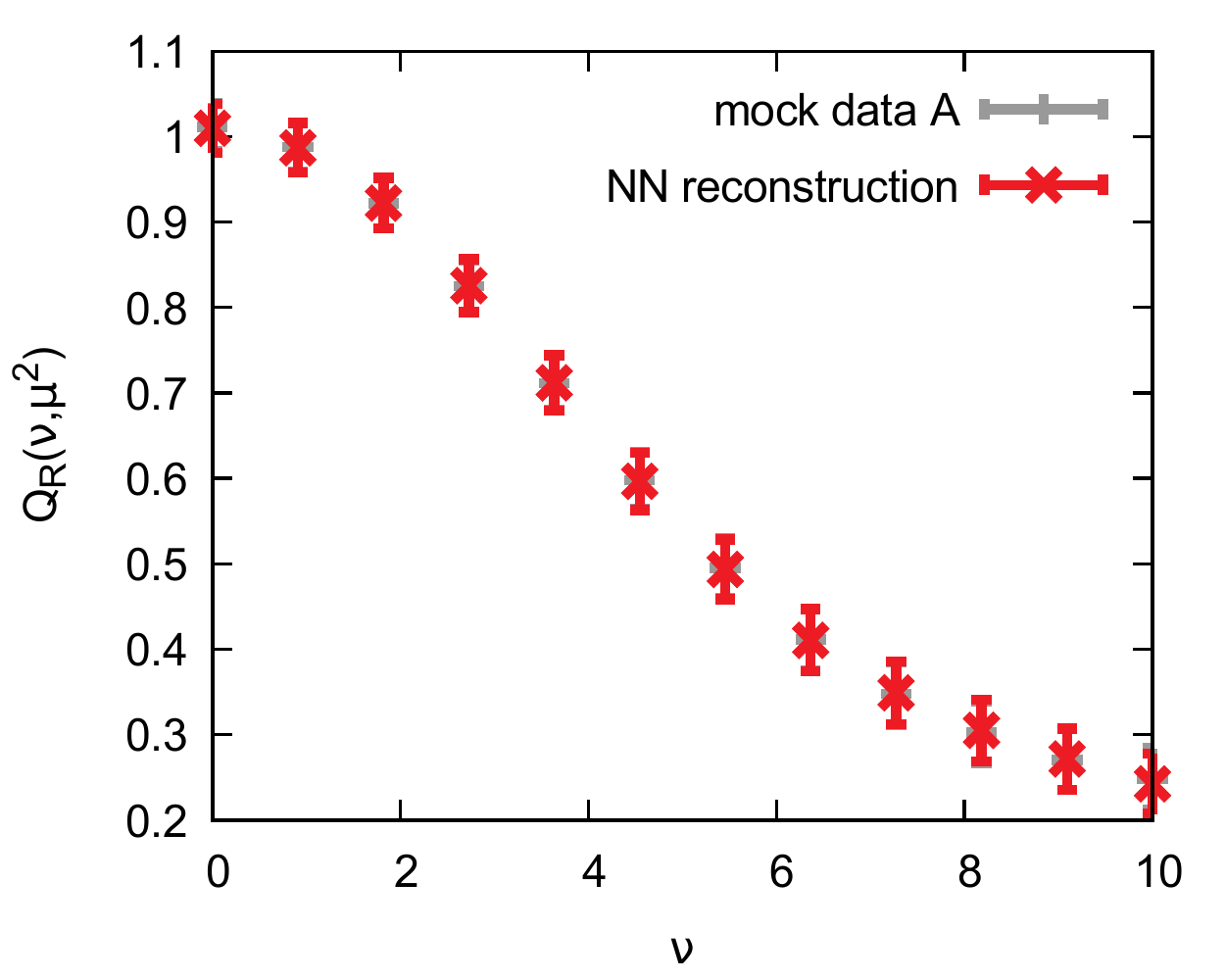}
\includegraphics[scale=0.58]{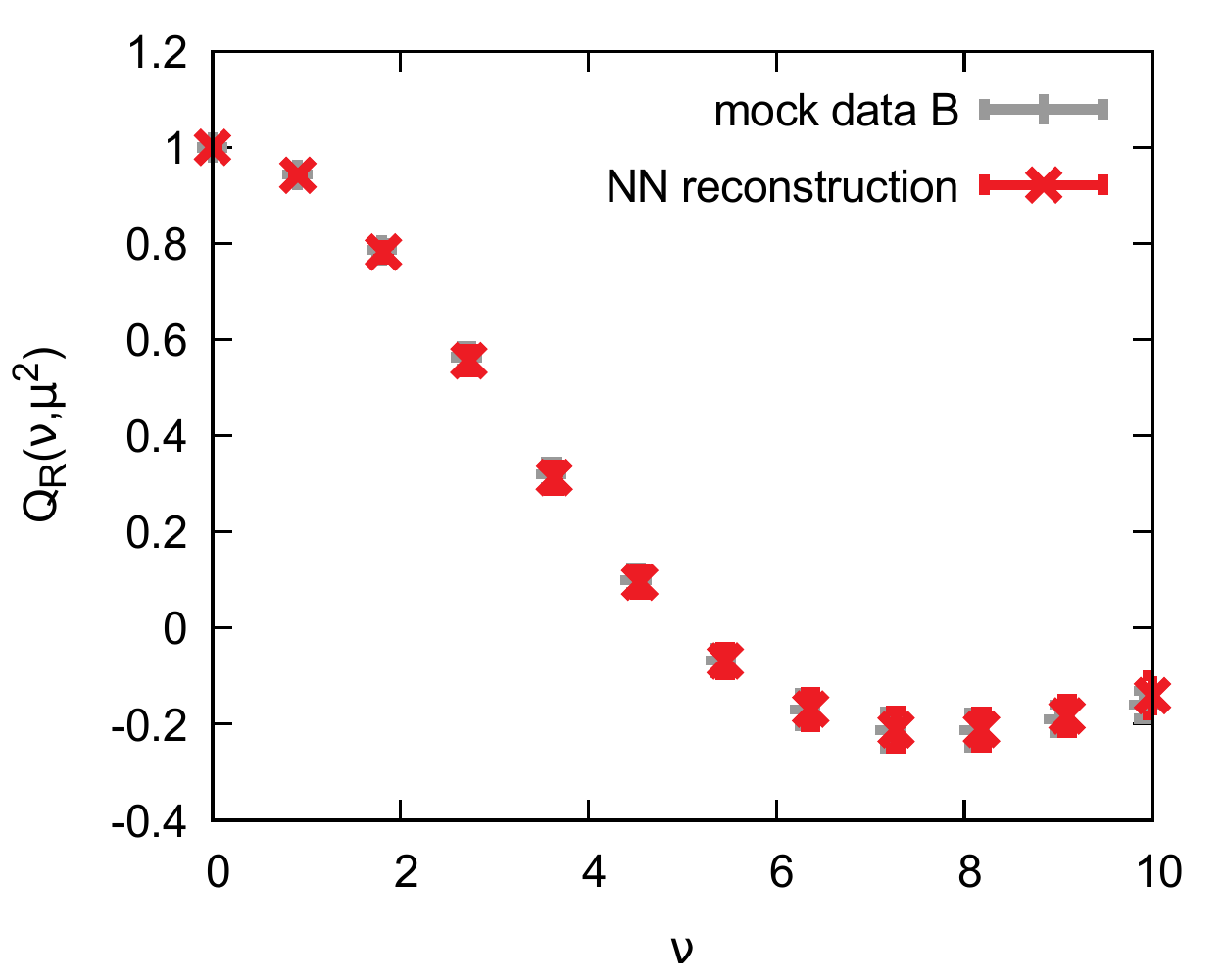}
\caption{The genetically trained neural nets. The grey points are the original data while the red points are the ones reconstructed by the NN. The left column is with NNPDF data. The right column is with modified data. The first row has a network geometry of 1-3-1. The second row has a geometry of 1-4-1. The third row has a geometry of 1-2-2-1.}\label{Fig:NNPDFResults_points}
\end{figure}

\begin{figure}[h!]
\centering
\includegraphics[scale=0.58]{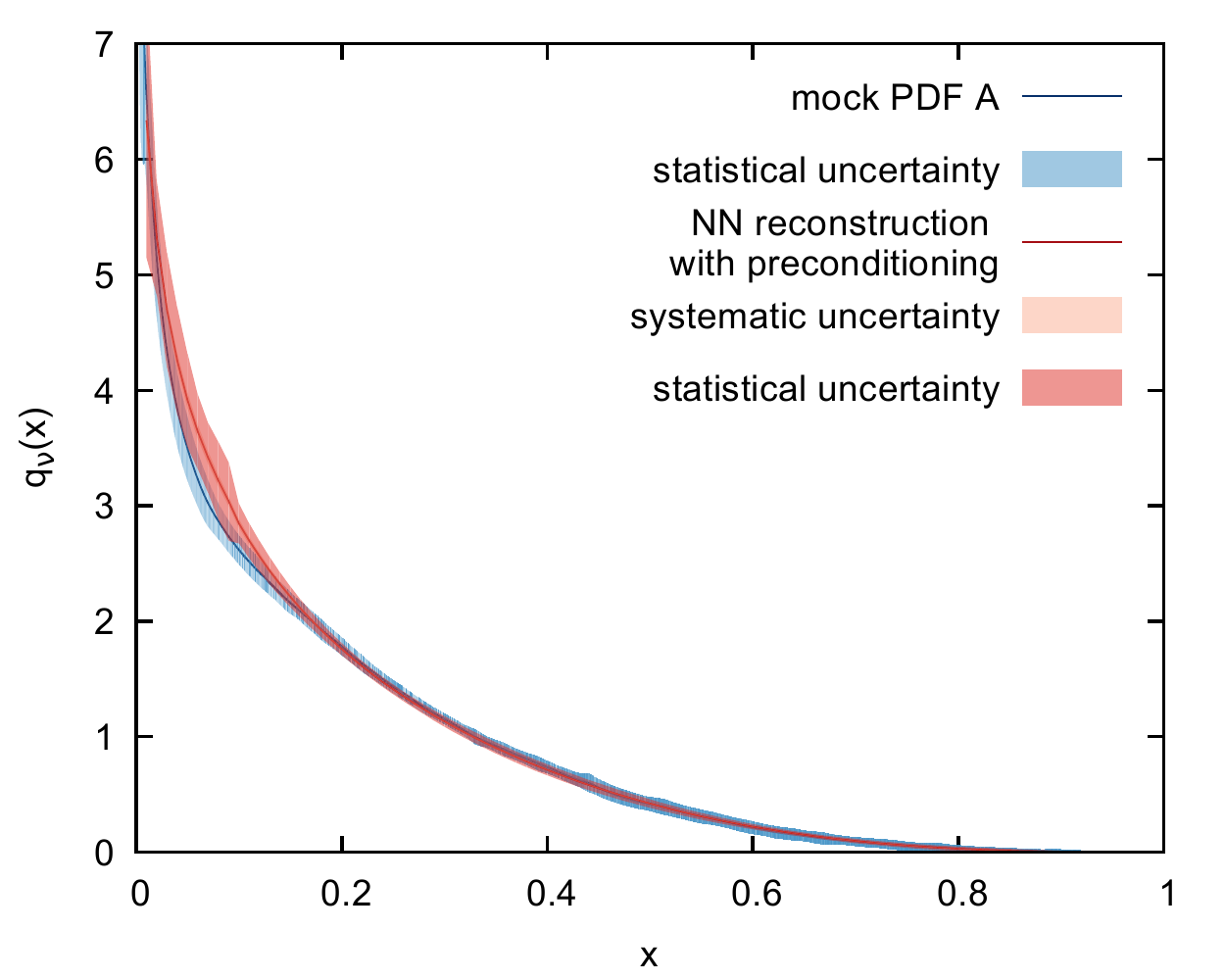}
\includegraphics[scale=0.58]{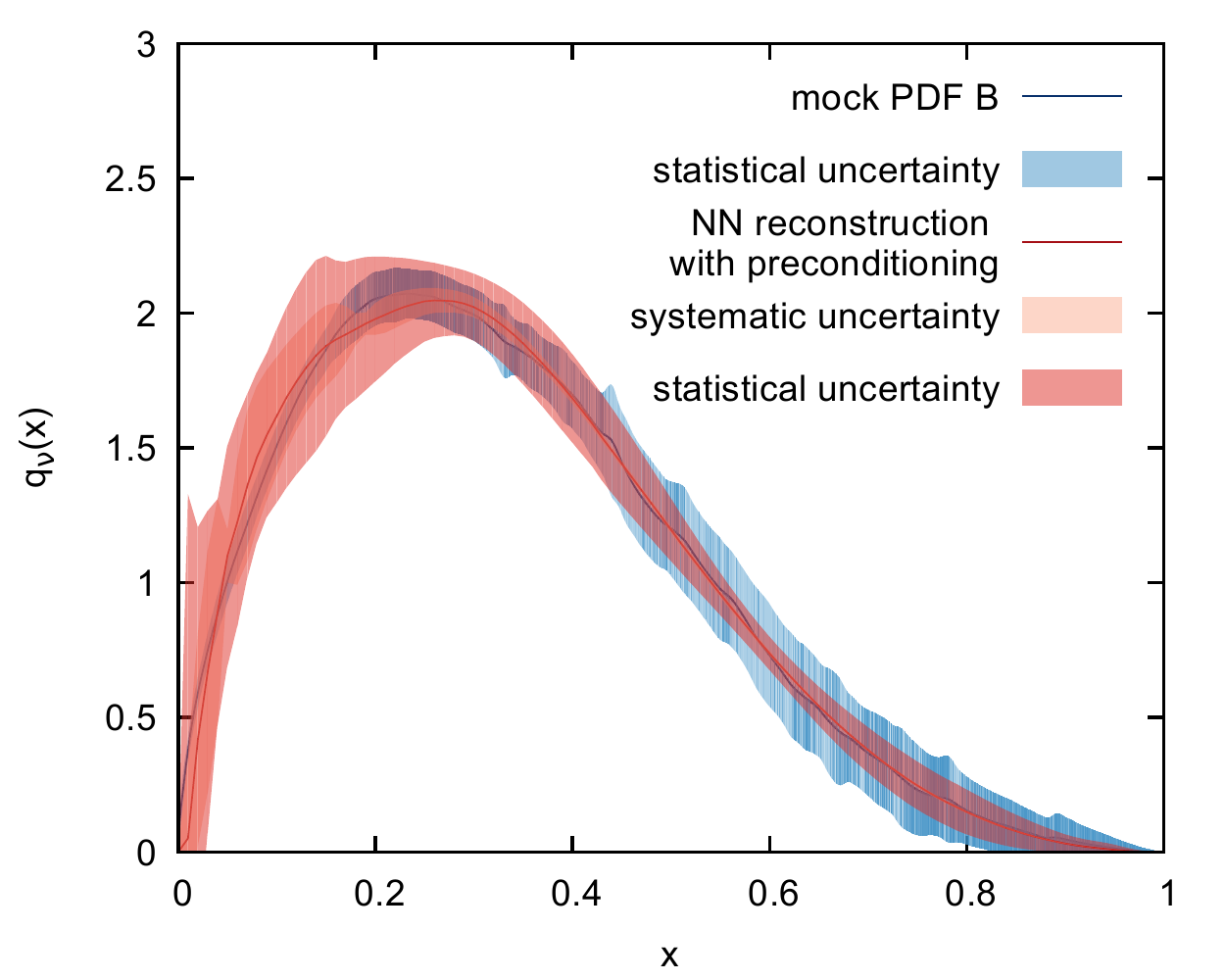}
\includegraphics[scale= 0.58]{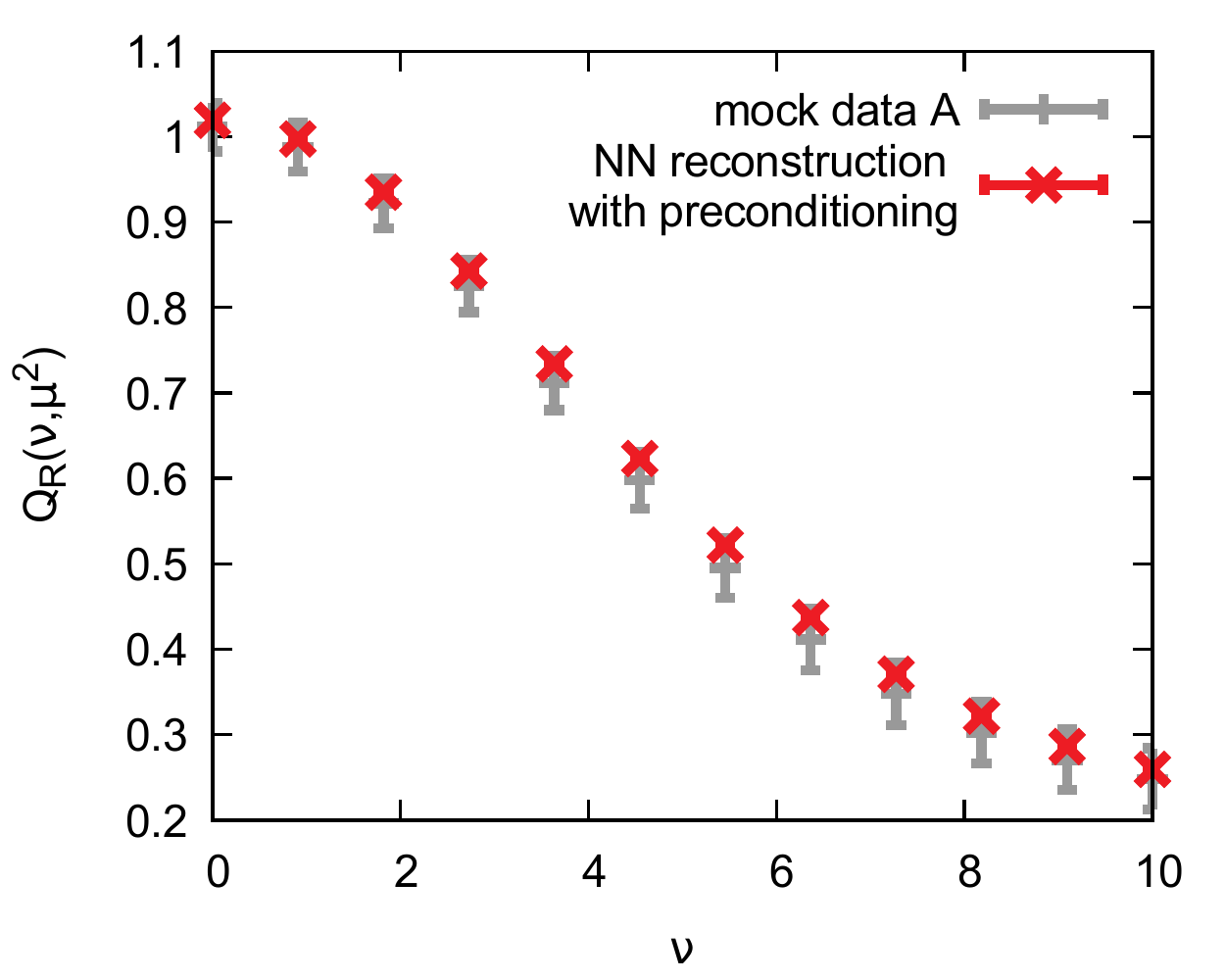}
\includegraphics[scale=0.58]{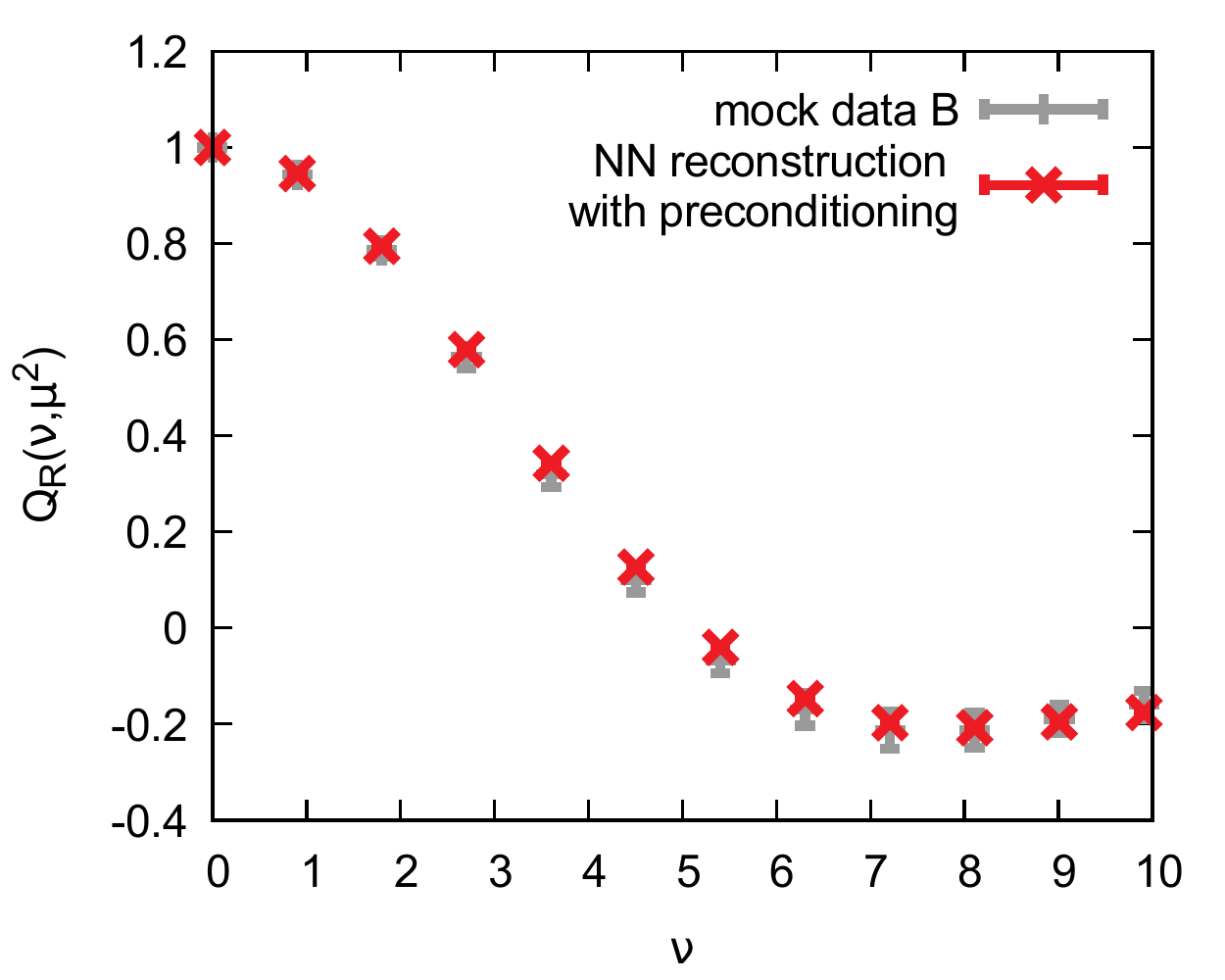}
\caption{The PDF reconstructions from preconditioned Neural Networks (top).
The grey points are the original data while the red points are the ones reconstructed by the NN (bottom). The left is scenario A and the right is scenario B.}
\label{Fig:NNPDFprecon}
\end{figure}

Next we consider how a preconditioning function affects the reconstructions. Using the function defined in Eq.~\eqref{eq:p-func}, the preconditioning function for scenario A with $a=-0.25$ and $b=2$ and for scenario B with $a=0.3$ and $b=2$. Both of these cases were tested on the 1-3-1 geometry. The reconstructed PDFs, shown in Fig.~\ref{Fig:NNPDFprecon}, are not significantly different for the majority of large $x$ region. For scenario A, the low $x$ behavior is slight improved compared to the case without preconditioning. This behavior is perhaps to be expected, because the preconditioned neural network is parameterizing a much slower varying function in that region compared to the divergent PDF.

Our conclusion is that for ranges of $\nu$ which are realistic in modern lattice calculations, a neural network is capable of reconstructing the PDF for a wide range of $x$ without the need of preconditioning, though the neural networks can benefit from it. This study of neural network parametrization of the PDF is by no means complete. There are a number of choices that were made which may not be optimal. It is clear that a more dedicated study of this approach is required in order to understand its full potential. However, we find the results obtained here very encouraging and thus we plan to further investigate this approach in a future publication. 

\subsection{Bayesian Analysis}

Finally, mock data tests of the Bayesian strategy outlined above are carried out to determine the feasibility of extracting the $x$-space PDF $q(x)$ from the Ioffe-time data $\mathcal{Q}(\nu)$ in realistic settings for applications to lattice generated data. The $x$ interval is discretized in $N_x=2000$ steps to have fine enough resolution of the possibly highly oscillating cosine. To remain as close as possible to the situation encountered with real data, we first carry out a simple fit on the noisy mock Ioffe-time data, based on the simple Ansatz Eq.~\eqref{eq:p-func} via the expression in Eq.~\eqref{Eq:FitFunc}. 

\begin{figure}[h!]
\centering
\includegraphics[scale=0.58]{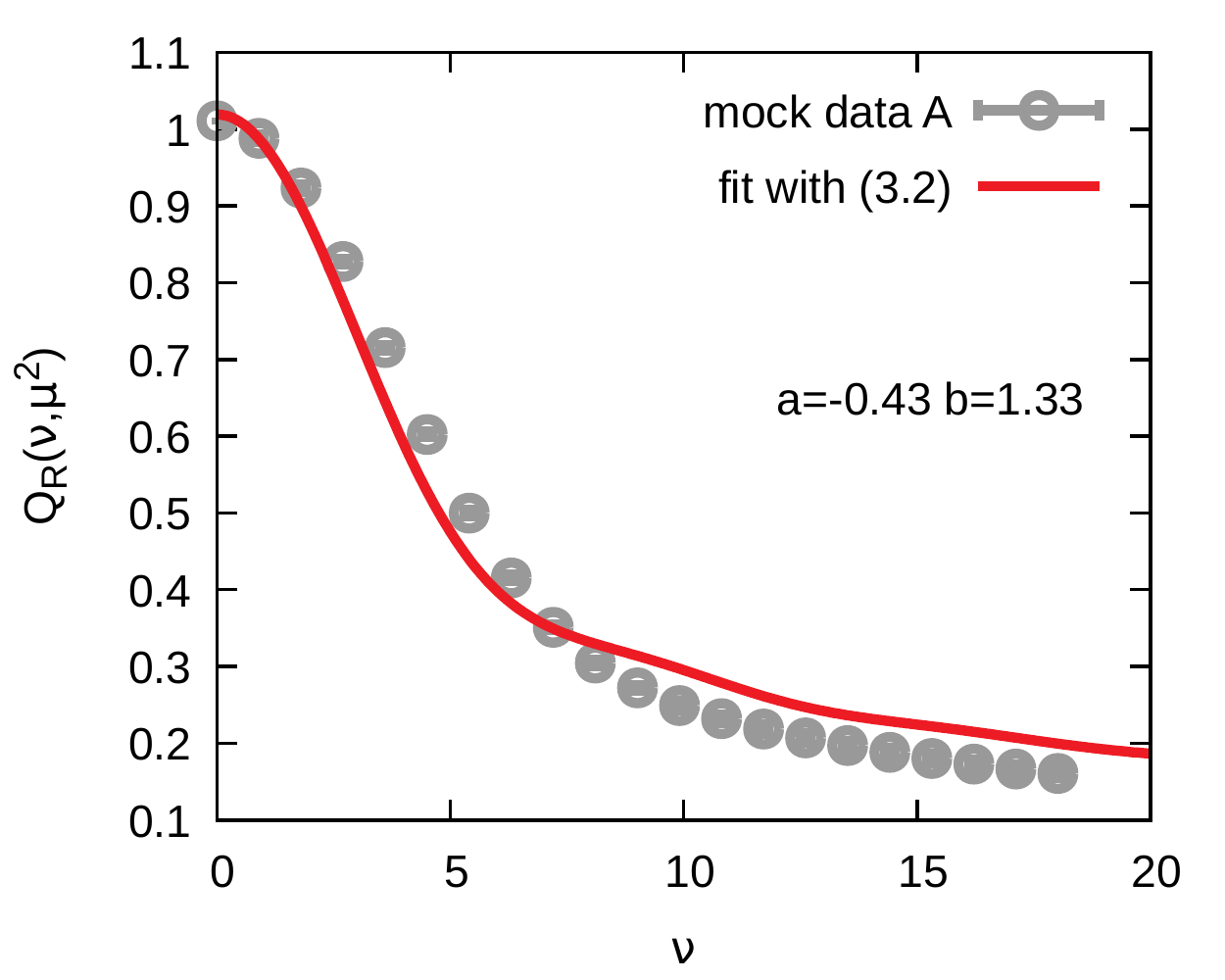}
\includegraphics[scale=0.58]{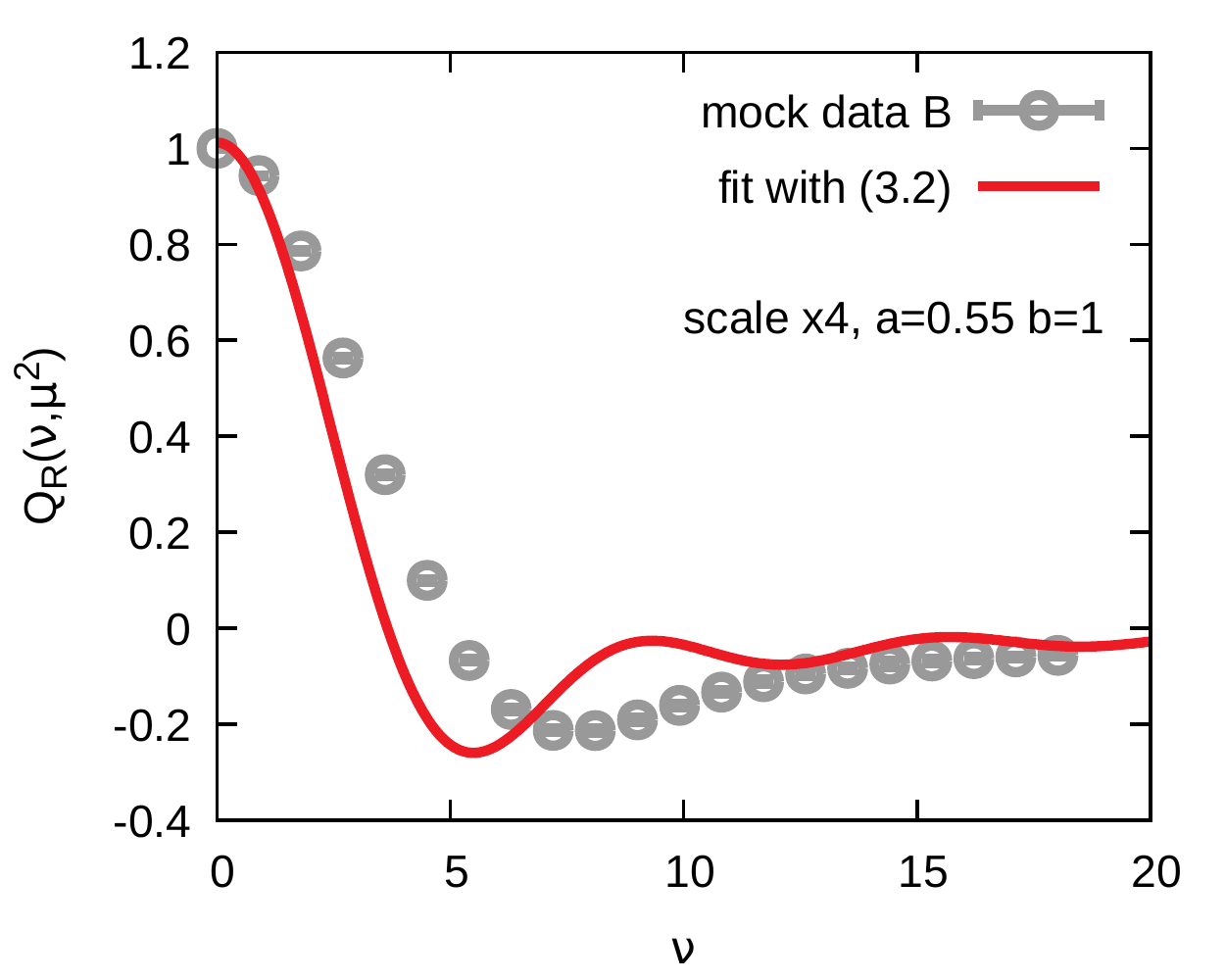}
\caption{Mock data $\mathcal{Q}(\nu)$ in the interval $I_3$ from (left) realistic PDF data [mock scenario A] and (right) a modified scenario with the PDF vanishing at the origin [mock scenario B]. In red we show the corresponding best fit of the noisy data with Eq.\eqref{Eq:FitFunc}.
 \label{Fig:FitsData}}
\end{figure}

\begin{figure}[h!]
\includegraphics[scale=0.6]{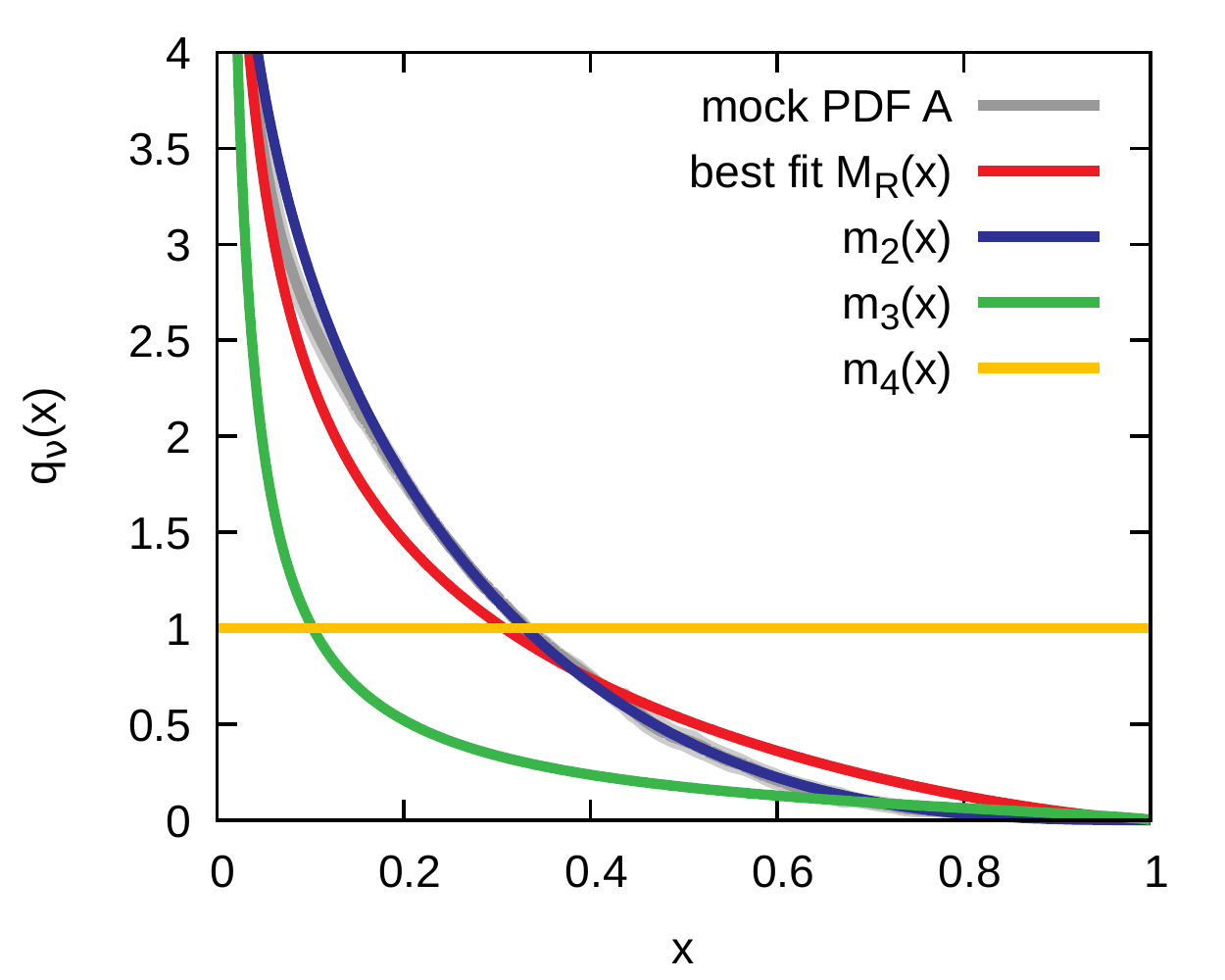}
\includegraphics[scale=0.6]{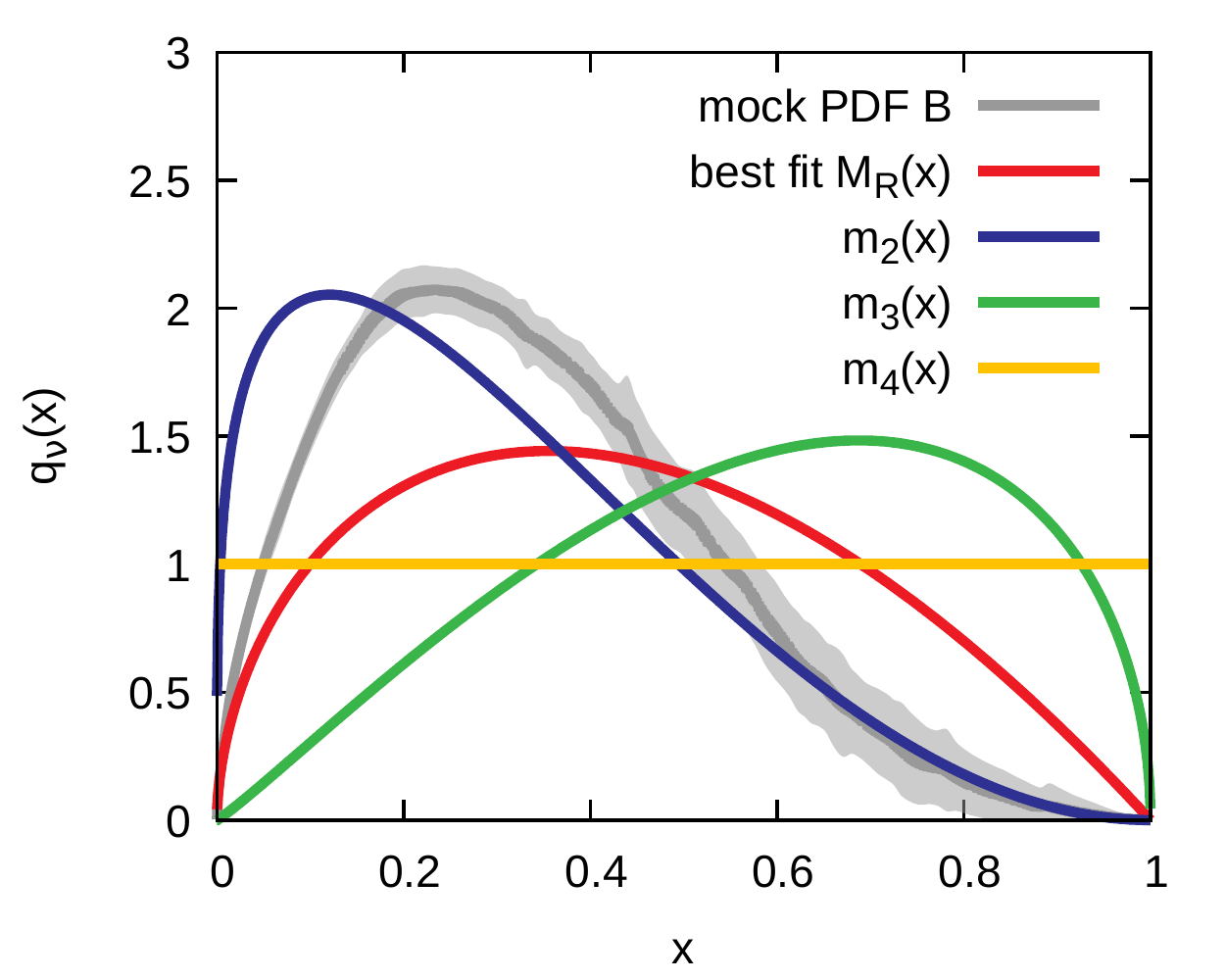}
\caption{Best fit PDF (red solid line) from (left) realistic PDF data [mock scenario A]  and (right) a modified scenario with the PDF vanishing at the origin [mock scenario B]. The actual mock PDF in the former cases is given as gray solid line. To determine the dependence of our results on the choice of default model, three further choices for $m$ are plotted, two arising from varying the best fit parameters by factors of 2, one being the constant default model m=1.}\label{Fig:FitsDM}
\end{figure}

In both scenario A and B we find that the simple functional form~\eqref{Eq:FitFunc} allows us to capture the overall features of the Ioffe-time data very well, as shown in Fig.~\ref{Fig:FitsData}. The best fit result for $q(x)$ is shown in Fig.\ref{Fig:FitsDM} as a red curve and compared to the actual mock PDF, given as gray dashed line. At intermediate-$x$ values, the best fit in both scenarios deviates from the correct $q(x)$ but close to $x=0$ provides a rather good description. This is exactly what we have in mind: the fit will provide us with prior information about the non-analytic behavior of the function, which we can use as default model in the Bayesian reconstruction. The Bayesian approach will then imprint the information encoded in the simulation data as deviations from the default model onto the end result.

Note that in order to use the best fit $q(x)$ as default model we need to modify its functional form at the boundaries of the $x$-interval in both cases. If there is e.g. a pole at the origin, then the value of $m$ is not well defined there. On the other hand if the function vanishes exactly at the origin, then the prior assumption of positive definiteness in some of the Bayesian methods is not fulfilled. Therefore we set $m(0)=q_{\rm best\,fit}(\Delta x)/2$. To further understand how strongly our end result depends on the choice of default model, we also carry out reconstructions with other functions than the best fit one, two of which are obtained from varying the best fit parameters by factors of two and one function simply being the constant $m(x)=1$.

\begin{figure}[t]
\centering
\includegraphics[scale=0.58]{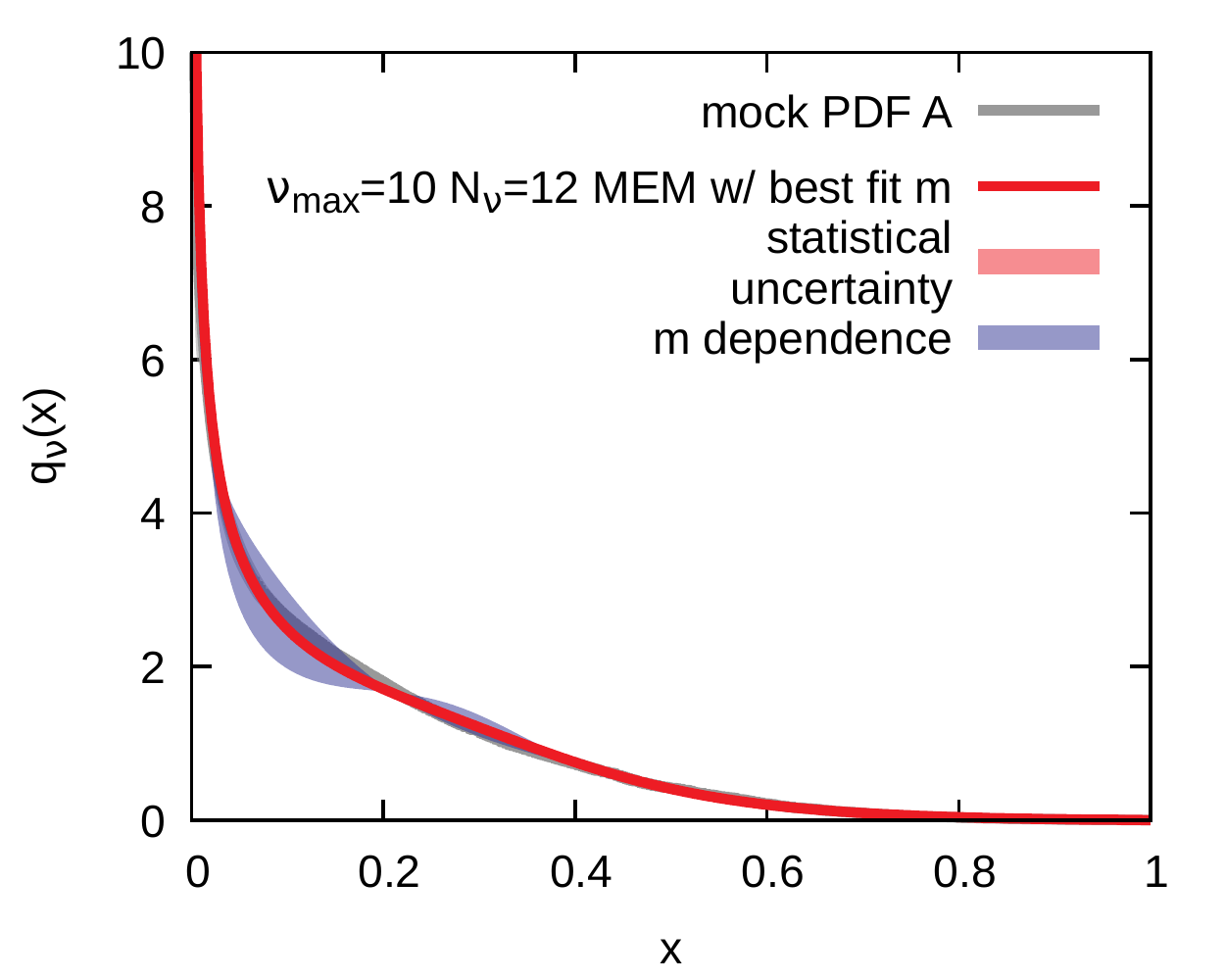}
\includegraphics[scale=0.58]{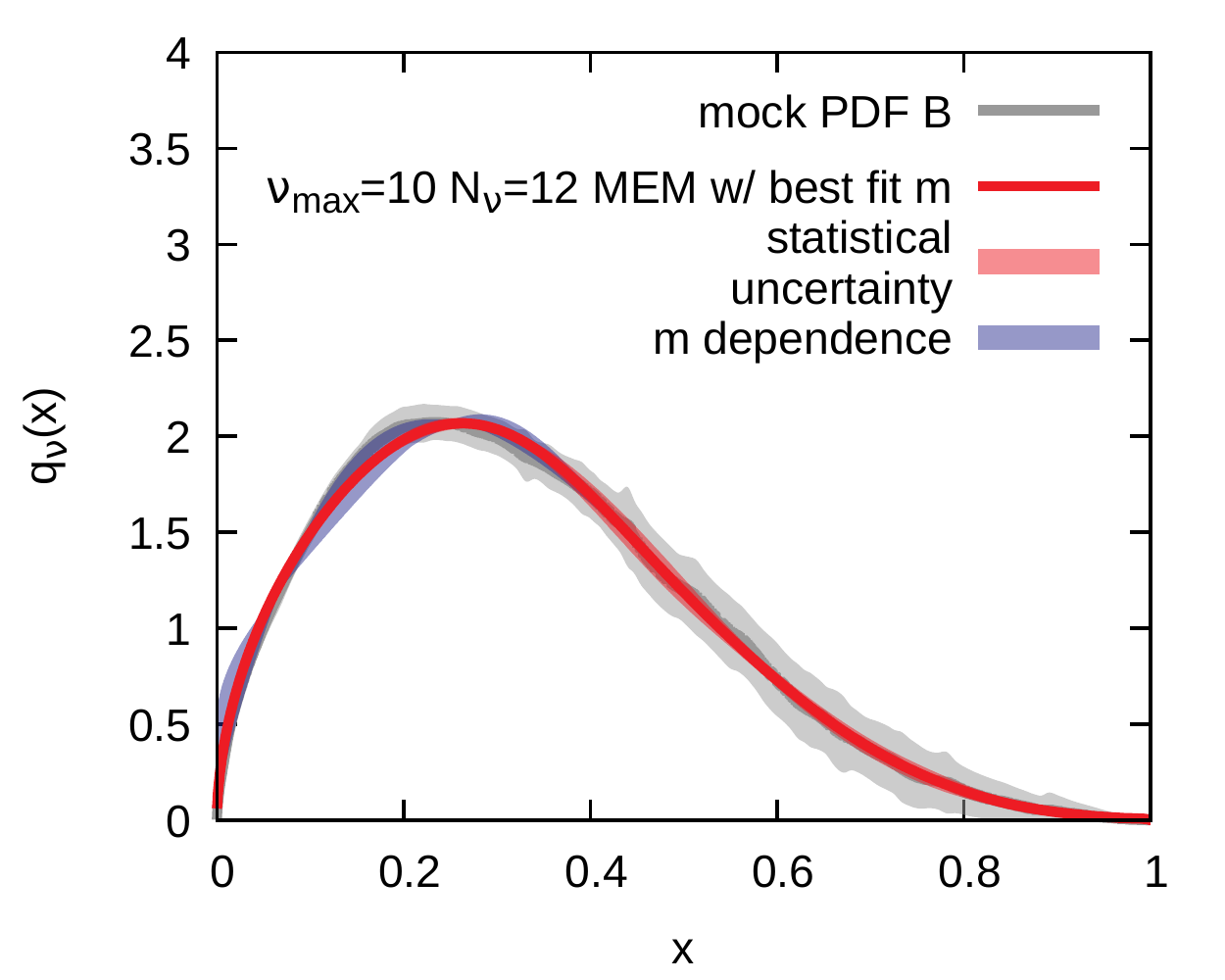}
\includegraphics[scale=0.58]{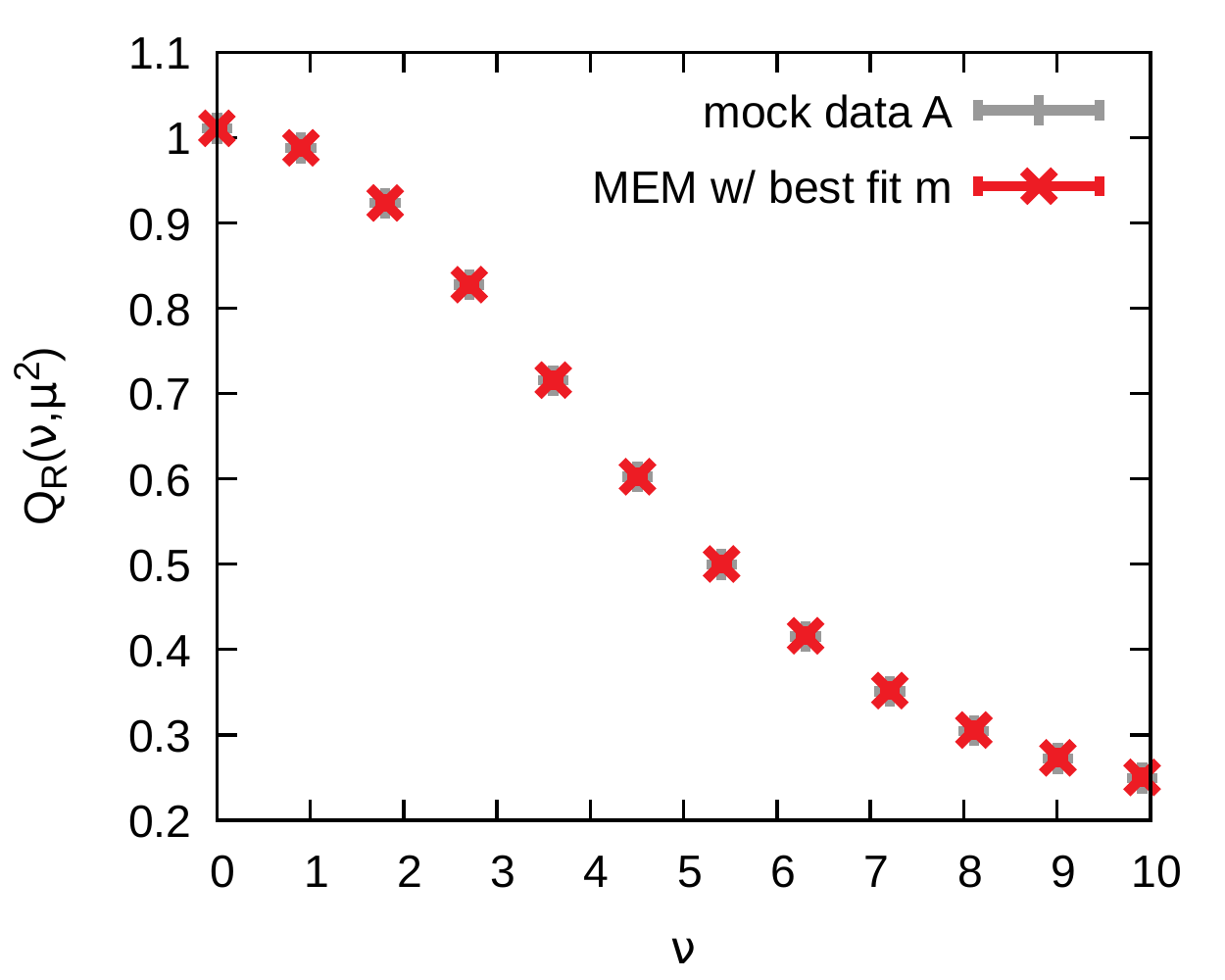}
\includegraphics[scale=0.58]{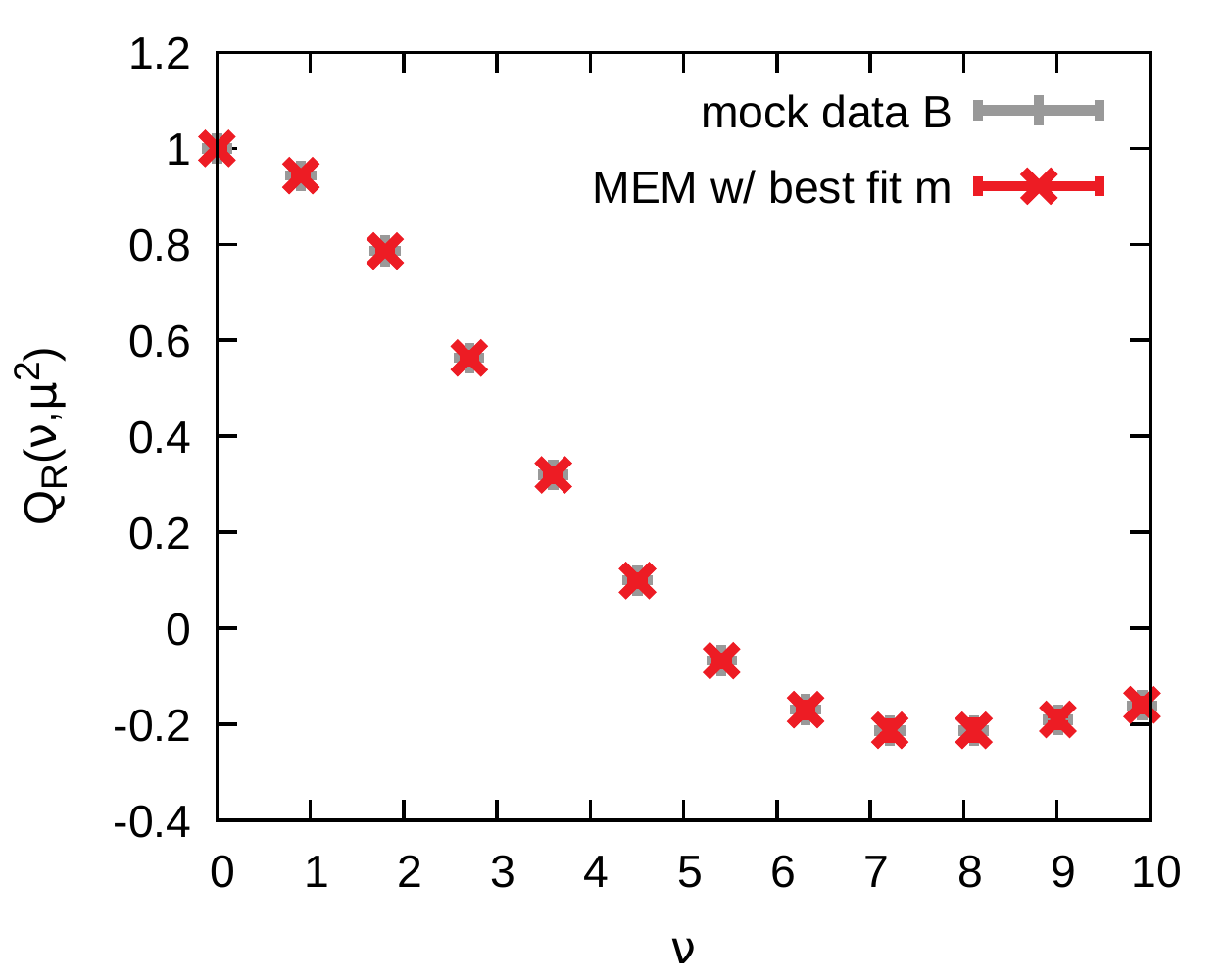}
\caption{$x$-space PDF's reconstructed using the Maximum Entropy method (MEM) from $N_\nu=12$ Ioffe-time data points on the interval $\nu=[0,10]$ (top) as well as the input data (gray crosses) compared to the data arising from the reconstructed PDF (red crosses) in the bottom panels. The plots in the left column denote the results for mock data based on a phenomenological PDF, while the right column from the modified scenario where the PDF vanishes at the origin. (top) The original mock PDF is given as gray dashed  line and the result based on the BR method with the best fit default model as solid red line. Statistical uncertainty is visualized by a red, systematic uncertainty from the choice of default model by a blue errorband respectively. (bottom) The red solid line corresponds to the relative deviation of the reconstruction from the input mock PDF, the red errorband encodes the combined uncertainty of the BR reconstruction, the gray errorband the uncertainty of the original mock PDF.}\label{Fig:ResultsMEM10}
\end{figure}

We start with the most realistic and most challenging setting, where the input data $\nu$ is in the range $I_1$, i.e. $\nu=[0,10]$ discretized with $N_\nu=12$ points, very similar to what is  currently available in lattice QCD simulations. Due to the availability of a good approximate solution obtained from the fit with eq.\eqref{eq:p-func}, provided as default model, we first consider the Maximum Entropy Method, whose results are presented in Fig.\ref{Fig:ResultsMEM10}

The original mock $q(x)$ is shown on the top of Fig.~\ref{Fig:ResultsMEM10} as a gray dashed line, and the MEM reconstruction based on the best fit default model in red. The red errorband denotes the statistical uncertainty, while the blue band arises from the default model dependence. In the bottom row panels the input Ioffe-time data (gray) as well as the data corresponding to the Bayesian reconstruction (red) are shown.  We find that the reproduction for both scenario A and scenario B is excellent, i.e. the MEM is able to utilize the approximate solution provided by the default model to regularize the inverse problem and reproduce the input PDF within uncertainties. 
If one inspects the reconstruction in scenario B carefully around the maximum one finds that the MEM solution slightly underestimates the mean below $x=0.25$ and overestimates it for $x>0.25$. The deviation however is minute. 

\begin{figure}[t]
\centering
\includegraphics[scale=0.58]{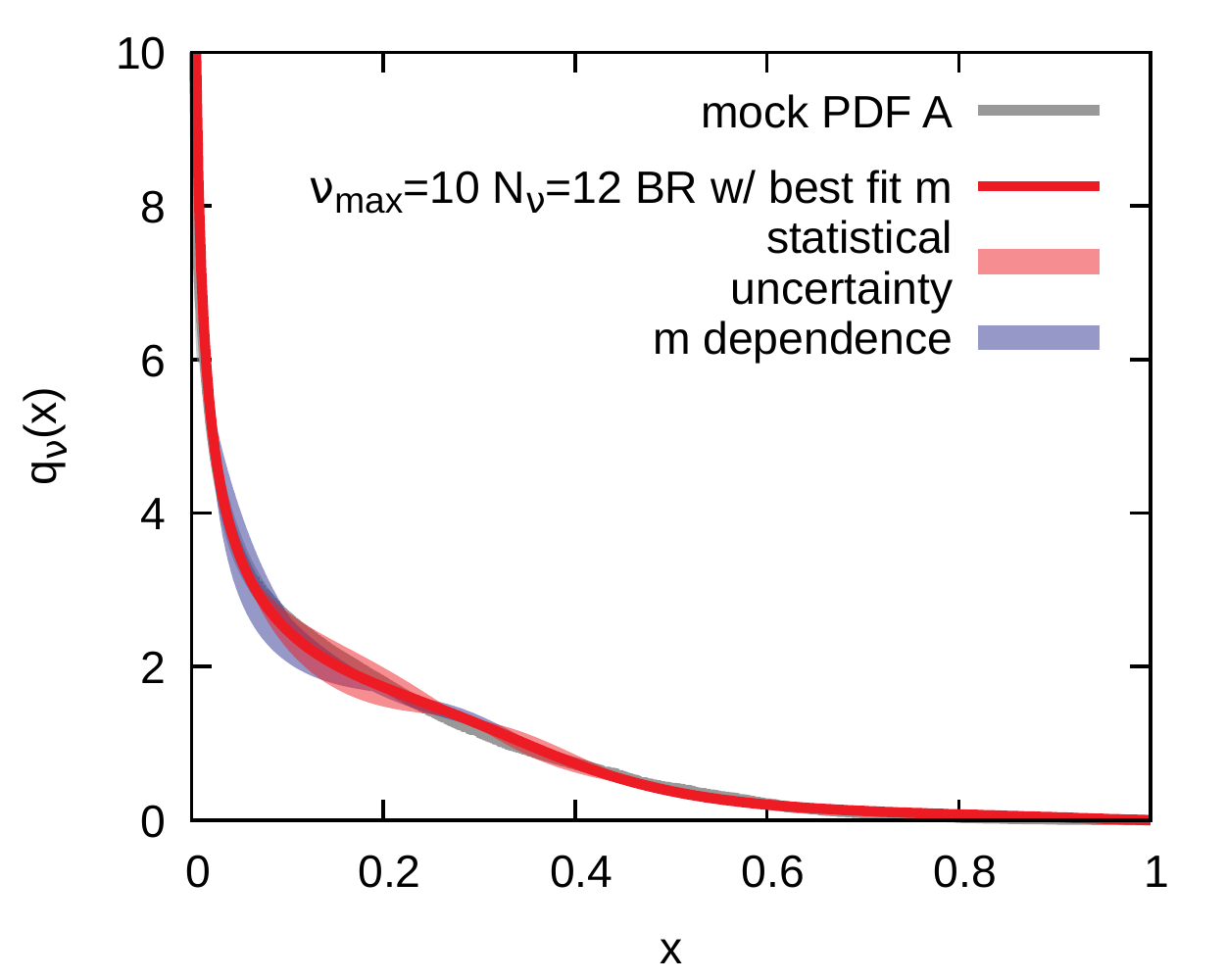}
\includegraphics[scale=0.58]{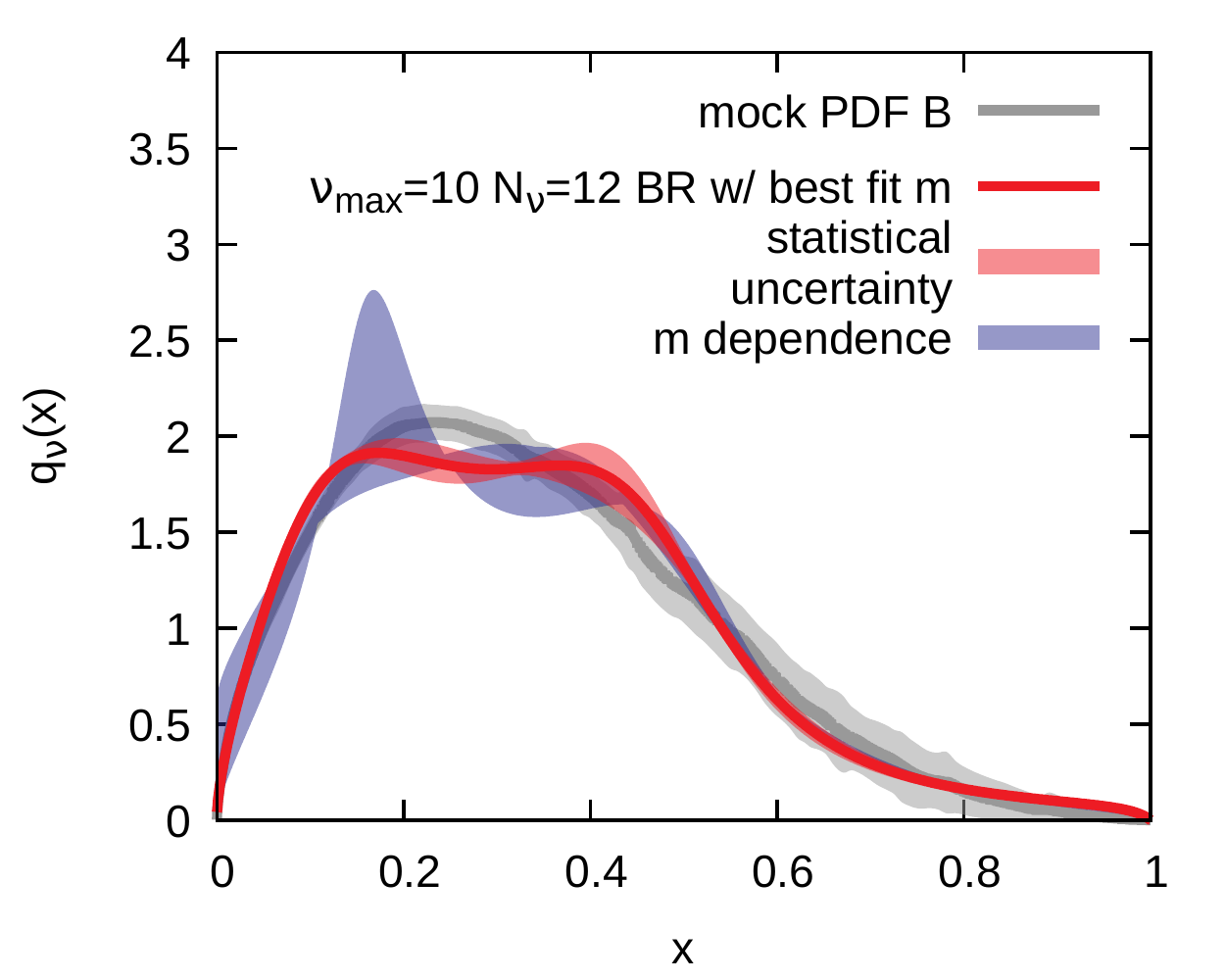}
\includegraphics[scale=0.58]{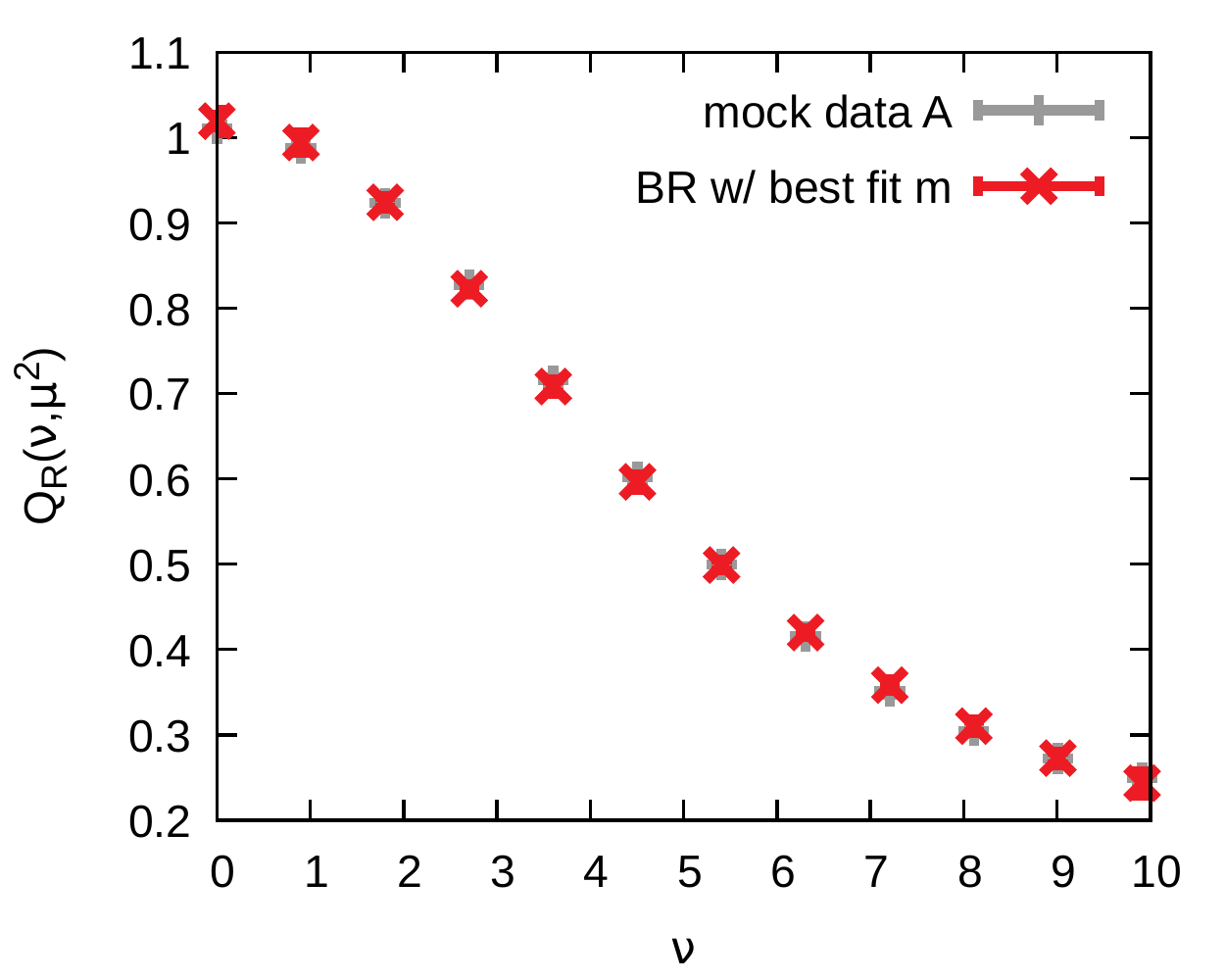}
\includegraphics[scale=0.58]{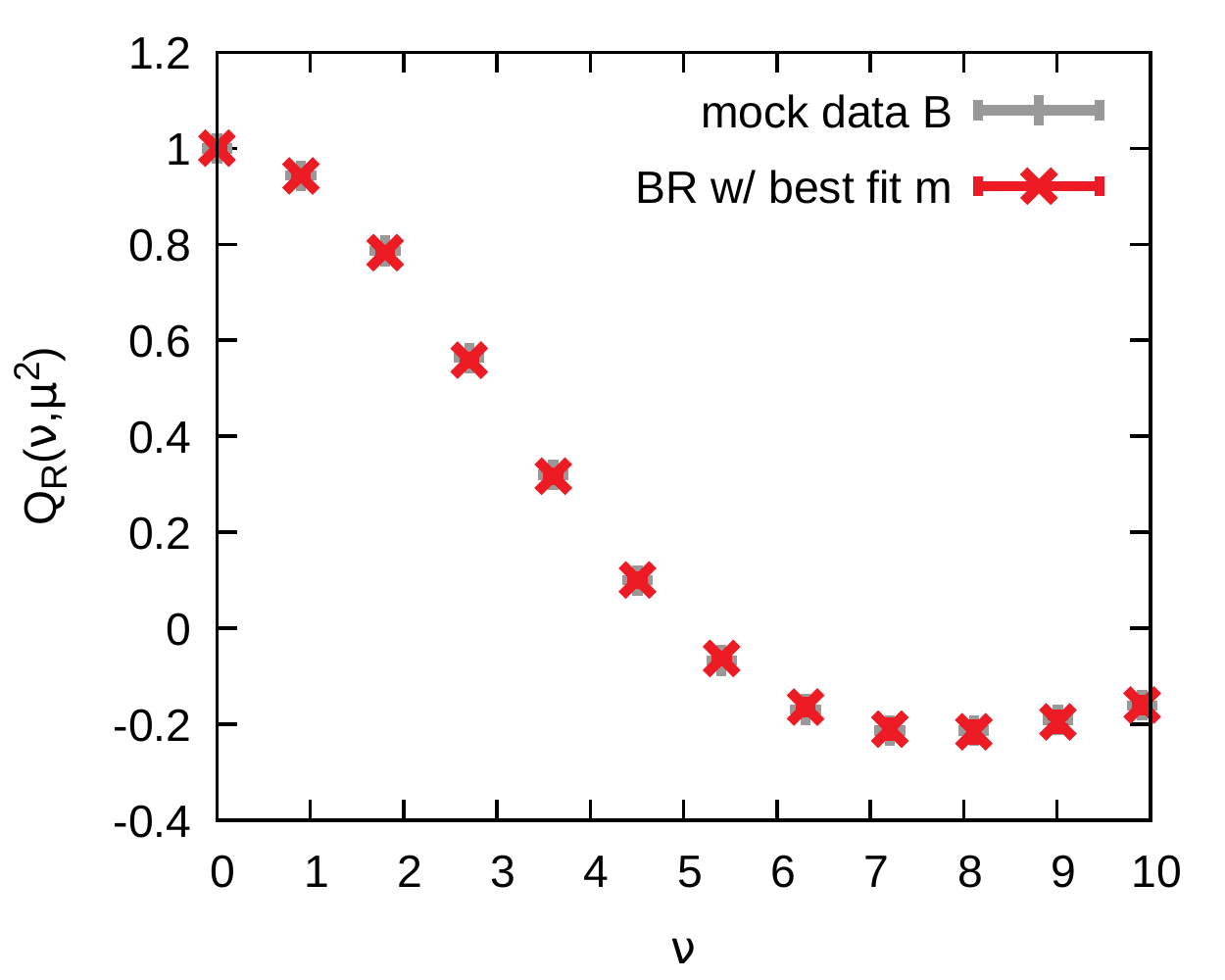}
\caption{$x$-space PDF's reconstructed using the Bayesian reconstruction (BR) method from $N_\nu=12$ Ioffe-time data points on the interval $\nu=[0,10]$ (top) as well as the input data (gray crosses) compared to the data arising from the reconstructed PDF (red crosses) in the bottom panels.}\label{Fig:ResultsBR10}
\end{figure}

Let us compare how the BR method fares. In this method the information of the default model is more weakly incorporated. The corresponding results are shown in Fig.\ref{Fig:ResultsBR10} and yield, as expected, a less accurate reproduction of the input data. While for scenario A it reproduces the PDF within uncertainties, the reconstruction of scenario B shows undulations around the correct result. Close to the maximum at intermediate $x$ the mean is found to be underestimated. 

\begin{figure}[t]
\centering
\includegraphics[scale=0.58]{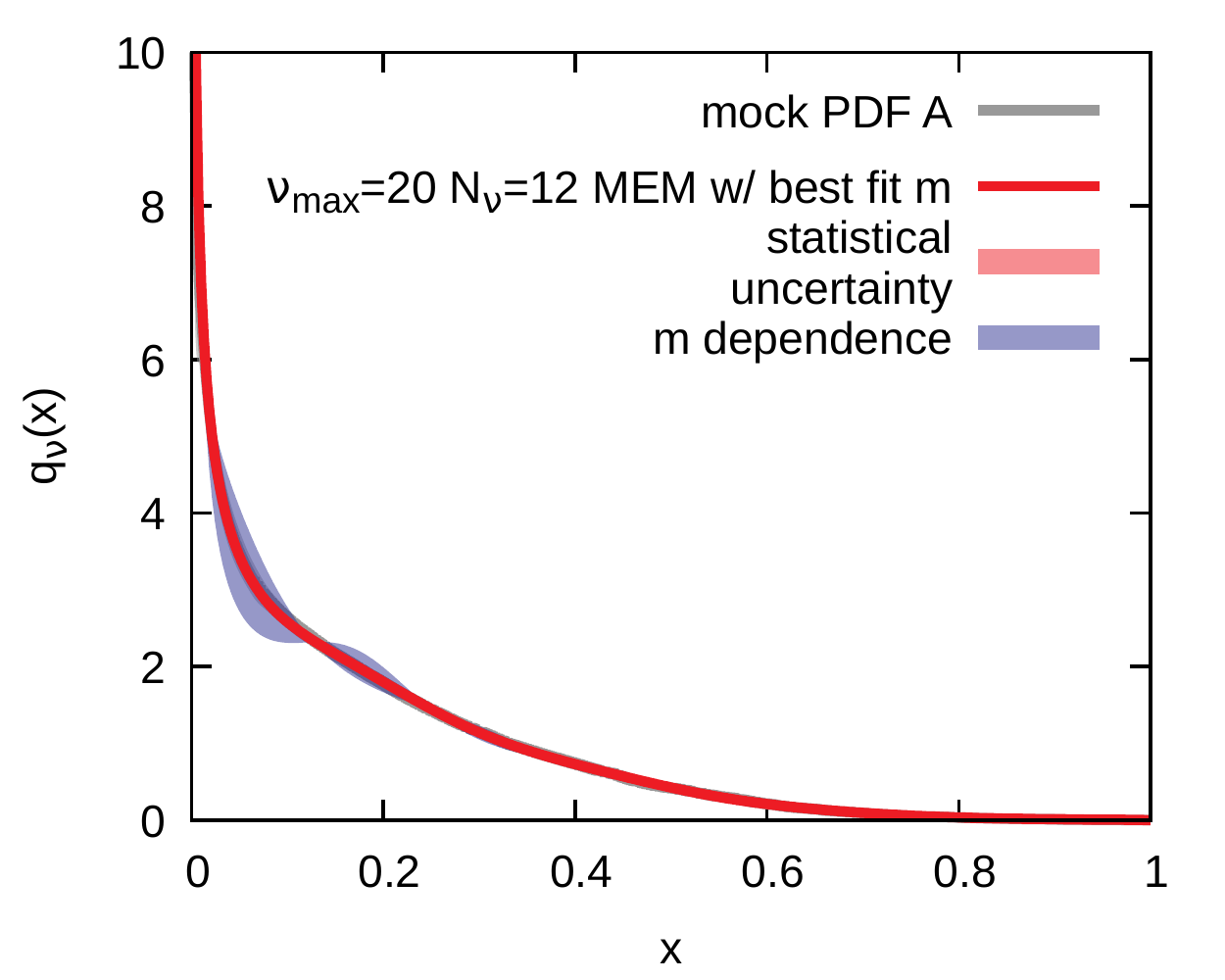}
\includegraphics[scale=0.58]{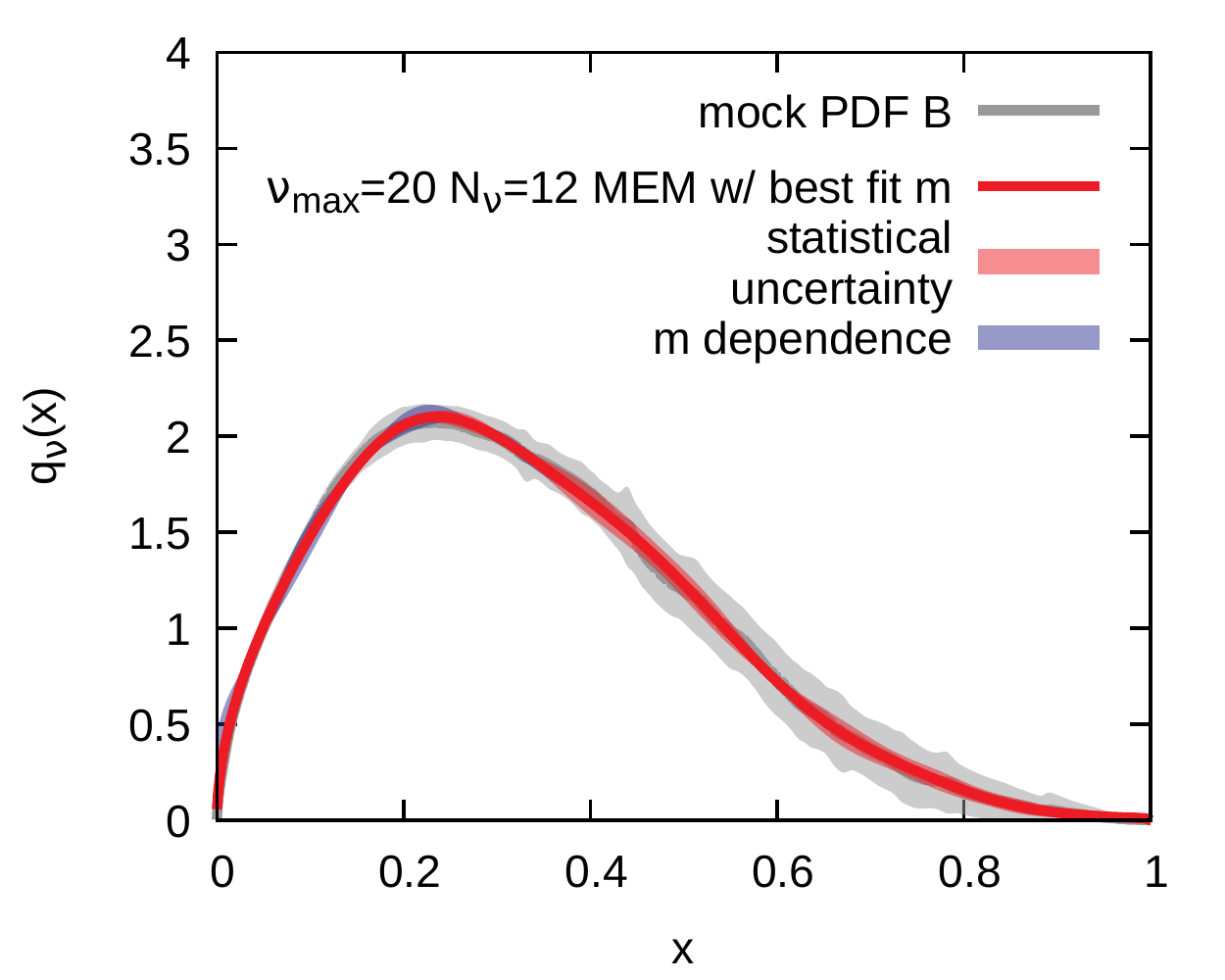}
\includegraphics[scale=0.58]{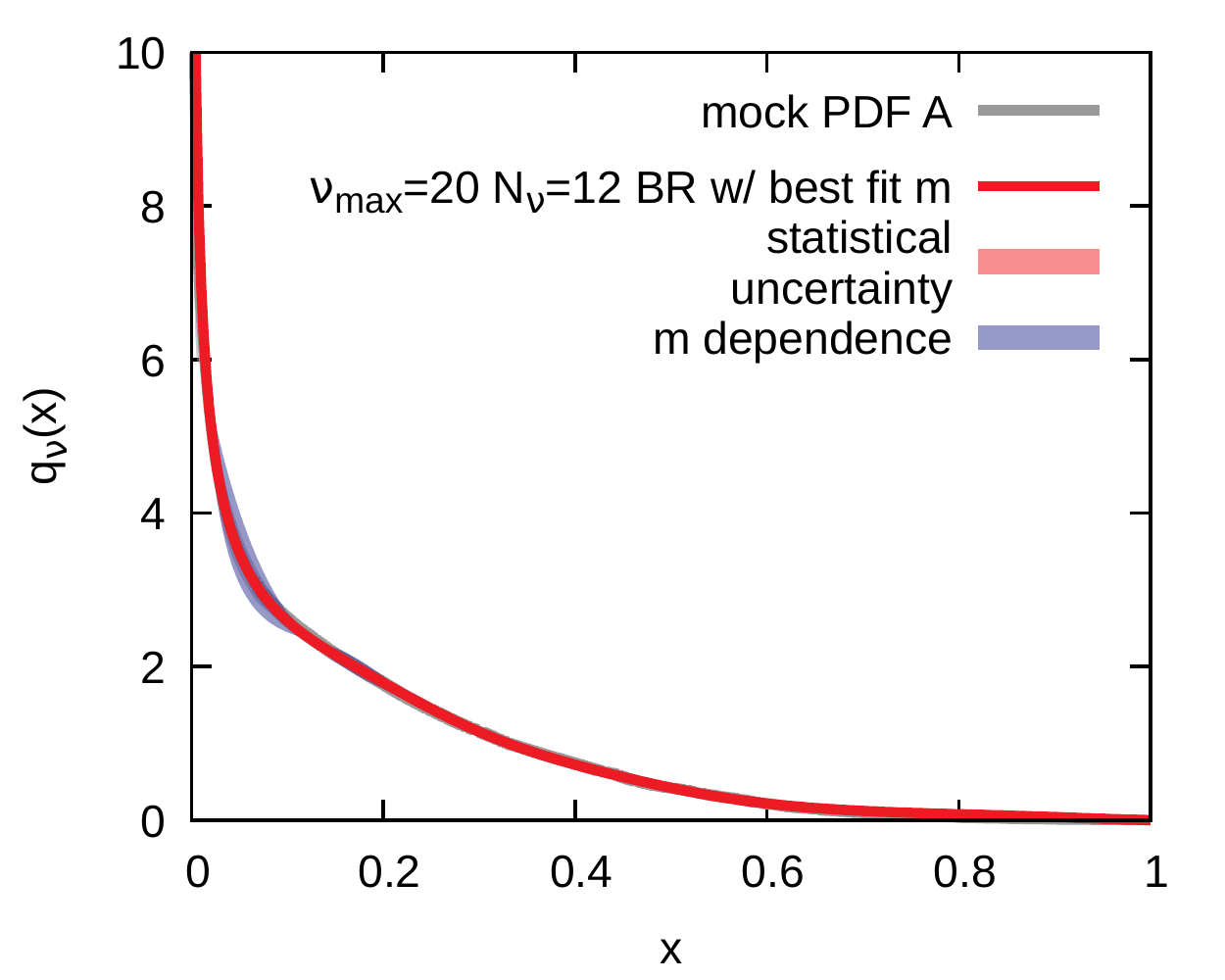}
\includegraphics[scale=0.58]{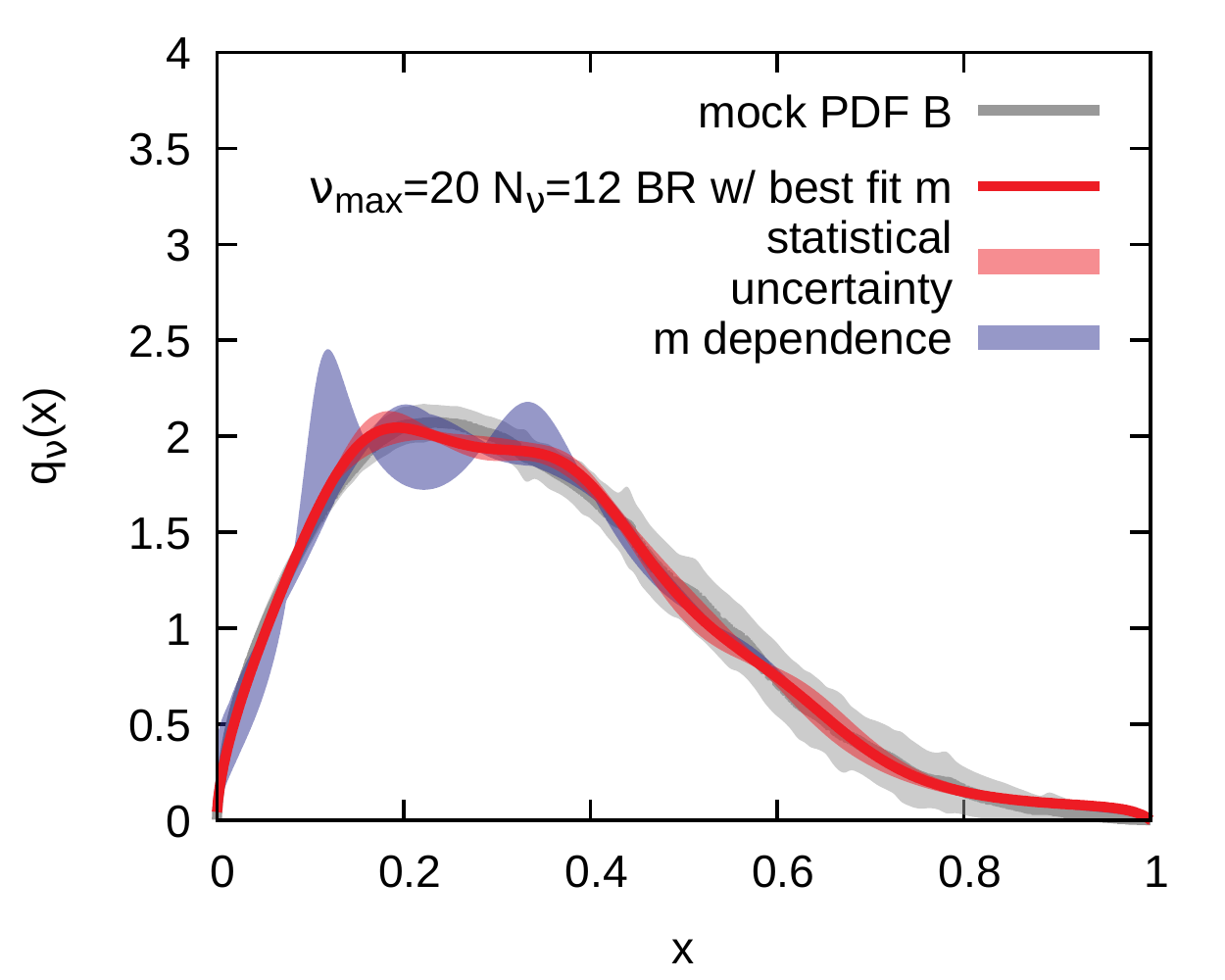}
\caption{$x$-space PDF's reconstructed using the Maximum Entopy Method (top) as well as the Bayesian reconstruction (BR) method (bottom) from $N_\nu=12$ Ioffe-time data points on the interval $\nu=[0,20]$.}\label{Fig:ResultsBRMEM20}
\end{figure}

This result is not surprising but the appearance of differences in the results from different Bayesian regulators urges us to figure out how far we have to improve the input data for both results to provide a reproduction of the correct result within uncertainties. To this end we have  also carried out reconstructions based on the larger interval $\nu=[0,20]$ with the same number of $N_\nu=12$ Ioffe-time data-points. The results of the MEM and BR method are shown in Fig.\ref{Fig:ResultsBRMEM20} and reflect that the improvement in input data has significantly reduced the errors of the reconstruction so that now both methods lie within the uncertainty of the original input data. It is still the case that the MEM utilizes the provided prior information to a stronger extent than the BR method and produces slightly more accurate results in this setting. A further extension of the interval to $\nu=[0,10]$ and increasing the number of available points to $N_\nu=100$ confirms this trend, improving the accuracy of the reconstructions in particular around the maximum of the PDF in scenario B. 

\begin{figure}[t]
\centering
\includegraphics[scale=0.58]{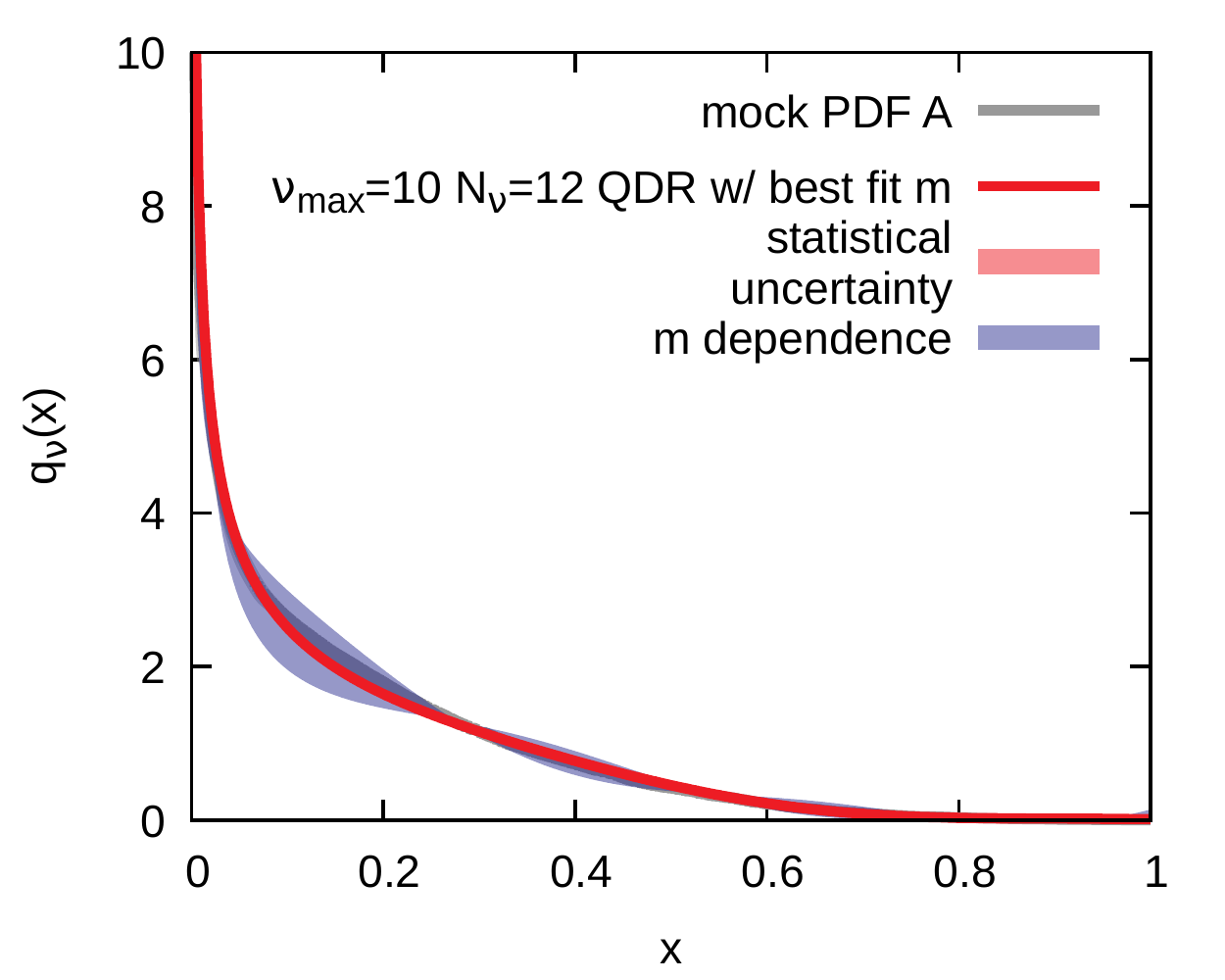}
\includegraphics[scale=0.58]{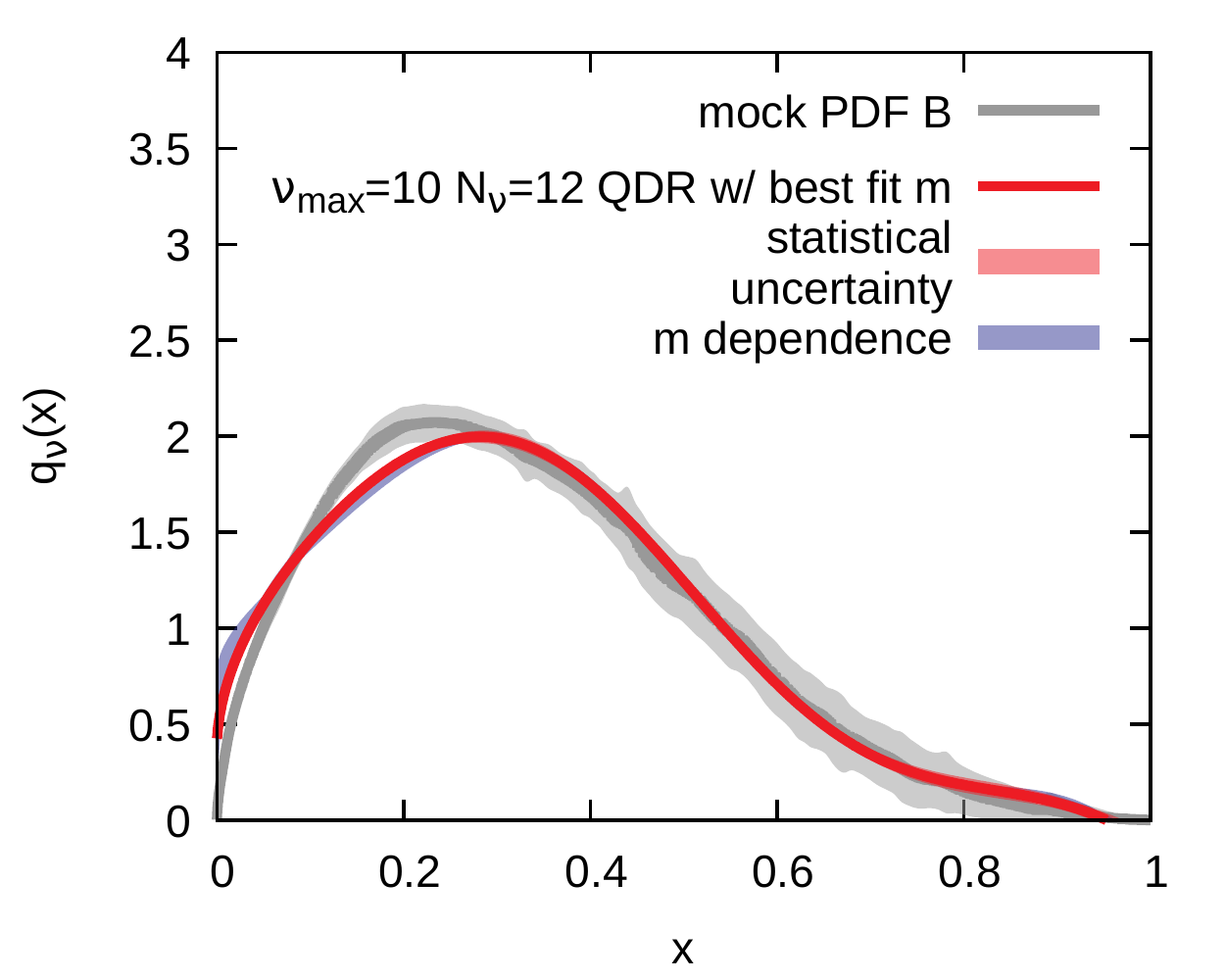}
\includegraphics[scale=0.58]{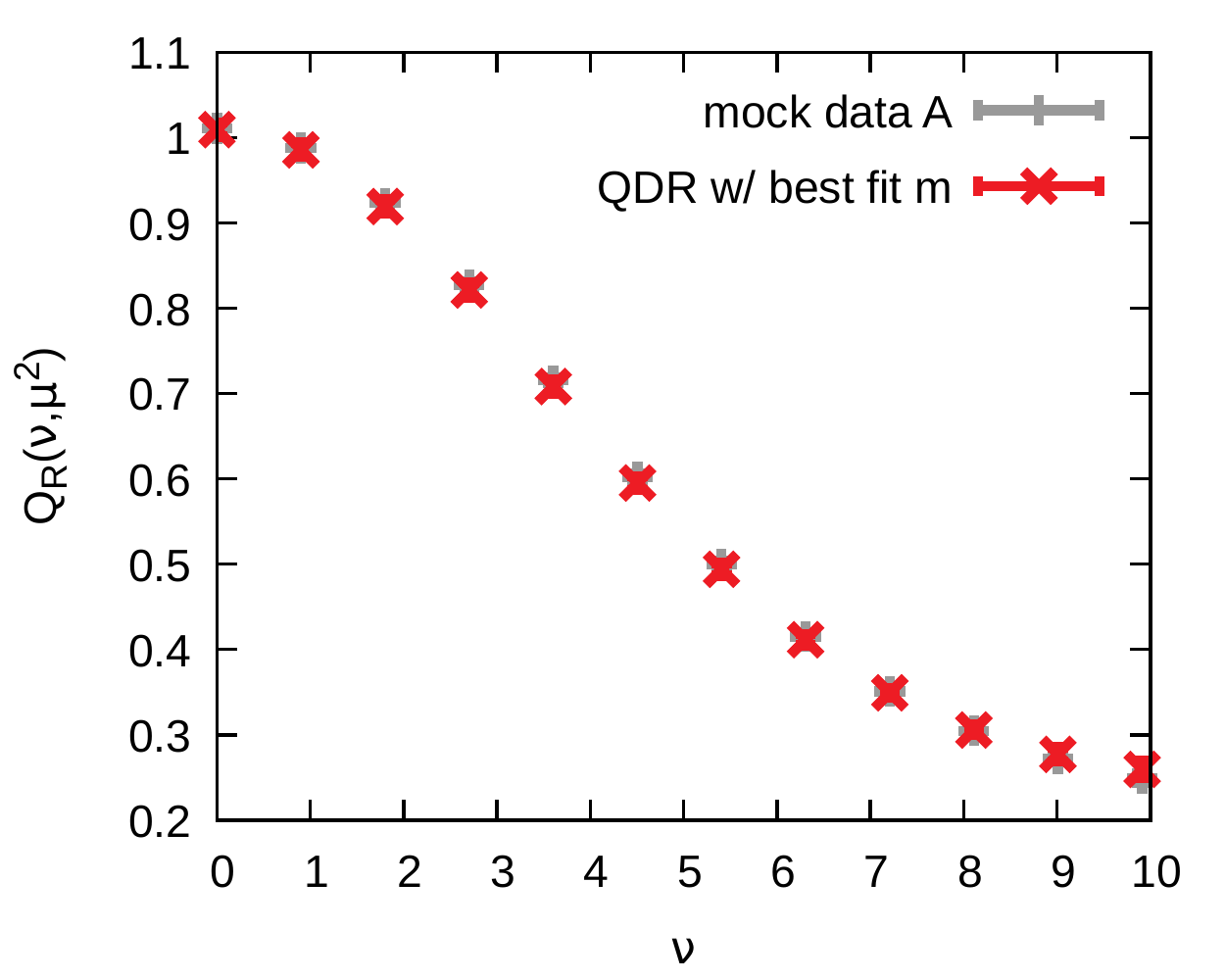}
\includegraphics[scale=0.58]{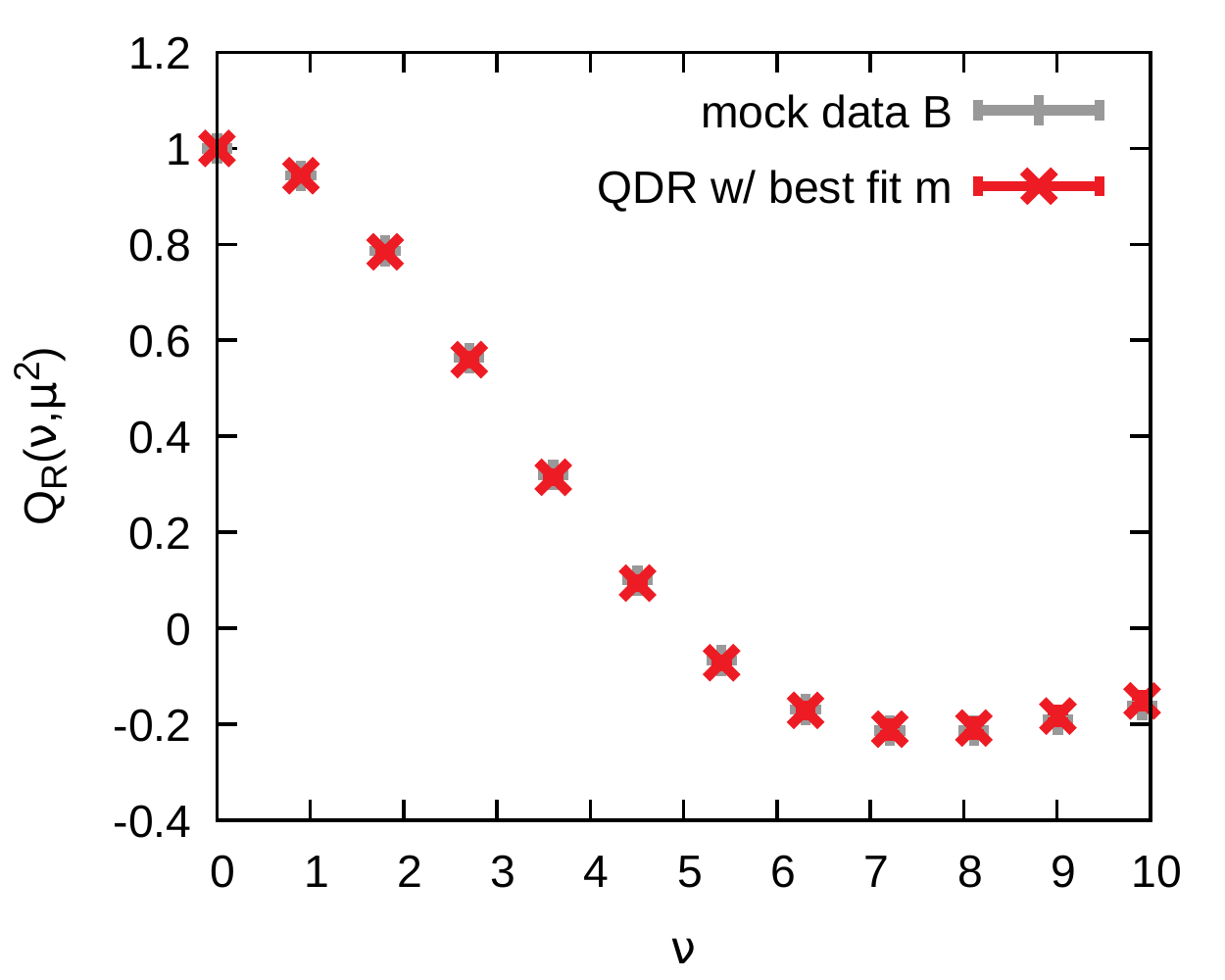}
\caption{$x$-space PDF's reconstructed using a quadratic prior Bayesian (QDR) method from $N_\nu=12$ Ioffe-time data points on the interval $\nu=[0,10]$ (top) as well as the input data (gray crosses) compared to the data arising from the reconstructed PDF (red crosses) in the bottom panels.}\label{Fig:ResultsQDR10}
\end{figure}

Let us turn our attention now to the Bayesian reconstruction methods, which do not require the positivity of the PDF. The first uses the quadratic prior combined with the "historic MEM" procedure to choose the hyperparameter $\alpha$. Results for both the realistic PDF mock scenario A (left column) and the mock scenario B (right column) are given in Fig.~\ref{Fig:ResultsQDR10}.
Since this method also imprints the default model information strongly on the end result, it is not surprising that the reconstruction of scenario A also works excellently here. For scenario B we find that while the overall shape has been reproduced in an acceptable manner, the small $x<0.1$ region is overestimated, while close to the maximum of the PDF the solution underestimates the correct value. It is reassuring to see that with only $N_\nu=12$, the presence of accurate prior information is able to efficiently suppress the many possible oscillatory solutions, which could also reproduce the provided input data.

\begin{figure}[t]
\centering
\includegraphics[scale=0.58]{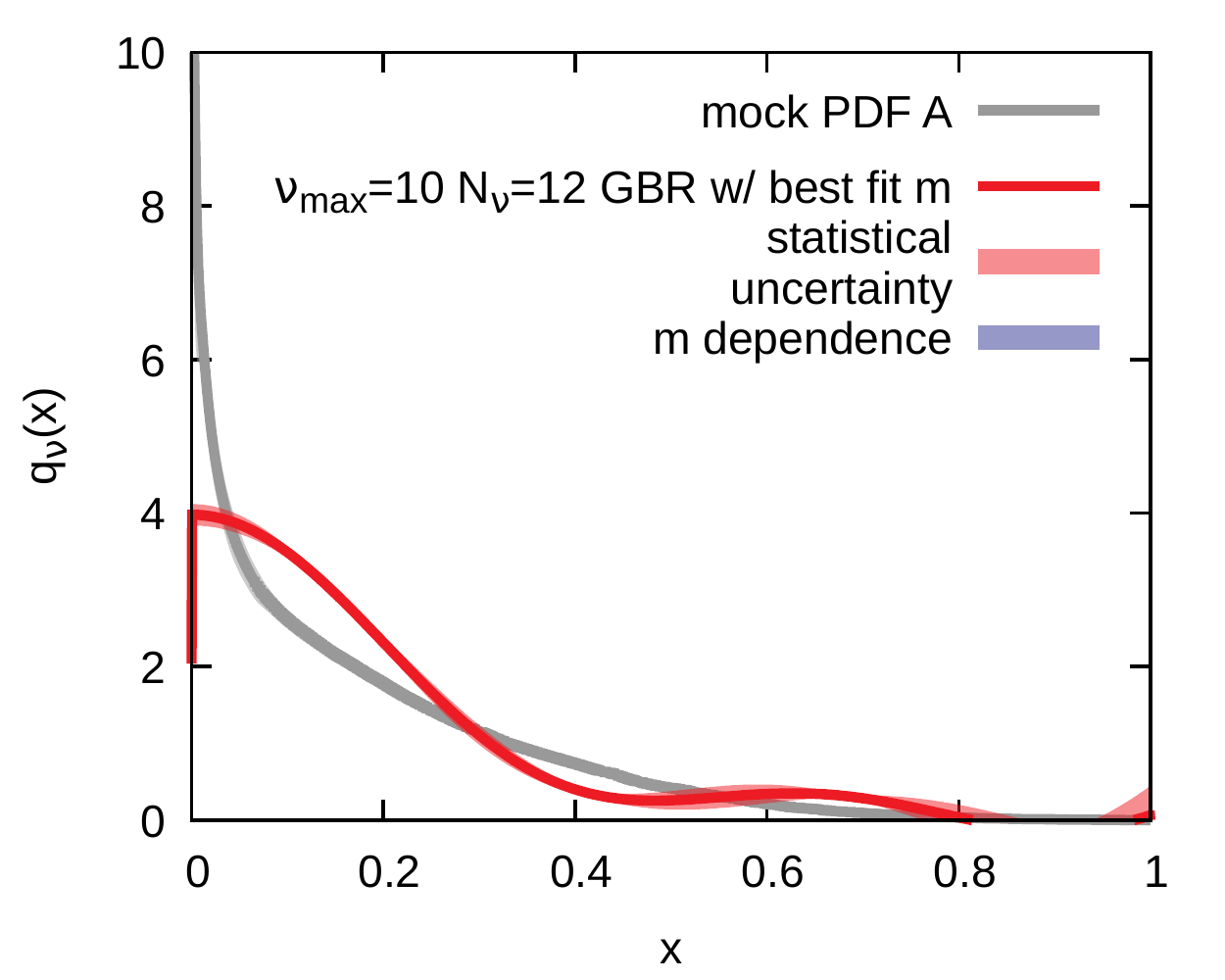}
\includegraphics[scale=0.58]{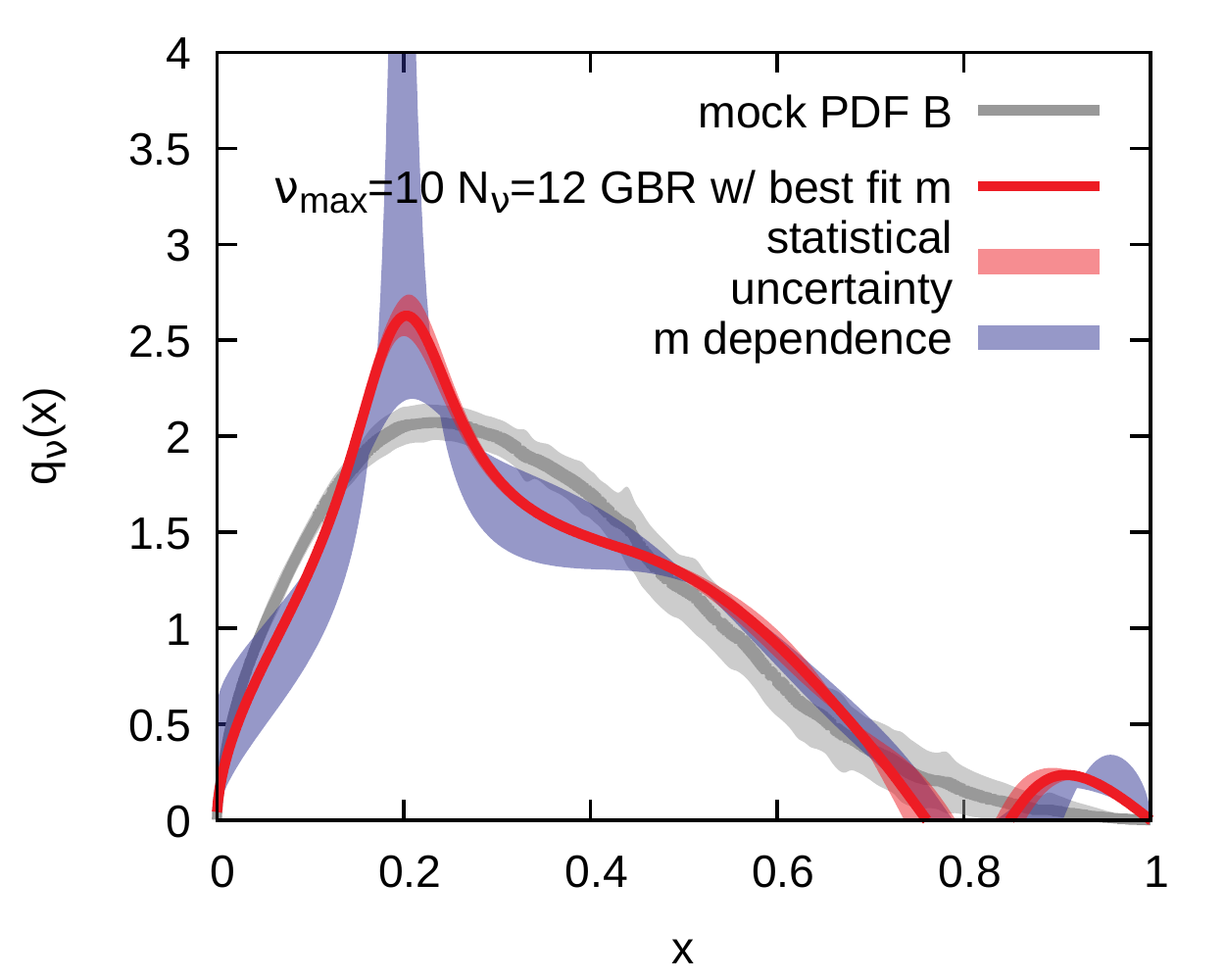}
\includegraphics[scale=0.58]{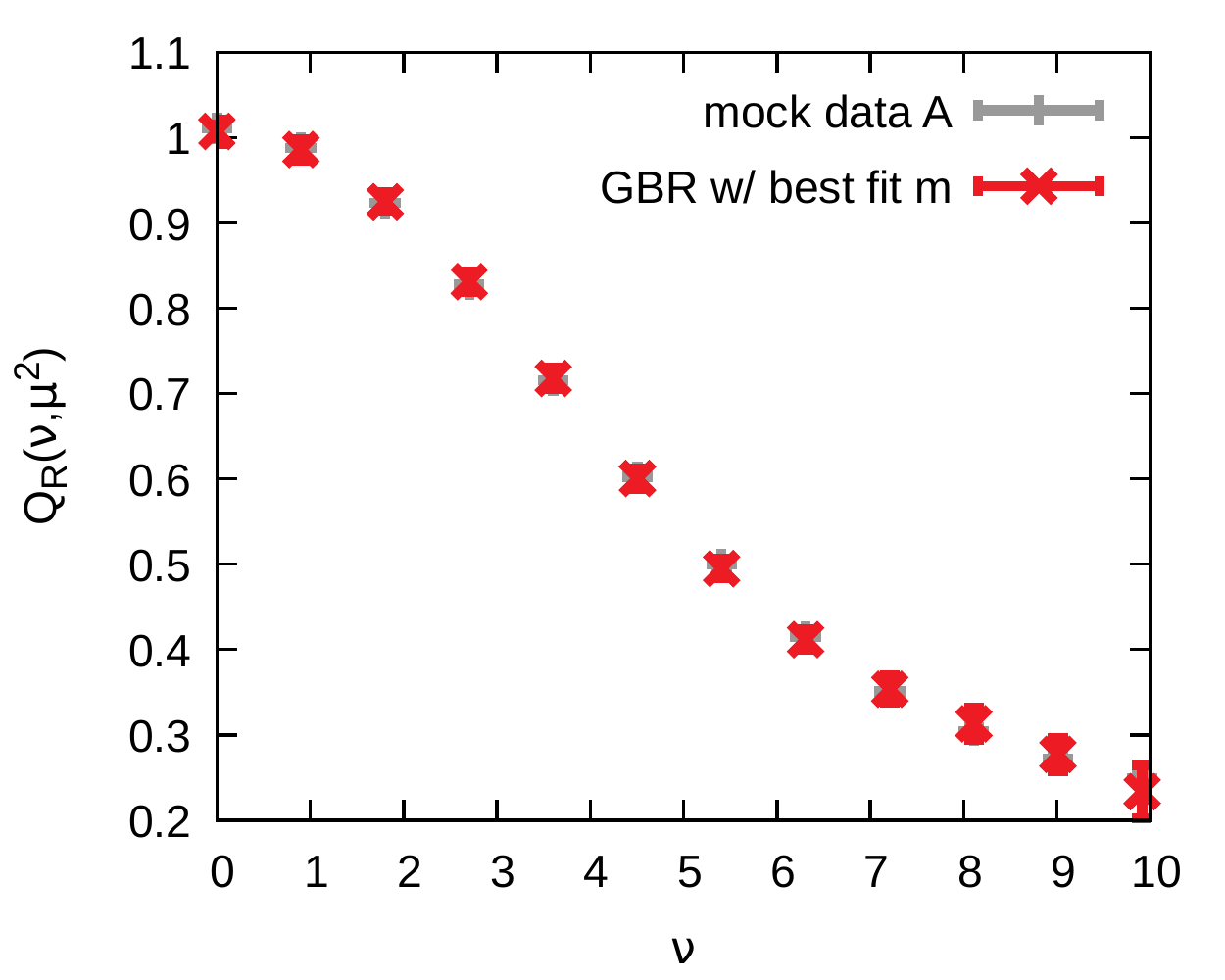}
\includegraphics[scale=0.58]{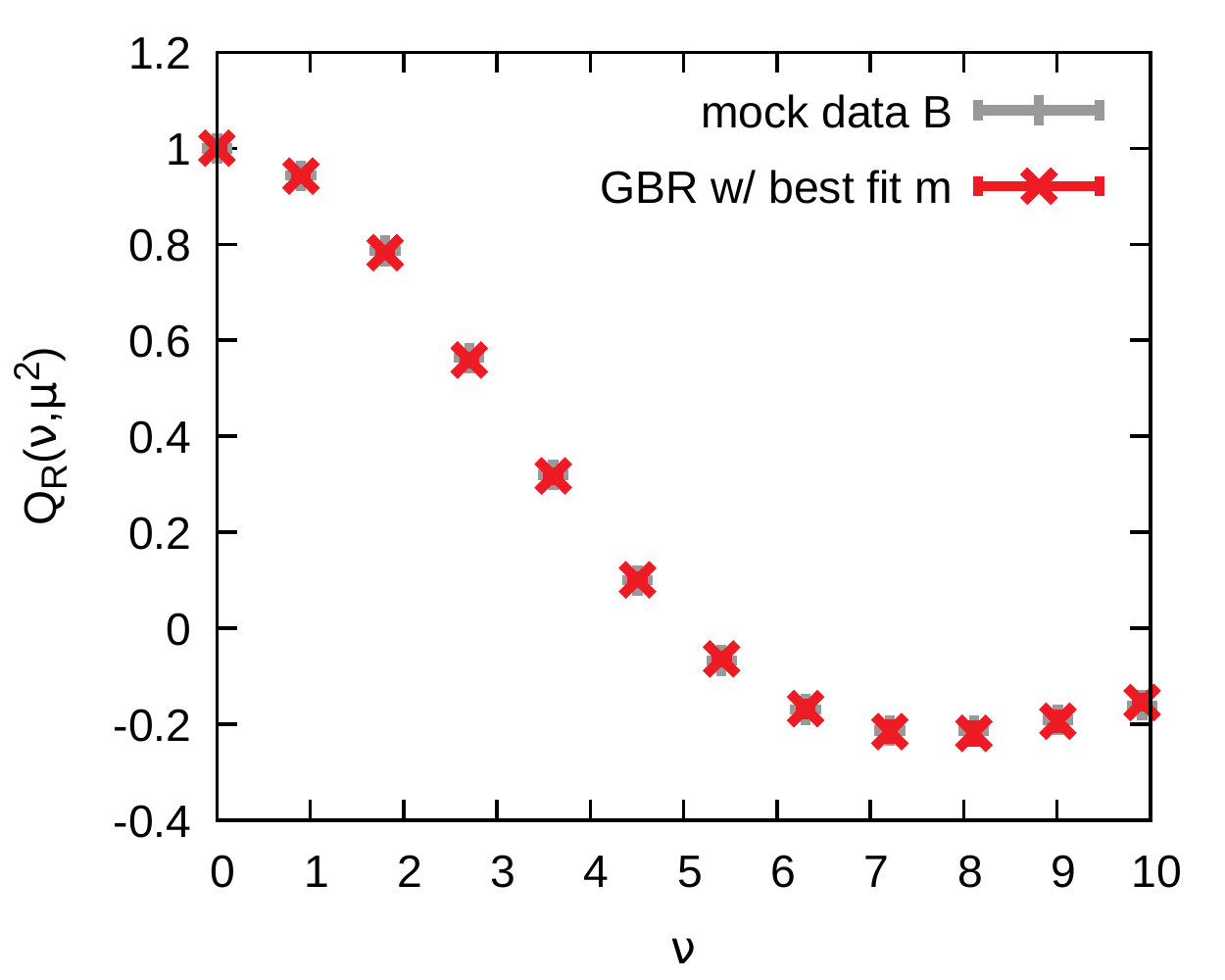}
\caption{$x$-space PDF's reconstructed using the generalized Bayesian reconstruction (BRg) method from $N_\nu=12$ Ioffe-time data points on the interval $\nu=[0,10]$ (top) as well as the input data (gray crosses) compared to the data arising from the reconstructed PDF (red crosses) in the bottom panels. }\label{Fig:ResultsBRg10}
\end{figure}

Let us consider as a last item the reconstruction based on the generalized BR method with $h=m$, whose results are given in Fig.~\ref{Fig:ResultsBRg10}. To understand the performance of this method, we note that it has been designed for cases, where no reliable prior model is available and where thus the influence of $m$ on the end result needs to be held as small as possible. We find that for both scenarios A and B the reconstructions are, as expected, less accurate than for the quadratic prior. In particular we see that excursions into negative values occur at some $x$, which in turn lead to overestimation of values at different $x$. Note that the obtained reconstructions however lead to Ioffe-time data that reproduce the input data very accurately.

It is possible to improve the accuracy of the generalized BR method by changing the value of the default model confidence function $h$, which essentially makes the regulator steeper and steeper, i.e. imprinting the default model more and more strongly. Our intention of showing the generalized BR method with a weak regulator is to emphasize the role accurate prior information has in approaching the correct reconstruction result. 

In summary, we find that due to the availability of accurate prior information on the true shape of the PDF via fits with eq.\eqref{eq:p-func}, Bayesian methods that are designed to imprint prior information strongly on the end result outperform those that minimize the influence of the default model. I.e. in case of a rather small interval $\nu=[0,10]$ and number of data points $N_\nu=12$, the Maximum Entropy Method provides the most accurate results for both scenario A and B (Fig.~\ref{Fig:ResultsMEM10}). The BR method is competitive in scenario A but shows a larger deviation from the true result for scenario B (Fig.\ref{Fig:ResultsBR10}). Increasing the extent of the $\nu$ interval and the number of data points consistently shows that both methods approach the correct solution in the "Bayesian continuum limit". 

When comparing methods without the requirement of positivity we find that the quadratic prior still fares excellently with scenario A but shows some deviations, in particular at small $x$ values in scenario B. As expected and by construction, the generalized BR method with the choice $h=m$ shows the weakest dependence on the used default model but at the same time also provides the least accurate reconstructions.

The success of the MEM relies on the availability of very good prior information. On the other hand the BR method is constructed to imprint the default model on the end result in a weaker fashion than the MEM (lower curvature of the prior functional). Hence we plan to deploy it together with the MEM when investigating actual lattice QCD data in the future. The reason is that as long as the asymptotics of the PDF at $x=0$ and $x=1$ are adequately provided, the BR method is expected to be able to maintain its reconstruction quality in case that in a genuine lattice data based reconstruction the quality of the default model at intermediate $x$ values will be worse than in the above tests.

If on the other hand a non-positive $q$ function is concerned and accurate prior information is available, then we can utilize the quadratic prior even in cases that only small number of data points are available. The generalized BR method to become competitive requires a more careful treatment of the default model confidence function $h$ in this context.

\subsection{Restricted $\chi^2$ sampling}

In order to clarify the role of different prior information on the
success of the reconstruction, we perform a further numerical experiment
on our mock data sets. Here we consider positivity as the only property
of the PDF known a priori. The ill-posedness of the inversion is related
to the many local extrema in the likelihood. If we work with a constant
prior probability, the resulting posterior probability will just be a
flat distribution, unable to provide meaningful insight. This fact has been explicitly tested and confirmed. It is interesting to then
ask, whether the restriction to positive PDF's limits the
number of local extrema in $L$, such that all of their contributions
taken together in a statistical fashion will lead to a meaningful posterior?

To explore this idea, the MC-Stan library, a modern tool for statistical inference, can be used.
It implements an efficient hybrid Monte-Carlo
algorithm, specifically a no-U-Turn sampler, which allows the sampling of a wide variety of
joint probability distributions of random variables. The applicability
of the machinery of HMC arises from identifying its Hamiltonian with the
logarithm of the joint probability distribution (for technical details
see Ref.~\cite{MCStan}). Using its high level programming language, the test can be
formulated by identifying each of the (appropriately decorrelated)
Ioffe-time mock data points with an individual Gaussian distribution.
Its mean is expressed in term of the PDF $q(x)$ and the cosine
integral Kernel, the spread by the corresponding data uncertainty.
Positivity of the PDF is enforced by restricting its sampling to values
larger than zero a priori. The complete STAN model reads

\begin{verbatim}
data {
    int NNu; int Nx;  matrix[NNu, Nx] Kernel;
    vector[NNu] Q;
    vector[NNu] Uncertainty;
}
parameters {
    vector<lower=0>[Nx] q;
}
model {
    Q ~ normal(Kernel * q, Uncertainty);
}
\end{verbatim}

\begin{figure}[t]
\centering
\includegraphics[scale=0.58]{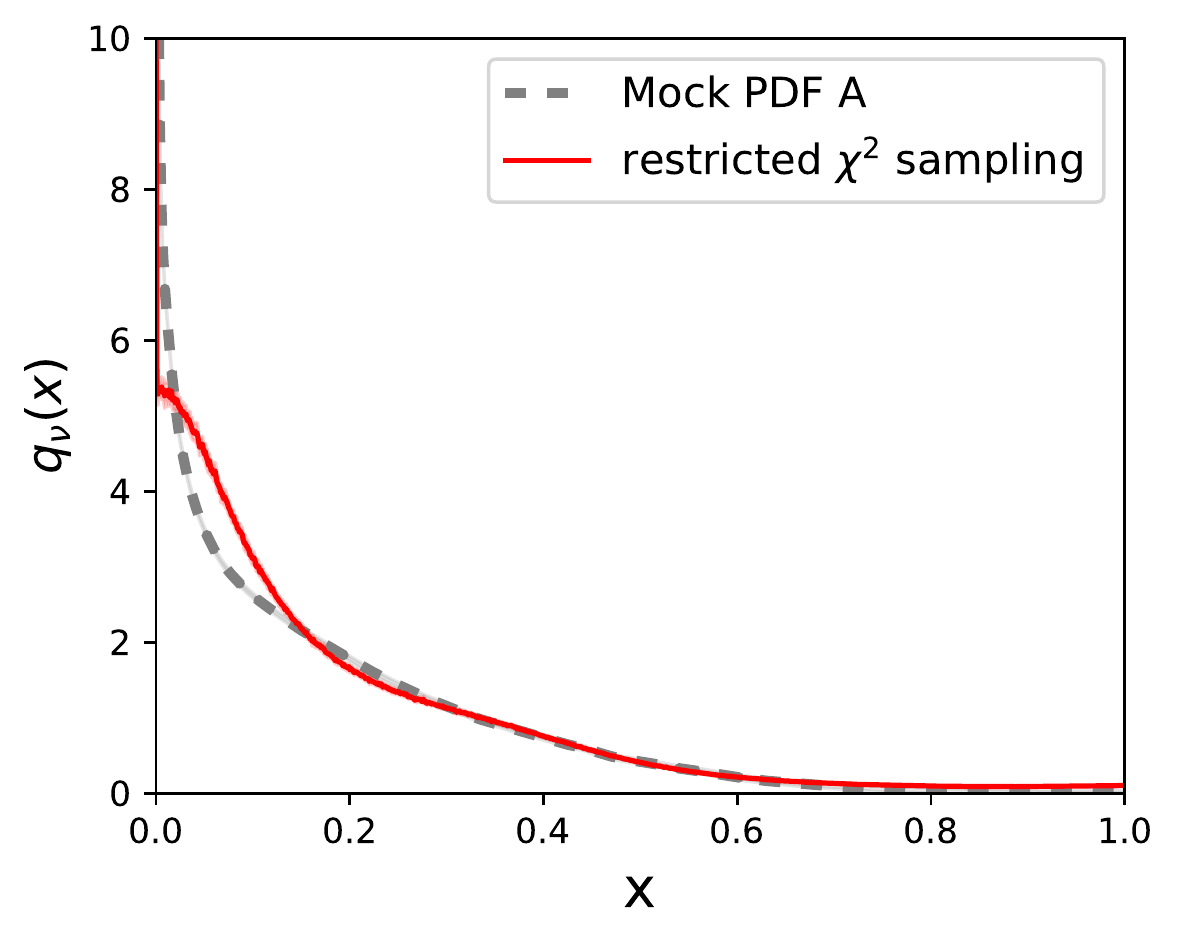}
\includegraphics[scale=0.58]{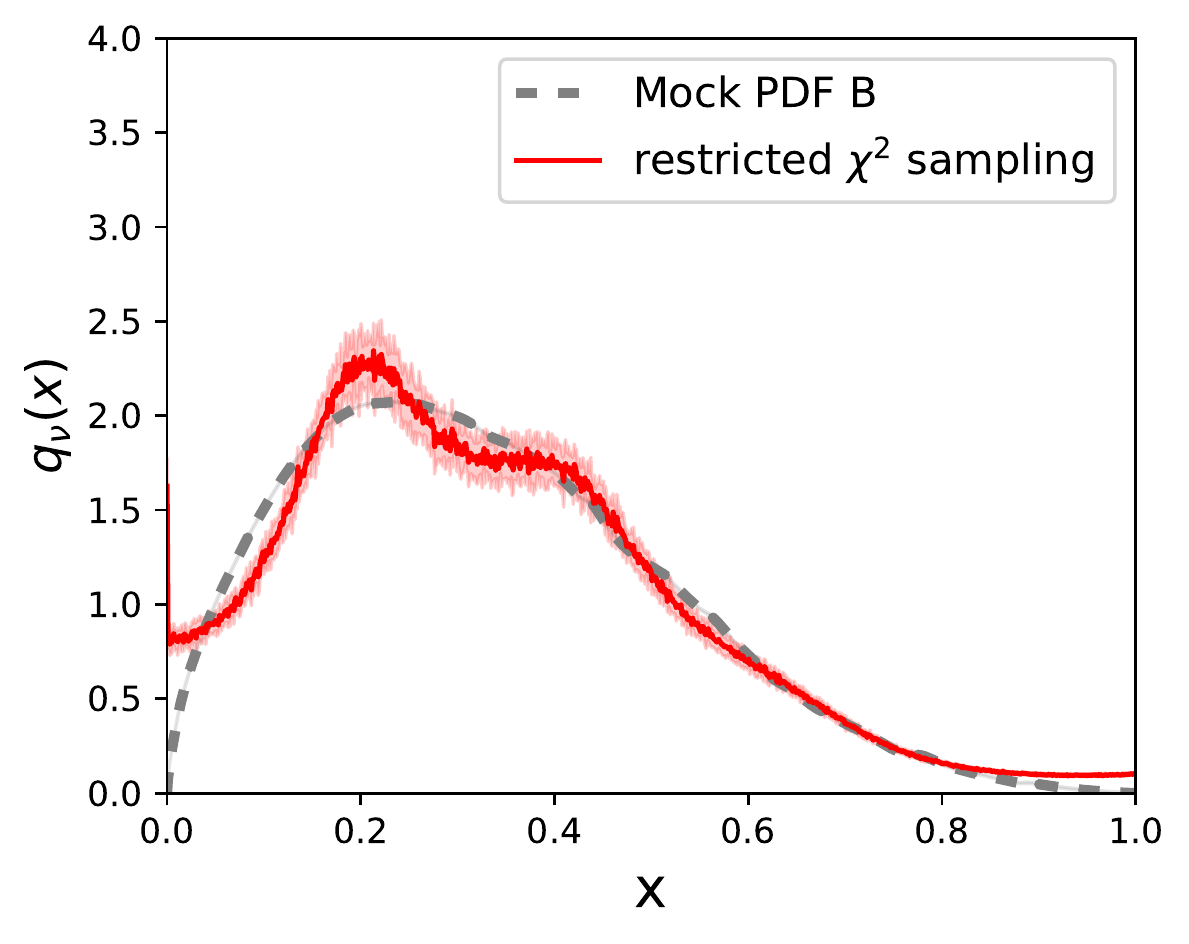}
\caption{Reconstructed PDF (red dashed) from restricted sampling of the $\chi^2$ functional based on $N_\nu=12$ Ioffe-time data points on the interval $\nu=[0,10]$. Mock scenario A is shown on the left, scenario B on the right. While positivity alone already provides a powerful regularization, for this realistic scenario we do observe deviations from the correct result, especially for scenario B.}
\label{Fig:StanResults2}
\end{figure}

\begin{figure}[t]
\centering
\includegraphics[scale=0.58]{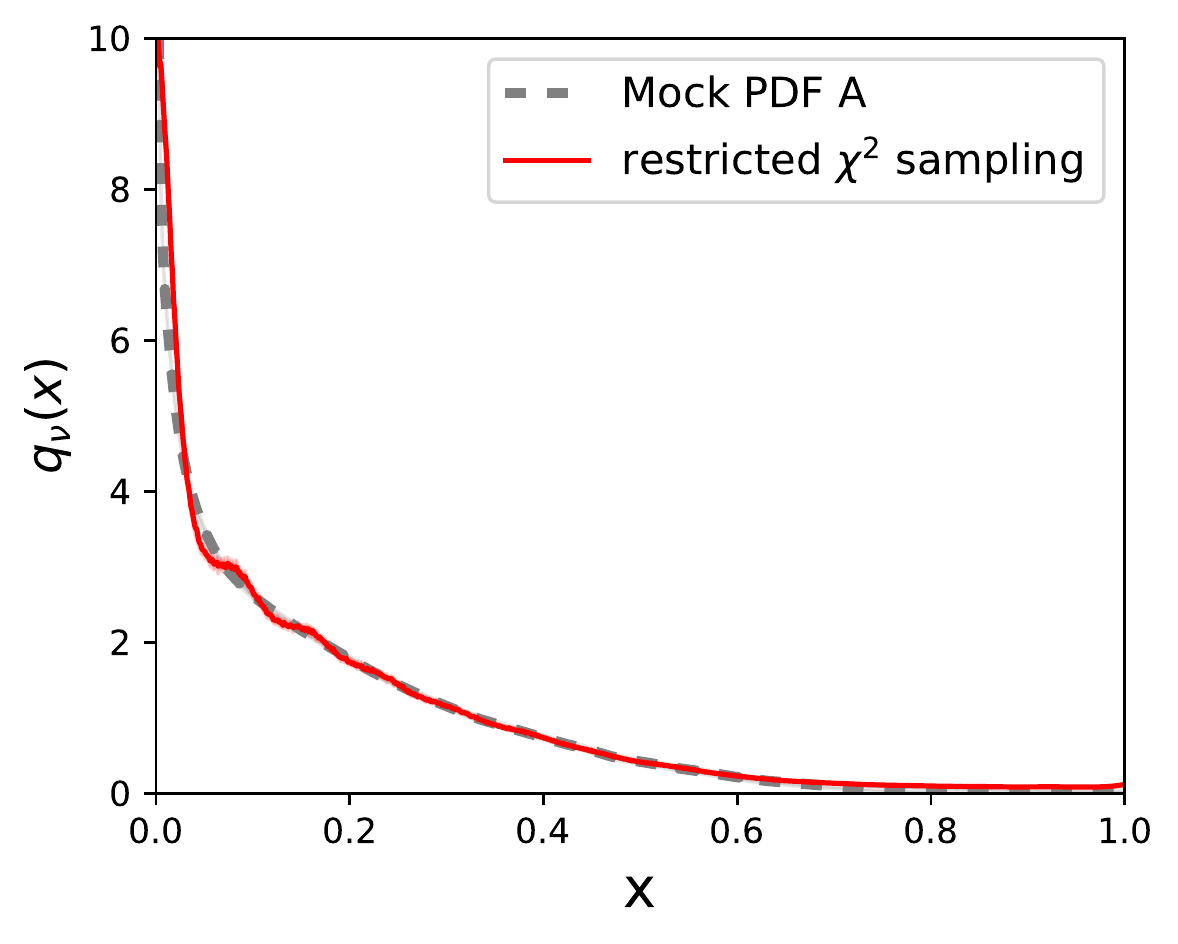}
\includegraphics[scale=0.58]{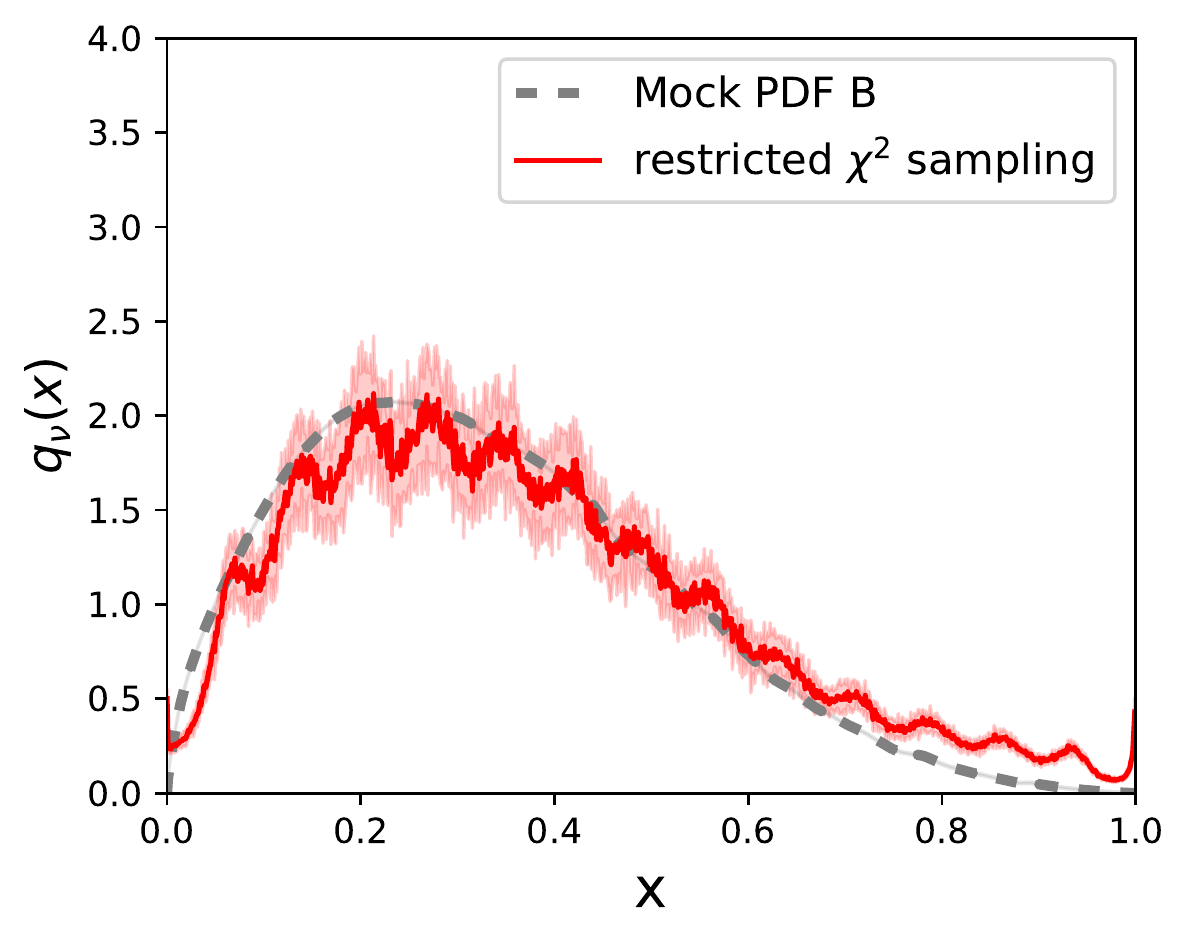}
\caption{Reconstructed PDF (red dashed) from restricted sampling of the $\chi^2$ functional based on $N_\nu=100$ Ioffe-time data points on the interval $\nu=[0,100]$. Mock scenario A is shown on the left, scenario B on the right. In this almost ideal setting, positivity alone provides a regularization that is sufficient for a reasonable determination of the PDF of scenario A, while it is unable to suppress ringing artifacts in scenario B.}
\label{Fig:StanResults1}
\end{figure}

For the Kernel we discretize the x-range with $N_x=1000$ points and
consider two sets of input data for each scenario A and B. As realistic test case we use again the a $\nu\in[0,10]$ with $N_\nu=12$ points (see Fig.~\ref{Fig:StanResults2}), exactly as has been used in the Bayesian reconstruction, as well as the ideal case of $N_\nu=100$ points with $\nu_{\rm max}=100$ (see Fig.~\ref{Fig:StanResults1}). The sampling is performed over 20 chains of HMC trajectories, each with 1500 steps in MC time, 500 of which are discarded as warmup phase. The mean value of $q(x)$ over the 20 chains is shown as the red data points, while its erroband is obtained from the usual variance among different chains. As we do not imprint any further prior information, execept positivity, and we assume that positive holds exactly, there is no further systematic error to be considered for this reconstruction.

We find that positivity alone is a powerful regulator, as sampling of
the restricted $\chi^2$ functional reveals. Even though we did not
specify any further constraints, the posterior probability for each
$q(x_i)$ leads to a well defined mean, its variance of course depends on
the number and quality of the provided input data. In the ideal case of
$\nu_{\rm max}=100$ we manage to reproduce the PDF for scenario A excellently,
for scenario B however ringing artifacts remain clearly visible. On the other
hand $\nu_{\rm max}=20$, as expected leads to results that are less 
accurate than those obtained by the advanced Bayesian reconstruction 
methods or the neural network.

We thus learn that the additional prior information encoded in the
Bayesian reconstruction manages to provide meaningful regularization to
the problem beyond positivity. At the same time it also tells us that
the neural network approach is able to extract more relevant prior
information for a regularization than just positivity, i.e. it goes
beyond a simple $\chi^2$ fit.

\section{Summary and Conclusion}

In this work, we studied the general problem of extracting the full $x$-dependence of PDFs or DAs which are defined by the Fourier transform of a position space hadronic matrix element computed by means of lattice QCD. We have identified the two main challenges that one encounters during the computation of the aforementioned Fourier transform. The first one is related to the range of integration being restricted compared to a genuine Fourier transform and the second issue is that one has to perform such a task with only a few data points at hand. We explicitly have showed direct inversion methods fail at performing their task and the same also holds for trivial modifications thereof. However, as we demonstrated, advanced reconstruction methods open new paths towards obtaining PDFs and DAs from lattice QCD data.  The methods that we tested here are the Backus-Gilbert method, a reconstruction which is based on neural networks, and a set of Bayesian reconstructions, including the MEM and the BR method.

We explicitly tested all these methods using mock input data computed using PDFs extracted from experiment that diverge at $x=0$ as well as an artificial variation of these PDFs where the behavior at small $x$ was modified to make the PDF vanish at $x=0$. It is clear that one can not probe the $x=0$ region employing calculations at finite lattice spacing, which imposes a natural cutoff to the highest available momentum, in plain analogy to a scattering experiment. However, one can use  advanced reconstruction methods that can determine the unknown PDFs with well defined estimates of the uncertainties. 

In order to obtain a realistic impression of the method efficiencies we carried out an analysis using $\mathcal{O}(10)$ points for $\nu \in [0,10]$. We find that such a limited set of data, which can be reached using current lattice QCD calculations, already leads to satisfactory reconstruction results for the underlying PDF for $x>0.1$.
The goal to double the available amount of Ioffe time input points to $\mathcal{O}(20)$ is realistic and achievable in the near future, given reasonable computational resources. However, tripling or quadrupling the maximum Ioffe time accessible would require a concerted effort by the lattice community working on PDFs. We have explicitly shown that an increase in the available range of $\nu$ values to $\nu_{\rm max }=20$ significantly further reduces the uncertainty of the reconstruction for all three tested approaches, the Backus-Gilbert method, the neural network reconstruction, as well as the Bayesian methods.

For the case of the Backus-Gilbert reconstruction we saw that preconditioning the ill-defined problem was mandatory for obtaining the correct result in the small-$x$ region. With a preconditioning that incorporated many of the relevant features of the final result, already a naive application of the algorithm was sufficient for obtaining a good reconstruction of the intermediate and large-$x$ region. An advantage of the Backus-Gilbert method is its simplicity (linear problem) and the fact that it is numerically the cheapest of all to implement. On the other hand the relative lack of freedom in terms of preconditioning due to restrictions being imposed by the requirement of convergence of the involved integrals represents a minor downside of this approach. 
 
The method of neural network parametrization also provided faithful reconstructions and has thus been shown to be a competitive candidate for solving the inverse problem for PDFs on the lattice. In this study, we have only explored a limited number of its facets, noticing that all three different geometries that we tested lead to fully equivalent results for the reconstructions of scenarios A and B. Furthermore, a small improvement of the reconstruction in the small x region can be achieved when preconditioning is employed as in the Backus-Gilbert method.
 
The Bayesian methods included the traditional Maximum Entropy Method (MEM) and the Bayesian reconstruction (BR) method. A fit of the input data with eq.\eqref{eq:p-func} already provides a very good estimate of the shape of the underlying PDF and this information can be incorporated into the Bayesian methods as a default model. Hence, as expected, those Bayesian methods that are designed to imprint the information of the default model more strongly onto the end result (e.g. MEM) outperform those designed to keep the influence of the default model to a minimum (e.g. BR). We found that already with $\mathcal{O}(10)$ points for $\nu \in [0,10]$ the MEM excellently reproduced the PDF of both scenario A and scenario B. The BR method is competitive in scenario A but showed larger deviations from the correct result for scenario B. Extending the range of $\nu$ to larger values and providing more input points consistently improves the reconstructions, approaching the correct result in the "Bayesian continuum limit".

When utilizing Bayesian methods that do not presuppose positive PDFs, the inverse problem is even more ill-posed. We have found that the quadratic prior Bayesian method performs remarkably well, as it also imprints the information provided by the default model strongly on the end result. It thus outperformed the generalized BR method for the case of a small number of input points $\mathcal{O}(10)$. Also in for general PDFs, increasing the range in available $\nu$ significantly reduced the uncertainties and made the results of different methods approach each other, as well as the correct result.

In order to further explore the role of prior information we performed an HMC sampling of the associated $\chi^2$ functional employing the MC-Stan library. These studies indicated that the positivity constraint alone is a powerful regulator. At the same time we conclude that the additional prior information included in Bayesian methods, the Backus-Gilbert method and also the neural network approach are essential in producing a faithful reconstruction on the underlying PDF. Further studies using the HMC sampling approach are underway both in BR-MEM and the neural network methods. 
  
We plan to apply the reconstruction approaches discussed in this paper in future publications that will employ realistic results from lattice calculations. We encourage other lattice practitioners to implement these methods, because in our opinion the systematic artifacts related to the Fourier transform among the numerous systematics that one has to face in the lattice studies of PDFs, are the ones that can not be dealt by brute force methods. This point has been discussed detail in~\cite{Rossi:2017muf,Rossi:2018zkn,Radyushkin:2018nbf,Rossi:2018stq,Karpie:2018zaz}, however our methods presented here completely avoid the difficulties that arise from the inverse Fourier transform that is required in other approaches. 

\acknowledgments
We thank  Anatoly Radyushkin for especially stimulating and enlightening discussions through out this work. Fruitful discussions with Luigi del Debbio are also acknowledged. 
 This work  has been supported
by the U.S. Department of Energy through Grant Number DE- FG02-04ER41302, and 
through contract Number DE-AC05-06OR23177, under which JSA operates the Thomas
Jefferson National Accelerator Facility. SZ acknowledges support by the DFG Collaborative Research Centre SFB 1225 (ISOQUANT).
K.O. acknowledges support in part by STFC consolidated grant ST/P000681/1, and the hospitality from DAMTP and Clare Hall at Cambridge University, where this work was performed. JK acknowledges support from the U.S. Department of Energy, Office of Science, Office of Workforce Development for Teachers and Scientists, Office of Science Graduate Student Research (SCGSR) program. The SCGSR program is administered by the Oak Ridge Institute for Science and Education for the DOE under contract number DE-SC0014664.  This work was performed in part using
computing facilities at the College of William \&
Mary which were provided by contributions from the National
Science Foundation (MRI grant PHY-1626177),
the Commonwealth of Virginia Equipment Trust Fund
and the Office of Naval Research. In addition, this
work used resources at NERSC, a DOE Office of Science
User Facility supported by the Office of Science of the
U.S. Department of Energy under Contract \# DE-AC02-
05CH11231 as well as computing resources provided by  
UNINETT Sigma2 - the National Infrastructure for High Performance Computing and Data Storage in Norway under project NN9578K-QCDrtX "Real-time dynamics of nuclear matter under extreme conditions"

\bibliographystyle{apsrev}
\bibliography{inversion}

 \end{document}